\documentclass[fleqn,usenatbib]{mnras}

\usepackage{newtxtext,newtxmath}

\usepackage[T1]{fontenc}

\DeclareRobustCommand{\VAN}[3]{#2}
\let\VANthebibliography\thebibliography
\def\thebibliography{\DeclareRobustCommand{\VAN}[3]{##3}\VANthebibliography}

\usepackage{graphicx}	
\usepackage{adjustbox}  
\usepackage{amsmath}	
\usepackage{wasysym}    
\usepackage{bm}         
\usepackage{xspace}
\usepackage{xcolor}
\usepackage{booktabs}
\makeatletter 
  \patchcmd{\NAT@citex}
    {\@citea\NAT@hyper@{%
      \NAT@nmfmt{\NAT@nm}%
      \hyper@natlinkbreak{\NAT@aysep\NAT@spacechar}{\@citeb\@extra@b@citeb}%
      \NAT@date}}
    {\@citea\NAT@nmfmt{\NAT@nm}%
    \NAT@aysep\NAT@spacechar\NAT@hyper@{\NAT@date}}{}{}

  \patchcmd{\NAT@citex}
    {\@citea\NAT@hyper@{%
      \NAT@nmfmt{\NAT@nm}%
      \hyper@natlinkbreak{\NAT@spacechar\NAT@@open\if*#1*\else#1\NAT@spacechar\fi}%
        {\@citeb\@extra@b@citeb}%
      \NAT@date}}
    {\@citea\NAT@nmfmt{\NAT@nm}%
    \NAT@spacechar\NAT@@open\if*#1*\else#1\NAT@spacechar\fi\NAT@hyper@{\NAT@date}}
    {}{}
\makeatother

\newcommand\Msun{\text{M}_{\astrosun}} 
\newcommand\HI{\ion{H}{I}\xspace} 
\newcommand\HII{\ion{H}{II}\xspace} 
\newcommand\HeI{\ion{He}{I}\xspace} 
\newcommand\HeII{\ion{He}{II}\xspace} 
\newcommand\colt{\mbox{\textsc{colt}}\xspace}
\newcommand\arepo{\mbox{\textsc{arepo}}\xspace}
\newcommand\areport{\mbox{\textsc{arepo-rt}}\xspace}
\newcommand\thesan{\mbox{\textsc{thesan}}\xspace}
\newcommand\thesanzoom{\mbox{\textsc{thesan-zoom}}\xspace}

\newcommand\orcid[1]{\href{https://orcid.org/#1}{\adjustbox{trim={-.15\width} {0\height} {-.15\width} {0\height},clip}{\includegraphics[height=12pt]{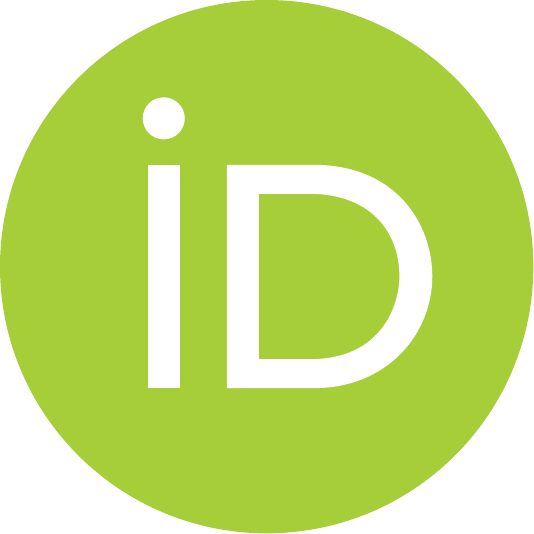}}}}

\title[Delayed LyC escape in bursty galaxies]{The \thesanzoom project: the delayed rise of Lyman continuum \\ escape fractions in bursty high-redshift galaxies}

\author[P. Mehrotra et al.]{%
Parth~Mehrotra\orcid{0009-0001-4403-892X}$^{1}$\thanks{E-mail: \href{mailto:m.26parth@gmail.com}{m.26parth@gmail.com}},
Aaron~Smith\orcid{0000-0002-2838-9033}$^{2}$,
Yuan~Bian\orcid{0000-0002-0210-946X}$^{2}$,
William~McClymont\orcid{0009-0009-5565-3790}$^{3,4}$,
\newauthor
Ewald~Puchwein\orcid{0000-0001-8778-7587}$^{5}$,
Rahul~Kannan\orcid{0000-0001-6092-2187}$^{6}$,
Laura~Keating\orcid{0000-0001-5211-1958}$^{7}$,
Xuejian~Shen\orcid{0000-0002-6196-823X}$^{8,9}$,
\newauthor
Mark~Vogelsberger\orcid{0000-0001-8593-7692}$^{8}$,
Sandro~Tacchella\orcid{0000-0002-8224-4505}$^{3,4}$,
Matthew~Hayes\orcid{0000-0001-8587-218X}$^{10}$,
Crystal~L.~Martin\orcid{0000-0001-9189-7818}$^{11}$,
\newauthor
Cody~Carr\orcid{0000-0003-4166-2855}$^{12}$,
Anand~Utsav~Kapoor\orcid{0000-0002-5187-1725}$^{13}$,
Shyam~Menon\orcid{0000-0001-5944-291X}$^{14,15}$,
and Alaina~Henry\orcid{0000-0002-6586-4446}$^{16}$
\\
$^{1}$Department of Physics, Indian Institute of Science Education and Research, Dr. Homi Bhabha Road, Pune, 411008, Maharashtra, India\\
$^{2}$Department of Physics, The University of Texas at Dallas, Richardson, TX 75080, USA\\
$^{3}$Kavli Institute for Cosmology, University of Cambridge, Madingley Road, Cambridge CB3 0HA, UK\\
$^{4}$Cavendish Laboratory, University of Cambridge, 19 JJ Thomson Avenue, Cambridge CB3 0HE, UK\\
$^{5}$Leibniz-Institut f\"ur Astrophysik Potsdam, An der Sternwarte 16, 14482 Potsdam, Germany\\
$^{6}$Department of Physics and Astronomy, York University, 4700 Keele Street, Toronto, ON M3J 1P3, Canada\\
$^{7}$Institute for Astronomy, University of Edinburgh, Blackford Hill, Edinburgh, EH9 3HJ, UK\\
$^{8}$Department of Physics, Kavli Institute for Astrophysics and Space Research, Massachusetts Institute of Technology, Cambridge, MA 02139, USA\\
$^{9}$Center for Astrophysics $|$ Harvard \& Smithsonian, 60 Garden St, Cambridge, MA 02138, USA\\
$^{10}$Department of Astronomy and Oskar Klein Centre for Cosmoparticle Physics, Stockholm University,\\
\hspace*{1.5em}AlbaNova University Centre, SE-10691 Stockholm, Sweden\\
$^{11}$Department of Physics, University of California, Santa Barbara, CA 93106, USA\\
$^{12}$Department of Astronomy, University of Michigan, 1085 S. University Avenue, Ann Arbor, MI 48109, USA\\
$^{13}$Department of Physics and Astronomy, Ghent University, Proeftuinstraat 86 N3, B-9000 Ghent, Belgium\\
$^{14}$Center for Computational Astrophysics, Flatiron Institute, 162 5th Avenue, New York, NY 10010, USA\\
$^{15}$Department of Physics and Astronomy, Rutgers University, 136 Frelinghuysen Road, Piscataway, NJ 08854, USA\\
$^{16}$Space Telescope Science Institute, 3700 San Martin Drive, Baltimore, MD 21218, USA
}

\date{Accepted XXX. Received YYY; in original form ZZZ}

\pubyear{2026}

\begin{document}
\label{firstpage}
\pagerange{\pageref{firstpage}--\pageref{lastpage}}
\maketitle


\begin{abstract}
The ionizing radiation that galaxies release into their surroundings depends on whether efficient Lyman continuum (LyC) escape coincides with strong photon production. Stellar feedback can open escape paths as the source population fades, linking the escaping radiation to a galaxy's recent history. We investigate this timing in thirteen high-redshift ($z\gtrsim3$) target haloes from the \thesanzoom radiation-hydrodynamic simulations, post-processed with \colt radiative transfer and sampled at $\approx10$\,Myr intervals. The clearest delayed relationship links $\text{SFR}_{10}/\text{SFR}_{100}$ to the LyC escape fraction. Selecting the nearest positive cross-correlation peak gives typical lags of $\sim16$\,Myr and peak correlations of $0.4$--$0.6$ in massive haloes, with longer and less coherent delays towards lower masses. The fitted delays scale approximately as $M_\star^{-0.19}$ and $M_\text{halo}^{-0.27}$, although the shortest lags approach the native snapshot spacing. Recurrent variations span multiple amplitude-dependent timescales, consistent with repeated episodes of star formation and feedback. Recent star formation closely tracks both intrinsic and escaping photon rates, while the escape fraction reflects the evolving gas transparency. The H$\alpha$-to-UV luminosity ratio retains a delayed association with escape, offering an observable connection between burst phase and leakage. High transparency during a fading phase can nevertheless yield little escaping radiation. The temporal overlap of photon production and transmission therefore shapes both the amount and timing of escaping radiation, informing the interpretation of observed leakers and their contribution to cosmic reionization.
\end{abstract}




\section{Introduction}

The Epoch of Reionization (EoR) connects the emergence of the first sustained galaxy population to a fundamental change in the intergalactic medium (IGM), as initially neutral hydrogen became ionized and the mean free path of ionizing radiation grew rapidly \citep{LoebFurlanetto2013,Wise2019,Gnedin2022}. The Thomson optical depth constrains the integrated ionization history and is compatible with relatively late reionization, while quasar absorption spectra and Ly$\alpha$ visibility favour a late and spatially inhomogeneous end to the transition \citep{Fan2006,Planck2018,Becker2021,Bosman2022,Zhu2024}. Reionization was neither an instantaneous global event nor a simple function of how many ionizing photons galaxies produced. The history depended on where and when galaxies released those photons, and how their radiation propagated through an evolving circumgalactic and intergalactic environment.

Star-forming galaxies are widely considered the main drivers of hydrogen reionization because they are abundant and produce large numbers of Lyman continuum (LyC; $h\nu>13.6\,\text{eV}$) photons \citep{Madau1999,Robertson2010,Stark2016,Atek2024, Lin2024, Pahl2025}. Their ionizing emissivity is commonly expressed through the product of the ultraviolet luminosity density and the luminosity-weighted mean of $\xi_\text{ion} f_\text{esc}^\text{LyC}$, where $\xi_\text{ion}$ is the ionizing-photon production efficiency and $f_\text{esc}^\text{LyC}$ is the LyC escape fraction. \textit{JWST} has begun to constrain the abundance, stellar populations, and nebular emission of galaxies well into the EoR, often finding efficient ionizing-photon production among young, low-metallicity systems \citep{Endsley2023,Endsley2024,Matthee2023}. Combining the high inferred photon production with escape fractions estimated from the UV-slope relation calibrated on low-redshift leakers can complete reionization earlier than allowed by IGM and cosmological constraints \citep{Chisholm2022,Munoz2024}. For example, integrating the fiducial model of \citet{Munoz2024} down to $M_\text{UV}=-13$ gives a Thomson optical depth of $\tau_\text{CMB}\simeq0.096$, above the Planck measurement of $0.054\pm0.007$ \citep{Planck2018}. The extent of this apparent photon-budget tension depends on the assumed escape fractions and the contribution from faint galaxies. Although a population-averaged escape fraction can reproduce the global emissivity, it can obscure how photon production and transmission evolve together in individual galaxies \citep{Papovich2026}. Their relative timing affects the mean escaping output because the time average of $f_\text{esc}^\text{LyC}\dot{N}_\text{int}^\text{LyC}$ depends on both the product of their separate means and their covariance.

Absorption by the intervening IGM makes LyC detections increasingly rare with redshift and effectively prevents direct measurements during the EoR \citep{Inoue2014,Grazian2017,Bassett2021,Bassett2022}. Low-redshift LyC leakers constrain the conditions associated with escape \citep{Izotov2018,Rivera-Thorsen2019,Flury2022,SaldanaLopez2022,Chisholm2022,Bait2024,Leclercq2024,Carr2025a,Carr2025b,Flury2025,Jaskot2025}, while Ly$\alpha$, ultraviolet absorption, and nebular emission offer indirect diagnostics at higher redshift \citep{Mason2019,Katz2022,Choustikov2024,Yuan2024,Jaskot2024}. In a sample of more than 1400 galaxies at $5<z<10$, \citet{Giovinazzo2025} inferred very low escape fractions for most objects but identified a small high-escape subset that raises the population average to approximately the level needed for galaxy-driven reionization. Such a distribution may reflect persistent differences among galaxies, intermittent escape phases, orientation effects, or a combination of all three. Distinguishing these possibilities requires a model for how an individual galaxy moves through observable and high-escape states.

Numerical studies provide the physical context for why LyC escape is strongly anisotropic and time-dependent. Massive stars produce ionizing photons almost immediately after they form, yet their dense birth clouds can absorb most of this radiation. Photoionization, stellar winds, radiation pressure, and supernova explosions subsequently redistribute the gas and open low-density channels. The intrinsic photon-production rate and the transmission of those photons can therefore evolve out of phase, producing order-of-magnitude changes in $f_\text{esc}^\text{LyC}$ over short intervals \citep{Kimm2014,Paardekooper2015,Cen2015,Trebitsch2017,Ma2020,Barrow2020,Katz2023}. An observed correlation with $f_\text{esc}^\text{LyC}$ can consequently depend on whether a galaxy is caught during an embedded burst, a feedback-cleared phase, or the re-assembly of its star-forming gas.

Cosmological radiation-hydrodynamic simulations connect this small-scale escape process to reionization as a whole. Large- and intermediate-volume suites, including CROC \citep{Gnedin2014}, CoDa \citep{Ocvirk2020}, SPHINX \citep{Rosdahl2018,Rosdahl2022}, \thesan \citep{KannanThesan2022,SmithThesan2022,GaraldiThesan2022}, and \textsc{lumina} \citep{Zier2026}, sample representative galaxy populations and the evolving ionization topology. Their finite resolution limits how directly they can represent the multiphase ISM and stellar birth environments that regulate LyC transmission. Smaller-volume and zoom-in calculations resolve more of the relevant galactic structure while sampling fewer independent environments. These include FIBY with radiative-transfer post-processing \citep{Paardekooper2013,Paardekooper2015}, the FIRE high-redshift suite \citep{Ma2015,Ma2016,Ma2018}, Renaissance \citep{Xu2016}, FirstLight \citep{Ceverino2017}, SERRA \citep{Pallottini2022}, SPICE \citep{Bhagwat2024}, and MEGATRON \citep{KatzMEGATRON2026}. Together these approaches establish that LyC escape is variable, but they still predict different normalisations, mass trends, and duty cycles because those results depend on how star formation, feedback, and unresolved absorption are modelled \citep{Gnedin2008,WiseCen2009,Ma2020,Ma2021,Kostyuk2023,Yeh2023}.

The temporal structure of LyC escape is inseparable from bursty star formation. Gas accretion, star formation, and feedback can approach a slowly evolving balance \citep{Bouche2010,Dave2012,Lilly2013}, but that balance breaks down when stellar feedback responds too slowly. \citet{FaucherGiguere2018} showed that cycles arise when the feedback delay approaches or exceeds the local dynamical time, allowing gas to collapse and form stars before supernova feedback can regulate it. This behaviour is especially pronounced at high redshift, where dynamical times are short, and in low-mass systems, where too few independent star-forming regions exist to average over individual events. FIRE simulations show the same broad regime, with strongly time-variable star formation across the galaxy population at $z\gtrsim1$ and smoother late-time growth emerging first in more massive systems \citep{Hopkins2014,Sparre2017,FaucherGiguere2018}. During each cycle, a burst consumes part of the accumulated dense gas, and the ensuing energy and momentum injection suppresses star formation until gas cools, returns, or is replenished \citep{Agertz2016,Dome2024}. The amplitude and regularity of the cycle depend on gravitational confinement, gas supply, and feedback efficiency. These quantities also determine how long cleared channels remain open and how rapidly dense, optically thick structures reform.

Comparisons of Balmer-line and UV emission have long provided evidence for variable star formation in nearby dwarfs and at intermediate and high redshift \citep{Weisz2012,Guo2016,Emami2019,Faisst2019}. \textit{JWST} observations increasingly expose both the short timescales and the selection effects associated with this variability. High-redshift galaxies show young stellar ages, strong nebular emission, and large specific star-formation rates, but they also populate phases of sharply reduced recent activity \citep{Endsley2023,Endsley2024,Matthee2023,Looser2024,Cole2025}. In particular, \citet{Endsley2025} find a wide distribution of H$\alpha$-to-UV ratios among $z\sim6$ galaxies and infer rapid SFR changes on $\sim10$--$30\,\text{Myr}$ timescales, with recent downturns especially common in the UV-faint population. FIRE-based mock observations show that galaxies with $M_\star\sim10^8$--$10^9\,\Msun$ can cross detection thresholds on short timescales, biasing flux-limited samples towards active phases with elevated gas fractions and specific star-formation rates \citep{Sun2023}. Comparing tracers with different response times, such as H$\alpha$ and the ultraviolet continuum, offers a statistical measure of burst phase and a possible route to connecting that phase with indirect LyC-leakage diagnostics.

The \thesanzoom suite provides a particularly useful setting in which to make this connection. Its resolved multiphase ISM and explicit stellar-feedback prescriptions naturally drive episodic star formation, causing galaxies to repeatedly move above and below the high-redshift star-forming main sequence \citep{Kannan2025,McClymontZoomSFMS2025}. Those excursions are linked to central starbursts and subsequent inside-out suppression of star formation \citep{McClymontZoomSize2025}, while the efficiency of turning cold dense gas into stars depends on both cloud-scale structure and the adopted early-feedback strength \citep{ShenZoom2026,WangZoomSFE2025}. The same burst phase changes the apparent clumpiness of recent-star-formation tracers even when the underlying stellar-mass distribution is comparatively smooth \citep{WangZoomClumps2026}. These results make the suite well-suited to testing whether the observability and escape of ionizing radiation follow a common feedback history.

That history also unfolds within a broader reionization environment. In \thesanzoom, an externally imposed ionizing field can leave long-lived signatures in low-mass galaxies \citep{ZierZoomUV2025}, metal-free star formation can continue in isolated pockets until the end of reionization \citep{ZierZoomPop2025}, and neutral oxygen absorbers do not necessarily reveal the nearest faint galaxy responsible for their enrichment \citep{PrutoZoom2026}. Repeated enrichment, dilution, and gas cycling also contribute to the unusual abundance patterns predicted for some early galaxies \citep{McClymontZoomMetal2025}. A time-resolved escape model links these results by determining when radiation from a changing source population reaches its circumgalactic environment.

Previous escape-fraction studies have established feedback-cleared channels and delayed or multi-phase leakage in individual systems and simulated populations \citep{Kimm2014,Paardekooper2015,Trebitsch2017,Ma2020,Barrow2020,Katz2023}. Simulations also connect efficient leakage to the simultaneous availability of young ionizing stars and low-opacity gas, motivating diagnostics of both stellar age and gas conditions \citep{Choustikov2024Diagnostics}. Within \thesanzoom, \citet{Summerfield2026} found that $\text{SFR}_{10}/\text{SFR}_{100}$ and the gas-to-stellar mass ratio carry the strongest predictive information for $f_\text{esc}^\text{LyC}$, whereas ultraviolet luminosity is the dominant predictor of the escaping LyC photon rate. That distinction points to a temporal problem. The source term responds immediately to young stars, while the escape fraction records the subsequent evolution of the gas. Snapshot predictors cannot by themselves determine the ordering, duration, or recurrence of these phases. We therefore examine the mass dependence of these temporal scales together with the extent to which observable tracers recover the burst phase.

Here, we follow the selected target haloes in the \thesanzoom regions through successive snapshots at approximately 10\,Myr intervals and treat bursty star formation and LyC escape as a delayed, feedback-regulated system. We measure the lag between burstiness and escape, identify recurrent phase-space evolution, and use peak--trough statistics to quantify variability at several logarithmic prominence thresholds. We determine how the lag and recurrence times vary with stellar and halo mass, separate source-regulated photon production from environment-regulated transmission, and ask whether an H$\alpha$-to-UV burstiness proxy retains the same delayed ordering. This time-domain view supplies the ingredients needed to translate a broad distribution of snapshot escape fractions into physical duty cycles and provides population-level predictions for interpreting early galaxies during different phases of their feedback history.

The paper is organised as follows. Section~\ref{sec:methods} describes the \thesanzoom simulations, radiative-transfer calculations, and analysis techniques. Section~\ref{sec:results} presents the temporal evolution of LyC escape, lag measurements, phase-space analysis, source-versus-environment diagnostics, and peak--trough statistics. It also discusses the physical interpretation of our results and compares them with previous theoretical and observational studies. Section~\ref{sec:conclusions} summarises our conclusions and outlines prospects for future work.

\section{Methods}
\label{sec:methods}

\subsection{Simulation overview}
\label{sec:simulations}

We analyse galaxies from the \thesanzoom simulation suite \citep{Kannan2025}, which refines selected regions of the larger \thesan radiation-magnetohydrodynamic simulation of cosmic reionization \citep{KannanThesan2022,SmithThesan2022,GaraldiThesan2022,Garaldi2024PublicRelease}. The parent calculation follows a $(95.5\,\text{cMpc})^3$ cosmological volume and supplies the large-scale density and reionization environment. The suite contains fourteen zoom regions selected to span target halo masses of approximately $10^8$--$10^{13}\,\Msun$ at $z=3$. We exclude the most massive region because its simulation does not reach $z=3$, leaving thirteen targets with broad mass coverage and more finely resolved internal gas structure than the parent simulation.

The simulations use \areport \citep{Kannan2019}, the radiation-hydrodynamic extension of the moving-mesh code \arepo \citep{Springel2010,Weinberger2020}. Hydrodynamics is solved on an unstructured, quasi-Lagrangian Voronoi mesh. The radiation solver evolves the first two moments of the transfer equation with an M1 closure and seven frequency groups spanning the infrared through ionizing energies \citep{Levermore1984,Kannan2019}. Radiation generated inside each refined region is coupled to the gas on-the-fly. A spatially and temporally varying radiation field interpolated from the parent \thesan volume is imposed at the zoom boundary, representing incoming ionizing radiation from outside the refined region \citep{Kannan2025,ZierZoomUV2025}.

The multiphase ISM, star formation, and stellar feedback are modelled with \textsc{smuggle} \citep{Marinacci2019,Kannan2020}. The thermochemistry includes non-equilibrium hydrogen, helium, and molecular hydrogen, together with metal-line cooling \citep{Vogelsberger2013}, and permits gas to cool to $10\,\text{K}$ \citep{Kannan2025}. Stars form probabilistically in Jeans-unstable gas above $n_\text{H}=10\,\text{cm}^{-3}$. Young stellar populations radiate according to BPASS version 2.2.1 spectral models with a Chabrier initial mass function \citep{Eldridge2017,Chabrier2003}. The radiative-transfer calculation couples photoheating, photoelectric heating, and radiation pressure from direct UV and reprocessed IR photons to the gas. Mechanical feedback includes continuous OB and AGB stellar winds and discrete supernovae. An additional early-stellar-feedback channel injects momentum around newly formed stars for their first 5\,Myr to represent unresolved cloud disruption and is calibrated as an effective model rather than identified with one unique physical mechanism \citep{Kannan2025}. This momentum input acts from the onset of star formation, with its strength and duration influencing how rapidly the birth environment is cleared. In particular, recent calculations suggest that resonant Ly$\alpha$ radiation pressure could enhance pre-supernova clearing in dense, dust-poor star-forming regions \citep{Smith2017,Nebrin2025,Nebrin2026,Menon2026, Kimm2018}, so the shortest escape delays may depend on physics represented only effectively here. The resulting star-formation efficiencies depend on the resolved cloud structure and feedback strength, as examined across the suite by \citet{ShenZoom2026} and \citet{WangZoomSFE2025}. \thesanzoom also follows the production, growth, and destruction of dust \citep{McKinnon2017,Kannan2020,Garaldi2026}. These processes set the burst sequence and the evolving neutral-gas geometry through which LyC photons propagate.

The 4$\times$, 8$\times$, and 16$\times$ tiers improve the spatial resolution of the parent simulation by the stated factors and its baryonic mass resolution by factors of 64, 512, and 4096, respectively. Their baryonic mass resolutions are $9.09\times10^3$, $1.14\times10^3$, and $1.42\times10^2\,\Msun$, with corresponding minimum gas-cell softening lengths of $69.2$, $34.6$, and $17.3\,\text{cpc}$ \citep{Kannan2025}. Haloes are first identified with a friends-of-friends search, and their bound substructure is separated with \textsc{subfind-hbt} \citep{Springel2021}. The catalogue uses the critical-density convention, with $R_{200\text{c}}$ enclosing a mean density of $200\rho_\text{crit}(z)$. The operational aperture $R_\text{vir}$ is a smoothed estimate of this radius, as described below.

The time-series analysis in this paper uses the evolutionary tracks of the thirteen designated target haloes listed in Table~\ref{tab:galaxy_sample}, which span nearly five orders of magnitude in halo mass. These tracks have sufficient temporal and spatial resolution to capture the repeated main-sequence excursions, central starbursts, and observable structural changes found in other \thesanzoom analyses \citep{McClymontZoomSFMS2025,McClymontZoomSize2025,WangZoomClumps2026}. We include all 27 available 4$\times$, 8$\times$, and 16$\times$ realizations of the targets. We pool the resolution levels because they cover different mass ranges. This preserves the available mass baseline, but the pooled relations do not establish numerical convergence, and different realizations of one target are not independent galaxies.

We follow the main progenitor of each $z\simeq3$ target through the merger-tree catalogue, with additional spatial and mass checks to verify the branch assignment. Starting at the linked subhalo position, we calculate a centre of mass using gas, stars, and dark matter within the tree-based $R_{200\text{c}}$. We recenter these selected elements, apply the same radial cut again, and calculate a second centre of mass. The procedure ends after these two passes, without a shrinking aperture or a convergence tolerance. We estimate $R_{200\text{c}}$ from the enclosed gas, stellar, and dark-matter mass profile, using the outermost crossing of $200\rho_\text{crit}(z)$. We smooth each centre coordinate and radius along the track with a cubic Savitzky--Golay filter spanning 15 outputs. The filter is applied to both the redshifts and the measured values, followed by linear interpolation back to the original redshifts. We denote the smoothed aperture by $R_\text{vir}$ throughout the analysis.

The catalogue stellar mass $M_\star$ sums the current masses of stellar particles within $R_\text{vir}$, accounting for stellar mass loss. The halo mass $M_\text{halo}$ includes gas, dark matter, and this stellar contribution within the same aperture. These definitions apply to all resolution levels, while SFRs use initial stellar masses as described in Section~\ref{sec:timeseries}.

\begin{table}
\centering
\begin{tabular*}{\columnwidth}{c@{\extracolsep{\fill}}cc}
\toprule
Halo & $M_\text{halo}\,[\Msun]$ & $M_\star\,[\Msun]$ \\
\midrule
m12.6 & $3.64\times10^{12}$ & $1.06\times10^{11}$ \\
m12.2 & $1.34\times10^{12}$ & $2.42\times10^{10}$ \\
m11.9 & $6.89\times10^{11}$ & $1.23\times10^{10}$ \\
m11.5 & $2.71\times10^{11}$ & $3.36\times10^{9}$ \\
m11.1 & $(1.25, 1.22)\times10^{11}$ & $(1.06, 0.502)\times10^{9}$ \\
m10.8 & $(6.43, 6.71)\times10^{10}$ & $(1.1, 0.359)\times10^{9}$ \\
m10.4 & $(2.41, 2.10)\times10^{10}$ & $(1.01, 0.328)\times10^{8}$ \\
m10.0 & $(9.39, 9.76)\times10^{9}$ & $(5.11, 5.42)\times10^{7}$ \\
m9.7 & $(4.89, 4.30, 3.46)\times10^{9}$ & $(2.45, 1.38, 0.252)\times10^{7}$ \\
m9.3 & $(1.73, 1.64, 1.39)\times10^{9}$ & $(2.64, 4.26, 1.04)\times10^{6}$ \\
m8.9 & $(5.82, 6.23, 6.35)\times10^{8}$ & $(1.95, 3.73, 3.24)\times10^{5}$ \\
m8.5 & $(3.00, 3.03, 3.17)\times10^{8}$ & $(3.28, 2.40, 2.21)\times10^{4}$ \\
m8.2 & $(7.02, 7.60, 7.60)\times10^{7}$ & $(5.67, 0.787, 1.12)\times10^{4}$ \\
\bottomrule
\end{tabular*}
\caption{The thirteen \thesanzoom targets and their available resolution realizations. Halo and stellar masses are evaluated within $R_\text{vir}$ at $z=3$. Single entries correspond to 4$\times$ resolution, pairs to (4$\times$, 8$\times$), and triples to (4$\times$, 8$\times$, 16$\times$). Names refer to the parent selection \citep{Kannan2025}.}
\label{tab:galaxy_sample}
\end{table}
\subsection{Radiative transfer and LyC escape fractions}
\label{sec:colt}

We quantify ionizing-photon escape by post-processing the \thesanzoom snapshots with the \colt Monte Carlo radiative-transfer code \citep{Smith2015,Smith2019,Smith2022disk,McClymontCOLT2025}. For this catalogue, \colt recalculates the hydrogen and helium ionization states in photoionization equilibrium with the radiation field sampled by the Monte Carlo packets. We retain the simulated thermal state rather than solving for thermal equilibrium. Recomputing the ionization and emission is motivated in part by insufficiently resolved \HII regions, whose cell-averaged ionization states can bias nebular luminosities. The resulting equilibrium solution sets the gas opacity used for the LyC escape calculation and supplies the ionization and recombination rates needed to predict nebular emission. This common treatment is particularly important for the present analysis because the LyC escape fractions and H$\alpha$ luminosities are derived from the same recomputed ionization structure. Photoionization equilibrium is expected to be most reliable in dense \HII regions, where ionization and recombination times are short compared with the evolution of the stellar radiation field. In the lower-density CGM, however, recombination times can approach or exceed the variability timescales studied here. The post-processing then erases non-equilibrium ionization memory and may bias the neutral opacity between bursts (see Section~\ref{sec:limitations}). The target calculation uses a spherical extraction extending to $4R_\text{vir}$ about the smoothed centre, drawn from the high-resolution zoom data. The LyC emission and escape statistics are evaluated within $R_\text{vir}$. Selection is spatial rather than based on bound-subhalo membership, so satellites inside this aperture contribute to the SFR, stellar mass, and intrinsic and escaping LyC photon rates.

Photon packets are emitted isotropically from high-resolution star particles using BPASS spectral energy distributions consistent with the simulation. The frequency grid resolves the \HI, \HeI, and \HeII ionization edges with 61 bins above $13.6\,\text{eV}$ and a maximum bin width of $0.05$ dex. Packets propagate through the native three-dimensional Voronoi tessellation, with continuous absorption by neutral hydrogen, helium, and dust. The adopted runs use the dust evolved on-the-fly, without an additional component assigned from gas-phase metals. Luminosity-biased packet sampling increases the statistical weight assigned to faint sources. We scale the number of equilibrium packets with the size and complexity of the snapshot, reaching up to $2\times10^9$ for the largest calculations, while individual target halo runs typically use $10^8$ line packets.

\begin{figure*}
    \centering
    \includegraphics[width=\textwidth]{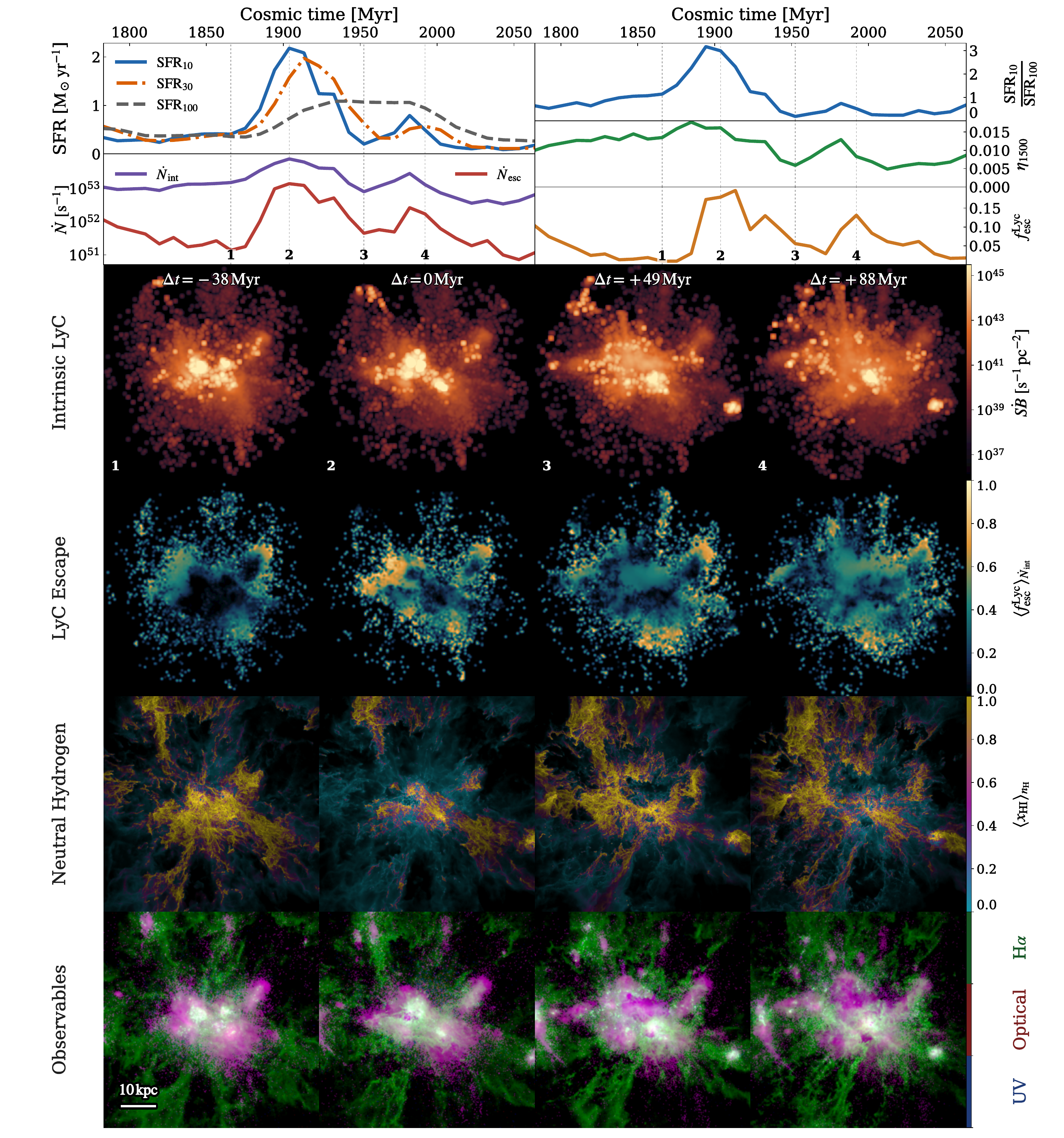}
    \caption{Temporal and spatial evolution of LyC production and escape in the m11.1 galaxy at 8$\times$ resolution. The upper panels show the 10, 30, and 100 Myr star-formation rates, intrinsic and escaped ionizing-photon rates, $\text{SFR}_{10}/\text{SFR}_{100}$, $\eta_{1500}$ from emergent H$\alpha$ and UV luminosities, and the global escape fraction $f_\text{esc}^\text{LyC}$. Vertical guides mark the four epochs mapped below, with times measured relative to the $\text{SFR}_{10}$ peak. From top to bottom, the maps show intrinsic LyC photon-rate surface density, intrinsic-photon-rate-weighted stellar escape fraction, neutral hydrogen, and mock optical, H$\alpha$, and UV observables. Each column shows the same epoch across all map rows. Neutral-fraction hue is weighted by hydrogen density, while opacity follows gas density. Each map spans $2R_\text{vir}$. The lower-left map includes a 10\,kpc scale bar. The histories and maps connect burst and escape episodes to the evolving neutral gas.}
    \label{fig:galaxy_rt}
\end{figure*}

Throughout this work, we use the halo-scale LyC escape fraction,
\begin{equation}
  f_\text{esc}^\text{LyC} = \frac{\dot{N}_\text{esc}^\text{LyC}}{\dot{N}_\text{int}^\text{LyC}} \, ,
\end{equation}
where $\dot{N}_\text{int}^\text{LyC}$ is the intrinsic LyC photon rate of sources within $R_\text{vir}$ and $\dot{N}_\text{esc}^\text{LyC}$ is the rate that crosses the virial sphere. Because escaped photons are counted only when they cross $R_\text{vir}$, this halo-scale definition includes attenuation through both the ISM and CGM. The LyC time series use the total escape integrated over solid angle. The spatial projections in Figure~\ref{fig:galaxy_rt} illustrate the distributions of sources and gas associated with this global escape fraction. Tracking $\dot{N}_\text{int}^\text{LyC}$, $\dot{N}_\text{esc}^\text{LyC}$, and their ratio separately allows us to distinguish variability in the ionizing sources from changes in transmission through the surrounding gas.

The same \colt calculation supplies the synthetic quantities used for the observable burstiness analysis. The H$\alpha$ emissivity is calculated from the recomputed gas state, including recombination cascades and collisional excitation, rather than inferred from the stellar ionizing luminosity with a fixed conversion. Both the on-the-fly \areport radiation solver and the \colt post-processing adopt the Case~B approximation, in which recombinations directly to the ground state are treated by on-the-spot re-absorption instead of propagating the resulting diffuse ionizing photons. For the observable burstiness analysis, we use the emergent H$\alpha$ luminosity and rest-frame $1500\,\text{\AA}$ continuum from within $R_\text{vir}$, including dust attenuation and averaging over viewing direction. These tracers weight young stellar populations over different timescales. The H$\alpha$ emission also reflects the absorption and escape of ionizing photons in the radiative-transfer calculation.

Figure~\ref{fig:galaxy_rt} places the SFR, photon rates, burstiness, and escape fraction on aligned time axes and shows the source and gas distributions at four selected epochs. The escaping photon rate and escape fraction do not necessarily peak together because the intrinsic photon supply also evolves. The projections associate these changes with a reorganising neutral medium and an evolving distribution of recent star formation. Our angle-integrated LyC measurements track the overall transparency of the halo while retaining the effect of its anisotropic gas structure.

\subsection{Time-series analysis}
\label{sec:timeseries}

The primary goal of this work is to investigate the temporal relationship between star formation and LyC escape. We therefore construct a time series for each galaxy by following its evolution through successive simulation snapshots. The matched SFR and LyC histories contain 102--189 native outputs per track, with median spacings of 9.6--9.9\,Myr and individual intervals of 5.3--19.8\,Myr. This cadence samples the longer feedback responses but cannot resolve the first few Myr of a burst or an individual massive star's evolution.

Star-formation rates (SFRs) are calculated using stellar particles younger than a specified averaging timescale,
\begin{equation}
  \text{SFR}_{t_\text{avg}} = \frac{1}{t_\text{avg}}\sum_{i:\,\text{age}_i<t_\text{avg}} m_{\star,i}^\text{init} \, ,
\end{equation}
where $m_{\star,i}^\text{init}$ is the initial stellar-particle mass. The sum includes stars younger than $t_\text{avg}$ within $R_\text{vir}$ of the smoothed target centre.

To probe different temporal scales, we compute both $\text{SFR}_{10}$ and $\text{SFR}_{100}$, which average star formation over 10 Myr and 100 Myr, respectively. The ratio of the SFRs averaged over short and long timescales,
\begin{equation}
  \text{sfr}^{10}_{100} \equiv \frac{\text{SFR}_{10}}{\text{SFR}_{100}} \, ,
\end{equation}
serves as a measure of recent star-formation variability and is used throughout this work as a burstiness indicator. Values $\text{sfr}^{10}_{100} > 1$ correspond to enhanced recent star formation relative to the longer-term average, while $\text{sfr}^{10}_{100} < 1$ indicates declining activity. This ratio describes recent activity relative to its longer-term baseline, so galaxies with the same ratio can have very different absolute SFRs and ionizing luminosities. We also use $\text{sfr}^{10}_{30}\equiv\text{SFR}_{10}/\text{SFR}_{30}$ to test sensitivity to the longer averaging interval.

In addition to the SFRs, we track the intrinsic LyC photon production rate, escaping LyC photon production rate, and LyC escape fraction. Together these quantities allow us to investigate the temporal coupling between star formation, ionizing-photon production, and photon escape.

\subsection{Lag analysis}
\label{sec:lagmethod}

We quantify temporal offsets with the normalised cross-correlation implemented in the \texttt{sunkit-image} Python library \citep{Viall2012,Barnes2019}. The simulation snapshots are irregularly spaced, so we first interpolate each time series onto a common grid of 2000 uniformly spaced cosmic times spanning each track, with a typical spacing of approximately 1\,Myr. This finer grid allows consistent numerical shifts, while the temporal information remains limited by the native snapshot cadence. We apply a numerical floor of $10^{-10}$ in the input units before interpolating linearly in the logarithms of time and each quantity, e.g. $\{\log t, \log \text{SFR}\}$. We then exponentiate the interpolated series back to linear values for the cross-correlation. Above the numerical floor, this piecewise power-law interpolation preserves the range between adjacent measurements without introducing additional extrema, which unconstrained higher-order splines can produce. At lags comparable to the native cadence, the heights and locations of correlation peaks can depend on the interpolation scheme.

For two time series $A(t_i)$ and $B(t_i)$, we subtract each mean ($\mu$) and divide by the corresponding standard deviation ($\sigma$), e.g. $\tilde{A}_i = (A_i - \mu_A) / \sigma_A$, before evaluating
\begin{equation} \label{eq:fft_crosscorr}
  C_{AB}(\tau) = N^{-1}\,\mathcal{F}^{-1}\big\{
  \mathcal{F}[\tilde{A}(-t)]\,\mathcal{F}[\tilde{B}(t)]
  \big\}(\tau) \, .
\end{equation}
Here $\mathcal{F}$ denotes the discrete Fourier transform, $N$ is the number of samples on the analysis grid, and $\Delta t_\text{grid}$ is its uniform cadence. Zero-padding the transforms to $2N-1$ samples gives the full linear cross-correlation, which the FFT evaluates efficiently. At lag $\tau_k = k\Delta t_\text{grid}$, we apply the linear overlap weight
\begin{equation} \label{eq:overlap_weight}
  w_k = 1 - |k|/(N-1) \qquad \text{such that} \qquad
  C^w_{AB}(\tau_k)=w_k C_{AB}(\tau_k) \, .
\end{equation}
This taper downweights large offsets where few samples overlap and reaches zero at the endpoints, $|\tau|=t_\text{final}-t_\text{initial}$. Recurrent structure in the histories can produce several local maxima in $C^w_{AB}(\tau)$. For the burstiness--escape comparisons, we select positive-valued local maxima at $\tau>0$ with a minimum prominence of $0.1$ in the weighted correlation and adopt the one nearest zero as $\tau_\text{peak}$. This identifies the earliest qualifying delayed association, while the global maximum can correspond to a later recurrence. If no peak qualifies, the intrinsic-burstiness routine falls back to the global maximum, which can have either lag sign. The observable-proxy routine instead uses the largest positive correlation at positive lag when available. The source--photon comparisons allow either lag sign and select the qualifying peak nearest zero. At the selected lag, $r_\text{peak}\equiv C^w_{AB}(\tau_\text{peak})$ measures the coherence of the response. Because the series are normalised over the full track and tapered at each lag, this coefficient differs from a Pearson coefficient recomputed using only the overlapping samples. We use $|r_\text{peak}|$ to weight the intrinsic lag--mass and lag--period fits, with the fitting procedure detailed in Appendix~\ref{appendix:polar_period}.

Throughout this paper, a positive lag indicates that variations in the first quantity precede those in the second. For star-formation activity and LyC escape, it therefore describes an escape-fraction response following a change in star formation. We apply the lag analysis to the SFRs, burstiness parameter, intrinsic and escaping LyC photon rates, and LyC escape fraction to measure their temporal offsets and dependence on galaxy mass. We denote the burstiness-to-escape delays by $\tau_{\text{peak},\,10/100}$ and $\tau_{\text{peak},\,10/30}$ for the 10/100 and 10/30 ratios, respectively.

\subsection{Peak--trough and prominence analysis}
\label{sec:peaktrough}

To characterise episodic variability, we identify peaks and troughs in the time series of the LyC escape fraction, burstiness parameter, SFRs, and LyC photon production rates. The peak-finding algorithm selects local extrema that exceed a minimum logarithmic amplitude difference between neighbouring peaks and troughs. This threshold is expressed in dex and controls the sensitivity of the analysis to low-amplitude fluctuations. Repeating the analysis over several thresholds allows us to test whether the inferred recurrence statistics are robust to the prominence required for an event to be identified.

For each threshold, the algorithm constructs an alternating sequence of significant peaks and troughs that partitions the temporal evolution into individual episodes. We use these extrema to measure peak-to-peak and trough-to-trough intervals, peak-to-trough amplitudes, and the variability within successive intervals. Together, these statistics describe the duration, strength, and regularity of the variability in each galaxy and enable us to examine how those properties depend on halo and stellar mass.

We also compare the logarithmic amplitudes of paired troughs and peaks in prominence diagrams. Their vertical displacement above the equal-amplitude diagonal measures the peak-to-trough contrast, allowing the amplitude distributions of the different quantities to be compared directly. Appendix~\ref{appendix:peak_finding} describes the selection algorithm and these diagrams in greater detail.

\subsection{Observable burstiness tracers}
\label{sec:observable}

Our analysis primarily relies on intrinsic simulation quantities. Observations cannot directly measure SFRs averaged over arbitrary timescales, so we use an observable burstiness tracer based on synthetic ultraviolet continuum (1500\,\AA) and Balmer (H$\alpha$) emission. The H$\alpha$ luminosity traces ionizing radiation produced by short-lived massive stars and is therefore sensitive to star formation over timescales of $\sim10\,\text{Myr}$. The UV continuum around 1500\,\AA\ traces a longer-lived stellar population and probes star formation over timescales of $\sim30$--$100\,\text{Myr}$. Motivated by these differing timescales, we define the observable burstiness proxy \citep{Sun2023, Asada2024}
\begin{equation} \label{eq:eta_1500}
  \eta_{1500} \equiv \frac{L_{\text{H}\alpha}}{\lambda_{1500} L_{\lambda,1500}} \, ,
\end{equation}
where $L_{\text{H}\alpha}$ is the H$\alpha$ luminosity (in erg\,s$^{-1}$) and $L_{\lambda,1500}$ is the spectral luminosity (in erg\,s$^{-1}$\,\AA$^{-1}$) at $\lambda_{1500} = 1500$\,\AA. We use emergent luminosities, including dust attenuation, as described in Section~\ref{sec:colt}.

These tracer timescales depend on the stellar population and star-formation history rather than corresponding to exact top-hat averages \citep{CaplarTacchella2019}. The observable burstiness indicator $\eta_{1500}$ captures recent star-formation variability in the same spirit as $\text{sfr}^{10}_{100}$ and allows a qualitative comparison with observationally inferred burstiness. Applying the same lag analysis to $\eta_{1500}$ and the LyC escape fraction tests whether the delayed-escape signature remains detectable when expressed in terms of observational diagnostics. We denote this delay by $\tau_{\text{peak},\,\eta_{1500}}$.

\section{Results and Discussion}
\label{sec:results}

\subsection{Recurrent phase-space evolution of burstiness and escape}
\label{subsec:quasi}
Stellar populations are the primary sources of ionizing radiation in galaxies. Episodes of enhanced star formation increase the abundance of young, massive stars and rapidly boost the production of LyC photons. During such starburst events, shorter-timescale SFR indicators consequently rise faster than longer-timescale indicators, increasing the burstiness parameter $\text{sfr}^{10}_{100}$. In the delayed mode resolved by these simulations, the LyC escape fraction responds more slowly because the gaseous environment requires time to develop sufficiently transparent pathways. An earlier escape mode may still occur below our temporal and spatial resolution.

Radiative feedback, stellar winds, and supernova explosions inject energy and momentum into the surrounding gas, disrupting the local reservoir and gradually suppressing star formation. LyC leakage during this declining phase may correspond to the ``remnant leaker'' mode identified in SPHINX, where feedback from an earlier burst has already cleared the ISM \citep{Katz2023}. Escape can also rise before the burst fully subsides, as occurs for active-burst leakers in SPICE, in which strong LyC emission and efficient escape occur while the galaxy is still undergoing an active episode of star formation \citep{Bhagwat2024}. These modes occupy different, sometimes overlapping parts of the same source--feedback sequence, and their relative importance depends on the timing and strength of the adopted feedback. As low-density channels persist, $\text{sfr}^{10}_{100}$ can decline while $f_\text{esc}^\text{LyC}$ remains elevated. Gas re-accretion eventually restores dense star-forming conditions and reduces the escape fraction as old channels close or reorient. The resulting modulation connects the burst--quench cycles and central starbursts previously identified in \thesanzoom to the transmission of ionizing radiation \citep{McClymontZoomSFMS2025,McClymontZoomSize2025}.

\begin{figure*}
    \centering
    \includegraphics[width=\linewidth]{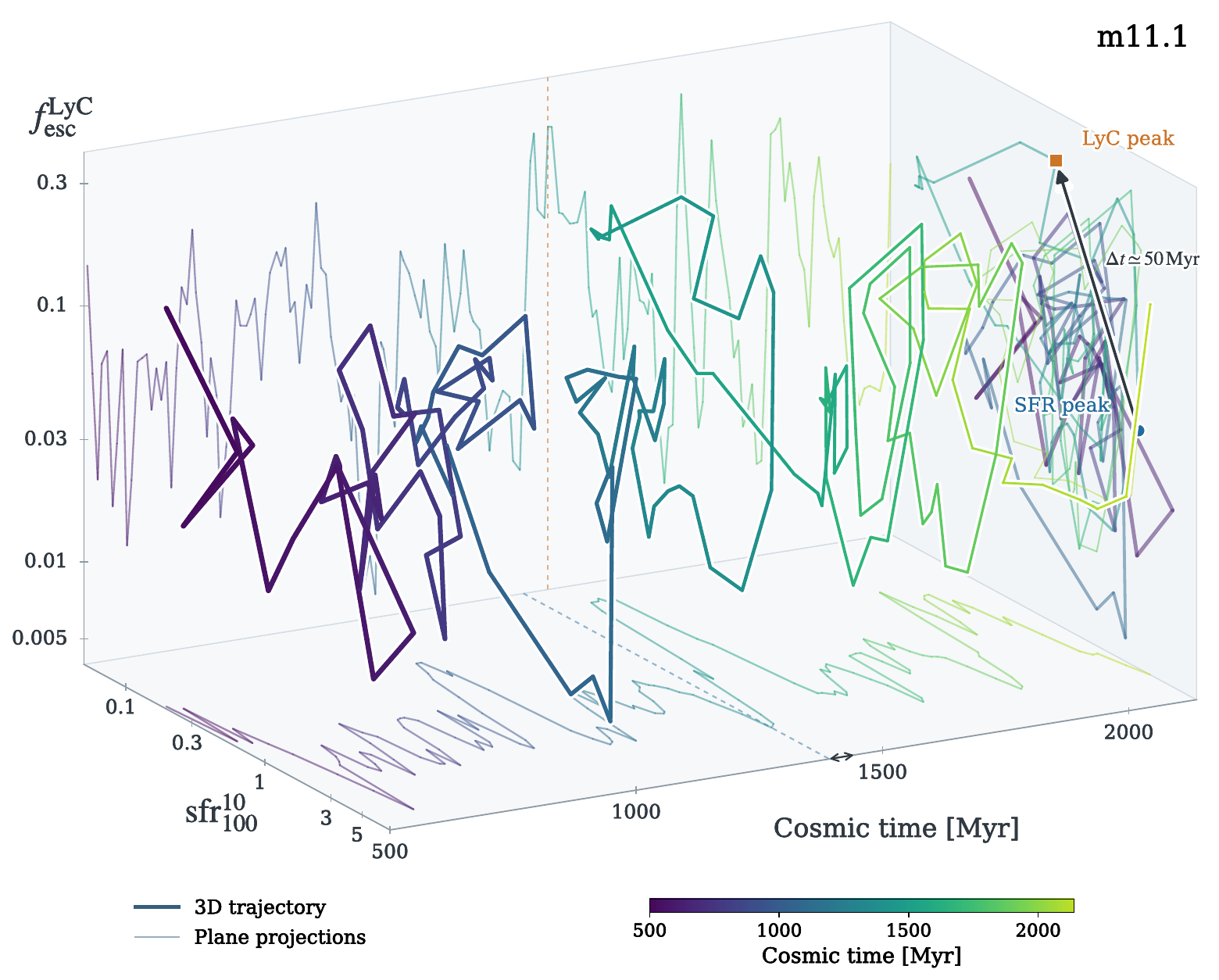}
    \caption{Three-dimensional burst--escape trajectory of the m11.1 galaxy at 8$\times$ resolution in cosmic time, $\text{SFR}_{10}/\text{SFR}_{100}$, and $f_\text{esc}^\text{LyC}$. The latter two coordinates are logarithmic and retain their physical values without time-mean subtraction. The white-edged curve joins all 163 finite states from $t=507$ to $2138\,\text{Myr}$, while the pale curves on the three faces are exact coordinate-plane projections of the same track. Colour gives cosmic time linearly. The main track and its SFR-ratio--escape projection become thinner towards later times. The annotated passage connects a local SFR-ratio peak at $t=1392\,\text{Myr}$ to a local LyC-escape peak at $t=1443\,\text{Myr}$, an illustrative delay of $\Delta t\simeq50\,\text{Myr}$. Repeated turns and partial returns are consistent with recurrent feedback-regulated evolution, but do not imply that the cycles repeat at regular intervals.}
    \label{fig:combined_phase_space}
\end{figure*}

To investigate whether this feedback-regulated picture produces a recurrent phase-space signature, we examine the joint evolution of $\text{sfr}^{10}_{100}$ and $f_\text{esc}^\text{LyC}$. Figure~\ref{fig:combined_phase_space} shows the m11.1 trajectory in these quantities and cosmic time, with projections onto the three coordinate planes. Colour encodes cosmic time, and the main track becomes thinner towards later times. The burstiness and escape coordinates retain their physical values on logarithmic axes. Repeated turns and partial returns indicate recurrent evolution, while rapid variations and secular drifts prevent a closed orbit. The annotated $\sim50$\,Myr separation links one pair of local peaks and is not a fitted winding period or a whole-track correlation lag. The aligned histories in Figure~\ref{fig:galaxy_rt} provide a complementary view of the temporal sequence. We use \emph{episodic} or \emph{recurrent} for general variability and reserve \emph{cycle} and \emph{period} for winding measured explicitly in the joint burstiness--escape phase space.

\begin{figure}
        \centering
        \includegraphics[width=\linewidth]{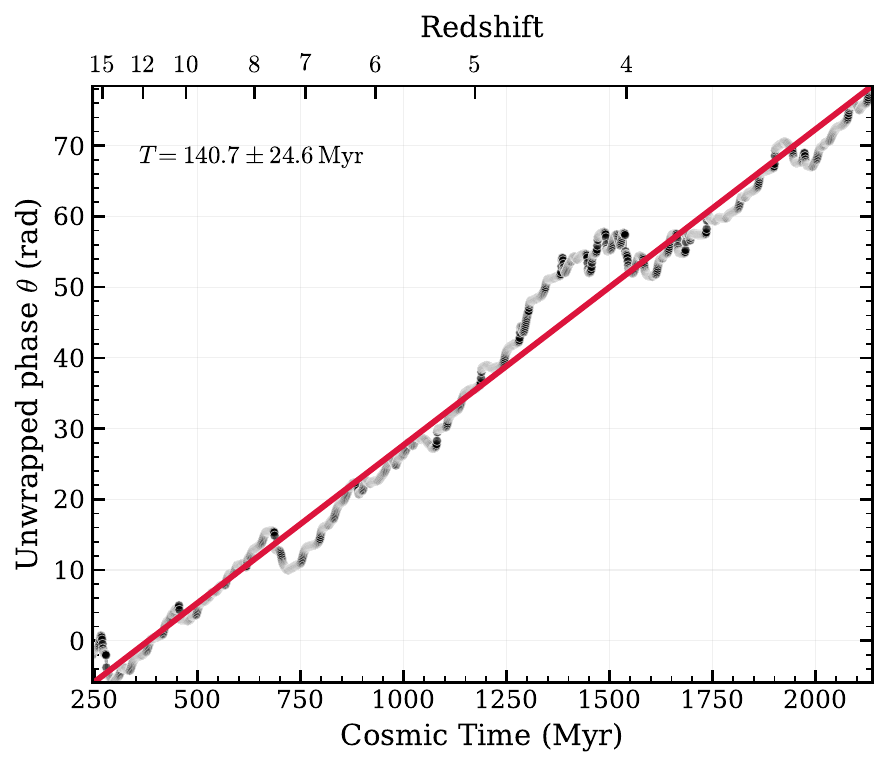}
    \caption{Unwrapped phase angle $\theta$ as a function of cosmic time for m12.6 at 4$\times$ resolution, using the fiducial 100\,Myr moving centre. Its net increase records repeated winding about the running phase-space centre, while the fitted slope defines the mean winding period through Equation~(\ref{eq:phase_period}). The fitted period is $T_{10/100} = 140.7\pm24.6\,\text{Myr}$, with a coefficient of determination $R^2 = 0.983$ for the linear phase fit.}
    \label{fig:theta}
 \end{figure}

To quantify this winding, we follow the polar angle of the trajectory about a moving centre. Appendix~\ref{appendix:polar_period} gives the full procedure, including the time grid, boundary treatment, uncertainty calculation, and comparisons with 50 and 150\,Myr windows and a fixed centroid.

We interpolate each halo's history onto a uniform cosmic-time grid and use logarithmic coordinates $x_i=\log\text{sfr}^{10}_{100}(t_i)$ and $y_i=\log f_\text{esc}^\text{LyC}(t_i)$. Logarithmic coordinates allow us to follow relative variations across the broad dynamic range of each quantity. A centred 100\,Myr boxcar average of each logged series defines the local centre $(\bar{x}_i,\bar{y}_i)$. The moving centre follows shifts in the typical burstiness and escape fraction, allowing us to measure recurrent motion relative to the galaxy's evolving state. We subtract the local centre and scale each coordinate by its standard deviation over the full track,
\begin{equation} \label{eq:phase_coordinates}
  X_i=\frac{x_i-\bar{x}_i}{\sigma_x} \qquad \text{and} \qquad
  Y_i=\frac{y_i-\bar{y}_i}{\sigma_y} \, .
\end{equation}
This puts the two axes on comparable scales before measuring the angle around the centre. We calculate the wrapped phase $\phi_i=\operatorname{atan2}(Y_i,X_i)$, which preserves the quadrant of the trajectory, and unwrap the time-ordered sequence. Unwrapping removes artificial jumps at the $-\pi/\pi$ boundary by adding integer multiples of $2\pi$, while retaining local reversals in the direction of winding. The resulting cumulative phase $\theta_i$ records the net angular motion, with a change of $2\pi$ in magnitude corresponding to one winding. We fit
\begin{equation} \label{eq:phase_period}
  \theta(t) = \theta_0 + \omega(t - t_0) \qquad \text{and} \qquad T = 2\pi / |\omega| \, ,
\end{equation}
where $\theta_0$ is the fitted phase at reference time $t_0$, $\omega$ is the mean winding rate, and $T$ is the corresponding period. We denote the periods measured with $\text{sfr}^{10}_{100}$ and $\text{sfr}^{10}_{30}$ by $T_{10/100}$ and $T_{10/30}$, respectively. The coefficient of determination $R^2$ measures the goodness of the linear phase fit.

The averaging window determines how closely the reference centre follows the trajectory. A shorter window follows more of the local motion, whereas a longer window leaves more of the slow drift in the centred trajectory. We use the same physical duration of 100\,Myr for all tracks. The 50 and 150\,Myr tests retain shorter periods at higher mass, although the fitted slopes and normalizations change. Appendix~\ref{appendix:polar_period} quantifies these changes and shows why a fixed centroid can give substantially different net winding for a drifting trajectory.

Figure~\ref{fig:theta} illustrates a sustained phase trend despite local reversals and changes in trajectory shape, with $T_{10/100}=140.7\pm24.6$\,Myr for m12.6 at 4$\times$ resolution. This period describes the joint evolution of burstiness and LyC escape. Projecting the motion onto one axis can merge different phases, and amplitude-based peak selection can skip windings. We measure the corresponding one-dimensional recurrence intervals separately in Section~\ref{sec:peak-trough}.

\begin{figure*}
    \centering
    \begin{minipage}{0.48\textwidth}
        \centering
        \includegraphics[width=\linewidth]{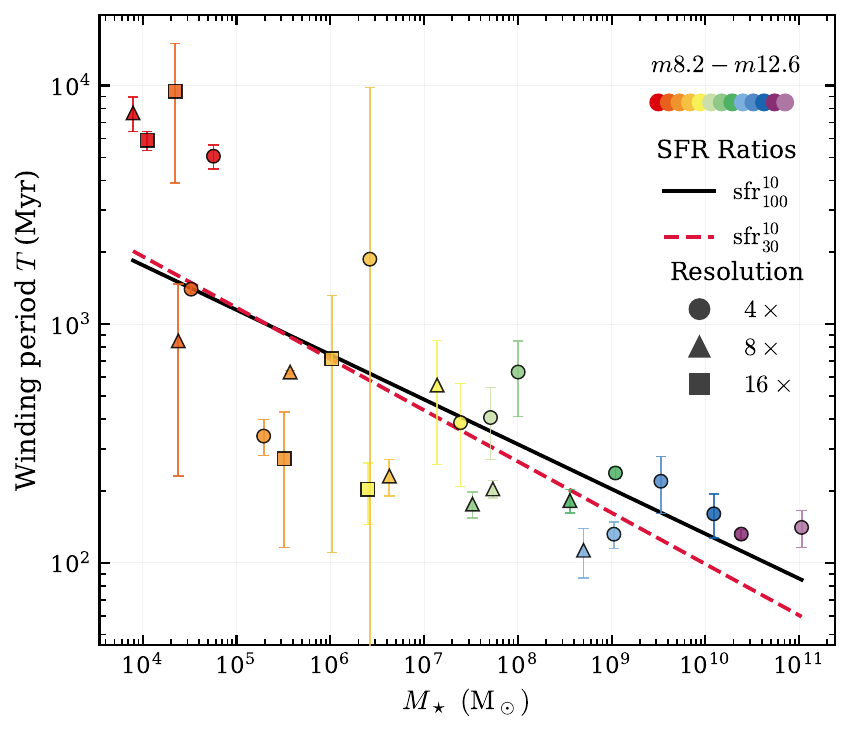}
    \end{minipage}
    \hfill
    \begin{minipage}{0.48\textwidth}
        \centering
        \includegraphics[width=\linewidth]{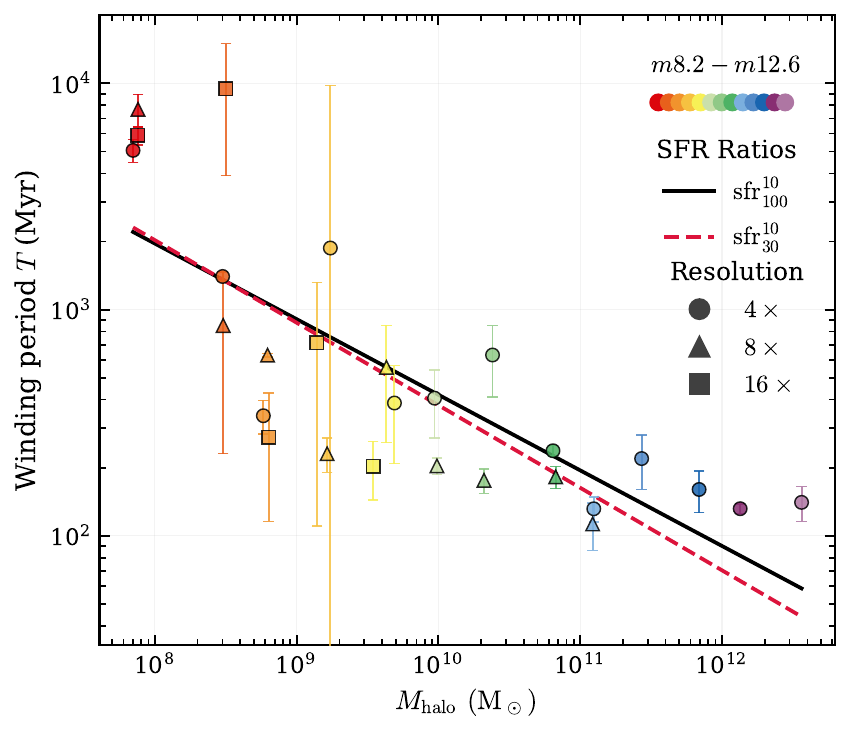}
    \end{minipage}
    \caption{Fitted winding period of the burstiness--escape trajectory as a function of stellar mass (left) and halo mass (right), with both masses evaluated at $z=3$. Points show $T_{10/100}$ for individual target--resolution pairs, with the period uncertainties described in Appendix~\ref{appendix:polar_period}. Colours identify the targets, and circles, triangles, and squares denote 4$\times$, 8$\times$, and 16$\times$ resolution. The black solid and red dashed lines show the pooled fits for $\text{sfr}^{10}_{100}$ and $\text{sfr}^{10}_{30}$, respectively, using the fiducial 100\,Myr moving centre. Both fits give shorter winding periods towards higher mass.}
    \label{fig:combined_period}
\end{figure*}

Extending the polar analysis to the full halo sample reveals a systematic dependence of the cycle period on galaxy mass (Figure~\ref{fig:combined_period}). The best-fitting power-law relations for the period are
\begin{equation} \label{eq:polar_relation}
  \frac{T_{10/100}}{\text{Myr}} \approx
  \begin{cases}
    313.9 \left(M_\star/10^{8}\Msun\right)^{-0.188 \pm 0.019} \\
    420.4\left(M_\text{halo}/10^{10}\Msun\right)^{-0.334 \pm 0.016} \, .
  \end{cases}
\end{equation}
The stellar- and halo-mass fits have $R^2 = 0.692$ and $0.778$, respectively. Using $\text{sfr}^{10}_{30}$ gives
\begin{equation} \label{eq:polar_relation_30}
  \frac{T_{10/30}}{\text{Myr}} \approx
  \begin{cases}
    265.8 \left(M_\star/10^{8}\Msun\right)^{-0.215 \pm 0.016} \\
    377.9\left(M_\text{halo}/10^{10}\Msun\right)^{-0.365 \pm 0.016} \, .
  \end{cases}
\end{equation}
The fits have $R^2 = 0.765$ and $0.819$, respectively.

The negative slopes indicate shorter fitted winding periods at higher mass. The stellar- and halo-mass relations describe the same population trend, with their different slopes reflecting in part the relation between stellar and halo mass. The $\text{sfr}^{10}_{30}$ fits have similar slopes and pivot normalizations to the $\text{sfr}^{10}_{100}$ fits in this analysis. In low-mass systems, feedback-driven outflows can travel farther through the shallow gravitational potential, and the gas requires more time to return and initiate another burst. The longer and less regular recycling process naturally produces extended burstiness--escape-fraction cycles. More massive haloes retain their gas more effectively and re-accrete displaced material more rapidly, allowing star formation to resume sooner after feedback suppresses a burst. This interpretation connects the period relation to the mass-dependent star-formation variability measured within \thesanzoom \citep{McClymontZoomSFMS2025} and to the short mini-quenched phases produced by bursty feedback in other high-redshift simulations \citep{Dome2024}. At fixed redshift, the virial dynamical time depends mainly on the enclosed overdensity and only weakly on halo mass, so gas retention and recycling offer a more direct interpretation of this mass trend.

\subsection{Time lag}
\label{subsec:lag}
\begin{figure}
    \centering
     \includegraphics[width=\linewidth]{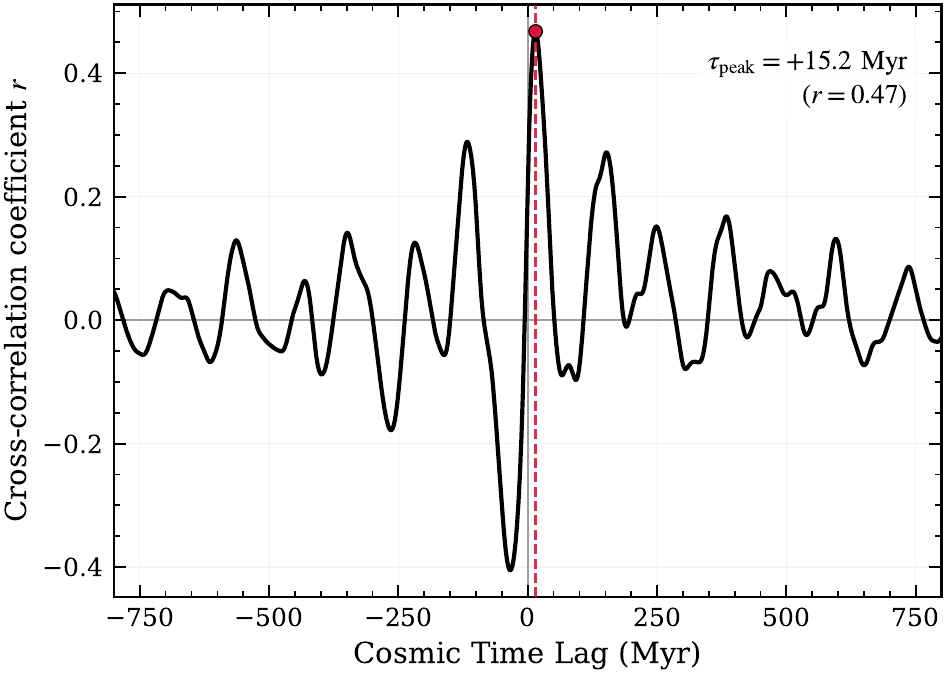}
    \caption{Lag-correlation function for halo m12.6 at 4$\times$ resolution computed using the FFT-evaluated cross-correlation and overlap weighting in Equations~(\ref{eq:fft_crosscorr}) and (\ref{eq:overlap_weight}). Recurrent structure produces multiple peaks across the starburst history. The red circle marks the local maximum nearest zero at positive lag, adopted as the characteristic delay. Its $15.2$\,Myr offset and $r_\text{peak}=0.47$ describe a short, moderately correlated association between burstiness and the escape fraction.}
    \label{fig:sunkit-time-lag-B}
\end{figure}

Within these simulations, the combined response to photoionization, stellar winds, and supernovae can produce a delay between a burst of star formation and the subsequent rise in LyC escape. Newly formed massive stars increase the intrinsic ionizing-photon production rate almost immediately, but much of this radiation can initially be absorbed by dense neutral gas around the young population. As feedback changes that gas distribution, the neutral covering fraction can fall and low-density escape paths can develop \citep{Kimm2014,Trebitsch2017,Kimm2019}. The relative contributions of the individual feedback channels are not isolated by the lag analysis. The selected offset instead summarises the delayed association in the integrated histories, which may combine several episodes. Its normalisation, and especially the unresolved earliest response, can therefore change if the strength or timing of radiative, early-momentum, wind, or supernova feedback is altered. We measure the offset between the burstiness parameter and the LyC escape fraction with the FFT-evaluated cross-correlation described in Section~\ref{sec:lagmethod}. Figure~\ref{fig:sunkit-time-lag-B} shows the resulting lag-correlation function for halo m12.6. Its nearest local maximum at positive lag gives the shortest delayed association between enhanced star formation and the subsequent rise in LyC escape, while $r_\text{peak}$ describes the coherence of that response.

Massive haloes typically show short cross-correlation lags of $\sim16\,\text{Myr}$, with peak strengths of $r_\text{peak}\sim0.4$--$0.6$. These values indicate a relatively coherent delayed response of LyC escape to recent star formation. Because the lags are comparable to the native $\approx10$\,Myr snapshot spacing, interpolation cannot determine the precise timing of the response between snapshots or rule out a prompt component.

Towards lower halo masses, the characteristic lag increases while $r_\text{peak}$ decreases, indicating a less consistent temporal relationship between burstiness and escape. The weak peaks in the lowest-mass haloes are consistent with feedback episodes that vary in strength, duration, and geometry from one burst to the next. Lags of $\sim100$\,Myr may reflect gas displacement over a larger fraction of the halo, changes in the orientation of transparent sightlines, or an escape maximum associated with a later star-forming episode. This offset measures an association across the galaxy's history rather than the age of the stars supplying the escaping photons. Continuing or renewed star formation can provide young ionizing sources after earlier generations have opened escape paths, as found in FIRE simulations \citep{Ma2020}. A high escape fraction contributes little ionizing radiation if that young source population is absent. The lag therefore describes the timing of transparency, while the escaping photon rate determines its radiative contribution (Section~\ref{subsec:source}).

\begin{figure}
    \centering
    \includegraphics[width=\columnwidth]{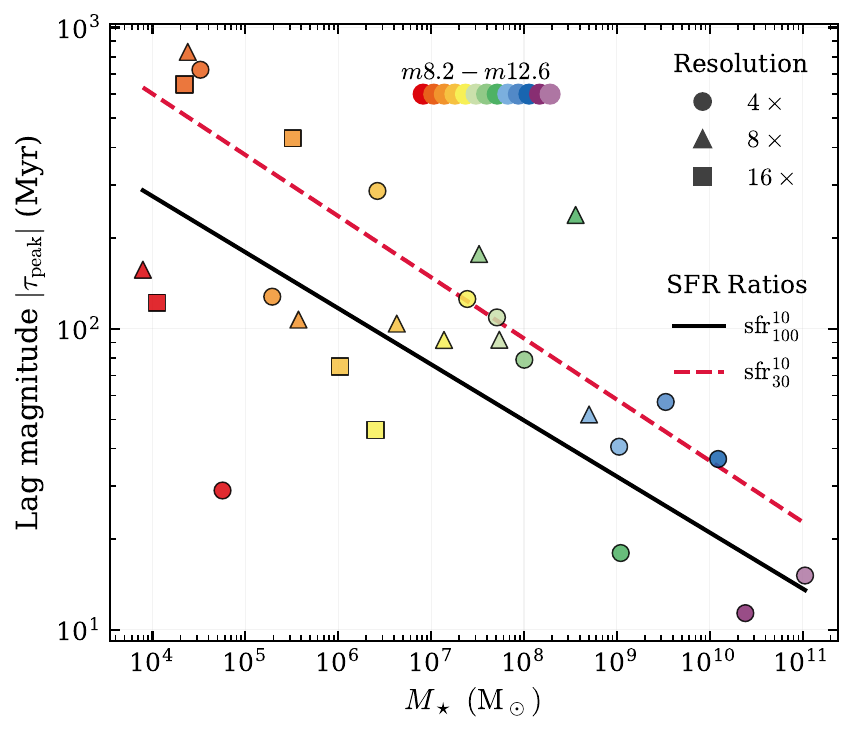}

    \includegraphics[width=\columnwidth]{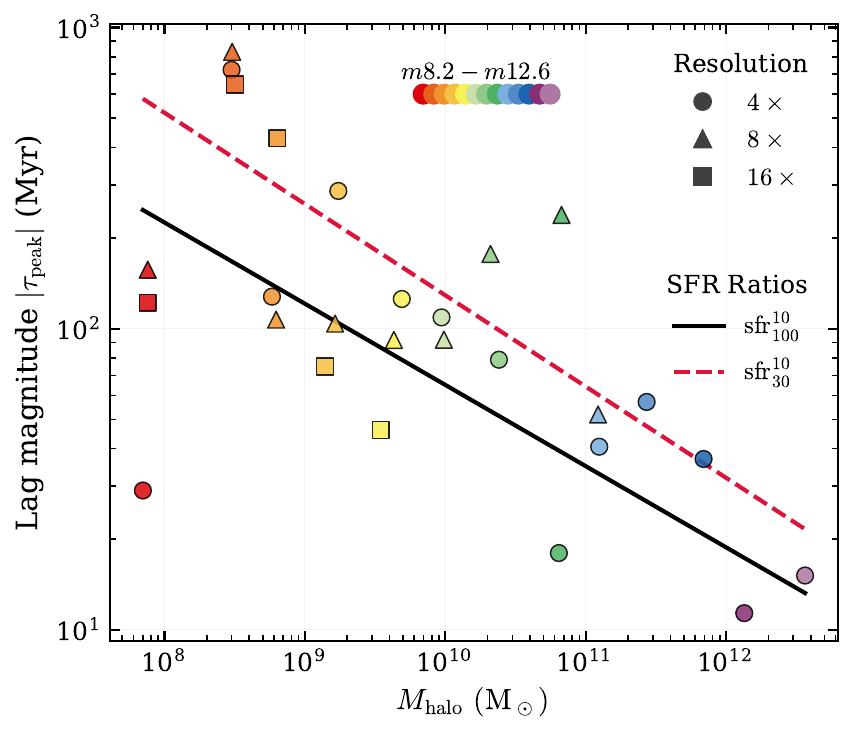}
    \caption{Magnitude of the selected lag between the burstiness parameter and the LyC escape fraction as a function of stellar mass (top) and halo mass (bottom). Points show $|\tau_{\text{peak},\,10/100}|$, with colours identifying targets and marker shapes denoting resolution as in Figure~\ref{fig:combined_period}. The black solid and red dashed lines show the $r_\text{peak}$-weighted fits for $\text{sfr}^{10}_{100}$ and $\text{sfr}^{10}_{30}$, respectively. Both fits retain the fallback estimates described in Appendix~\ref{appendix:polar_period}. The fitted lag magnitudes decrease towards higher mass, with a larger normalization for the 10/30 ratio.}
\label{fig:lag_power}
\end{figure}

Across the halo sample, the measured lag magnitudes decrease systematically with both stellar and halo mass. Weighting each measurement by $r_\text{peak}$ gives the power-law relations shown in Figure~\ref{fig:lag_power},
\begin{equation} \label{eq:lag_relation}
  \frac{|\tau_{\text{peak},\,10/100}|}{\text{Myr}} \approx
  \begin{cases}
    49.59\left(M_\star/10^{8}\Msun\right)^{-0.186 \pm 0.028}\\
    65.03\left(M_\text{halo}/10^{10}\Msun\right)^{-0.270 \pm 0.039} \, .
  \end{cases}
\end{equation}
The stellar- and halo-mass fits have $R^2 = 0.592$ and $0.581$, respectively. Using $\text{sfr}^{10}_{30}$ gives
\begin{equation} \label{eq:lag_relation_30}
  \frac{|\tau_{\text{peak},\,10/30}|}{\text{Myr}} \approx
  \begin{cases}
    92.81\left(M_\star/10^{8}\Msun\right)^{-0.203 \pm 0.034}\\
    129.0\left(M_\text{halo}/10^{10}\Msun\right)^{-0.303 \pm 0.048} \, .
  \end{cases}
\end{equation}
The fits have $R^2 = 0.434$ and $0.467$, respectively.

The negative slopes are consistent with gravitational confinement regulating the gas response to feedback. Using $\text{sfr}^{10}_{30}$ instead of $\text{sfr}^{10}_{100}$ gives similar mass slopes but approximately twice the lag normalization at the adopted mass pivots. Both definitions therefore give smaller characteristic offsets at higher mass, while the normalization depends on the interval used to define the star-formation baseline.

\begin{figure}
    \centering
    \includegraphics[width=\linewidth]{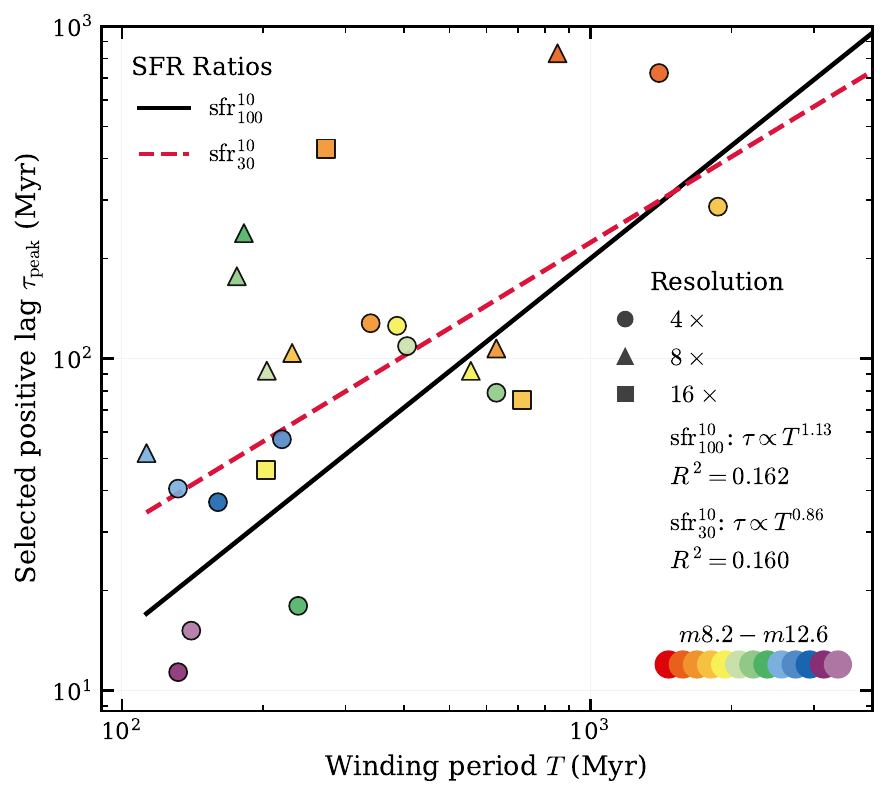}
    \caption{Relationship between the selected positive lag and winding period. Points show $\tau_{\text{peak},\,10/100}$ against $T_{10/100}$, with target colours and resolution symbols as in Figure~\ref{fig:combined_period}. The black solid and red dashed lines show the $r_\text{peak}$-weighted fits for the 10/100 and 10/30 ratios, respectively. Shorter lags tend to accompany shorter winding periods, with substantial scatter. The fits have $R^2=0.162$ and $0.160$, respectively.}
    \label{fig:lag_vs_polar}
\end{figure}

\subsection{Relationship between the lag and the cycle period}
\label{sec:lag_vs_period}

The lag and the winding period probe different aspects of the feedback-regulated evolution. The lag describes the offset between enhanced recent star formation and increased halo transparency, while the period follows a full winding of the joint burstiness--escape trajectory. This recurrent evolution can include gas clearing, suppression of star formation, larger-scale displacement, re-accretion, and renewed activity. The two diagnostics thus connect the response of LyC escape to the longer sequence of gas removal and replenishment.

Systems with shorter lags tend to have shorter winding periods (Figure~\ref{fig:lag_vs_polar}). Gas retention offers a physical link between the two timescales: displacing gas from the star-forming region can open escape paths, while the distance travelled by that gas affects when it returns to fuel another burst. The association has substantial scatter, and both timescales depend on mass. A comparison at fixed mass is therefore needed to establish how much of their association extends beyond this shared dependence.

With each measurement weighted by its selected peak cross-correlation, $r_\text{peak}$, the power-law relations shown in Figure~\ref{fig:lag_vs_polar} are
\begin{equation}
    \label{lag_period_relation}
    \frac{\tau_{\text{peak},\,10/100}}{\text{Myr}}
    \approx 14.87\left(\frac{T_{10/100}}{100\,\text{Myr}}\right)^{1.129\,\pm\,0.282}
\end{equation}
and
\begin{equation}
    \label{lag_period_relation_s30}
    \frac{\tau_{\text{peak},\,10/30}}{\text{Myr}}
    \approx 31.00\left(\frac{T_{10/30}}{100\,\text{Myr}}\right)^{0.858\,\pm\,0.214} \, .
\end{equation}
The 10/100 and 10/30 fits have $R^2 = 0.162$ and $0.160$, respectively.

The scatter around Equations~(\ref{lag_period_relation}) and~(\ref{lag_period_relation_s30}) can arise from stochastic variations among burst episodes and from the different aspects of the recurrent histories captured by the two estimators. Both fits describe a population-level tendency, with a broad range of lags at a given winding period.

\subsection{Source- and environment-regulated quantities}
\label{subsec:source}
\begin{figure}
\centering
\includegraphics[width=\columnwidth]{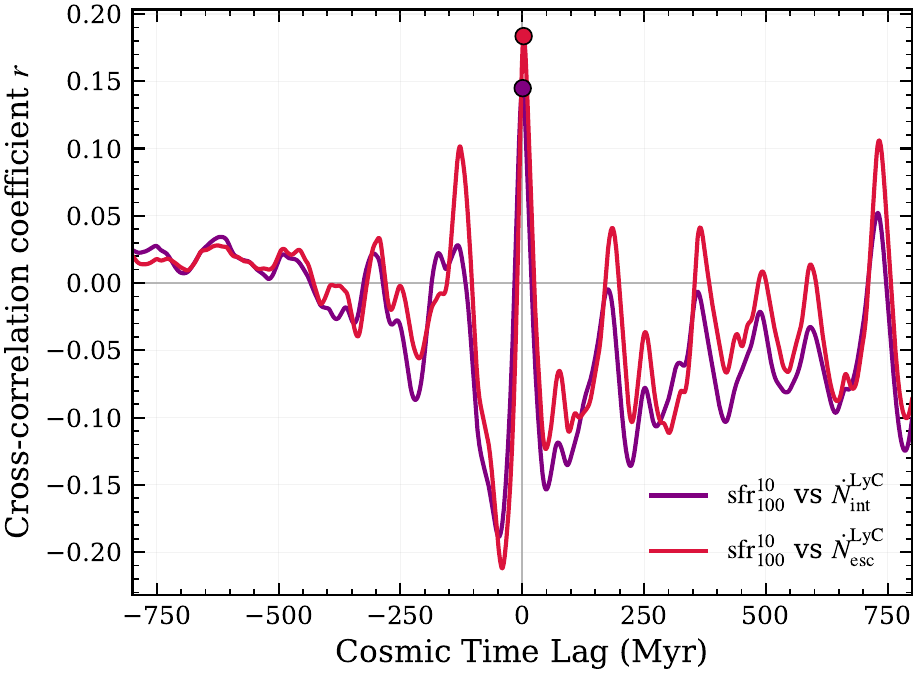}

\vspace{0.4em}

\includegraphics[width=\columnwidth]{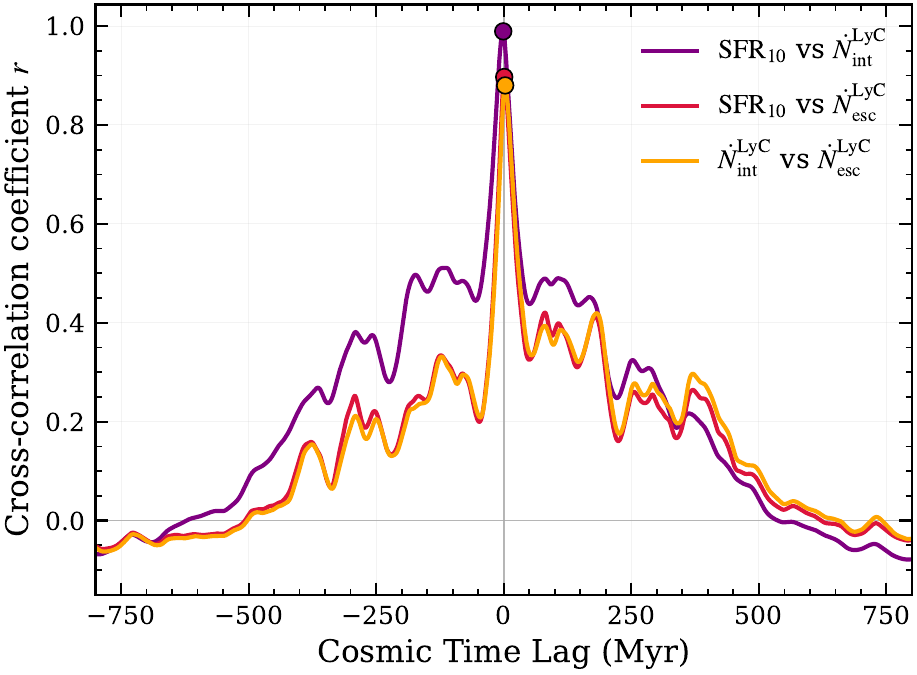}

\caption{Lag-correlation functions for selected combinations of star-formation and LyC-related quantities in m12.6 at 4$\times$ resolution. The upper panel compares $\text{sfr}^{10}_{100}$ with the intrinsic and escaping LyC photon rates, showing weak correlations. The lower panel compares $\text{SFR}_{10}$ with each photon rate and the intrinsic with the escaping photon rate. Circles mark the selected peaks. The strong peaks are consistent with no delay resolved by the native snapshot cadence.}

\label{fig:heatmaps_p1}
\end{figure}

The temporal correlation structure separates quantities governed by the stellar sources from those regulated by the surrounding ISM. Direct tracers of young stars should correlate strongly with little or no temporal offset, whereas quantities sensitive to feedback and gas geometry may correlate more weakly or respond after a delay. We test this distinction using SFRs averaged over 10 and 100 Myr, the burstiness parameter, the intrinsic and escaping LyC photon production rates $(\dot{N}_\text{int}^\text{LyC}$ and $\dot{N}_\text{esc}^\text{LyC})$, and the LyC escape fraction. The combinations shown in Figure~\ref{fig:heatmaps_p1} reveal whether variability in the ionizing output that leaves a halo is inherited primarily from the stellar source or from changes in the intervening gas.

The clearest source-regulated behaviour appears in the correlations involving $\text{SFR}_{10}$. For m12.6 at 4$\times$ resolution, $\text{SFR}_{10}$ has a sharp near-zero-lag peak with both $\dot{N}_\text{int}^\text{LyC}$ and $\dot{N}_\text{esc}^\text{LyC}$, with $r_\text{peak}=0.99$ and $0.90$, respectively. The intrinsic and escaping production rates are similarly well correlated ($r_\text{peak}=0.88$) with a lag of only $2.8\,\text{Myr}$, consistent with no delay resolved by the snapshot cadence. This close correspondence links the variability of the escaping radiation to the evolving population of young, massive stars. The absolute escaping photon rate is set by $\dot{N}_\text{esc}^\text{LyC}=f_\text{esc}^\text{LyC}\dot{N}_\text{int}^\text{LyC}$, so the relative timing of photon production and gas transparency determines the contribution of each burst phase. The aligned histories in Figures~\ref{fig:galaxy_rt} and~\ref{fig:peak_overplot} illustrate this distinction: a high escape fraction during a fading burst can coincide with a small escaping photon rate. The top panel of Figure~\ref{fig:heatmaps_p1} shows that the burstiness parameter correlates much more weakly with either photon production rate ($r_\text{peak}=0.14$--$0.18$). As a ratio of recent to longer-term star formation, $\text{sfr}^{10}_{100}$ tracks the phase of the star-formation history, while the photon rates respond to the absolute number of ionizing stars.

The escape fraction exhibits a different temporal signature. The burstiness--escape curve for m12.6 reaches $r_\text{peak}=0.47$ at $15.2\,\text{Myr}$ (Figure~\ref{fig:sunkit-time-lag-B}). These individual-track correlations illustrate the distinction between source strength and burst phase. The population-level lag--mass relations in Section~\ref{subsec:lag} describe how the delayed association between burstiness and escape varies across the sample. This temporal ordering supplies the physical context for the snapshot predictors measured by \citet{Summerfield2026}. Burstiness and gas content carry the strongest information about $f_\text{esc}^\text{LyC}$ because they encode the stage and gaseous state of the feedback cycle, whereas UV luminosity and SFR-related quantities predict the escaping photon rate mainly through the strength of the stellar source. The delayed rise in escape traces the gas response to the combined radiative and mechanical feedback in \thesanzoom. Section~\ref{sec:nearby_comparison} compares this sequence with the early and delayed escape phases inferred for nearby galaxies.

\subsection{Observable signatures of delayed LyC escape}
\label{sec:obs_signatures}

\begin{figure}
    \centering
    \includegraphics[width=\linewidth]{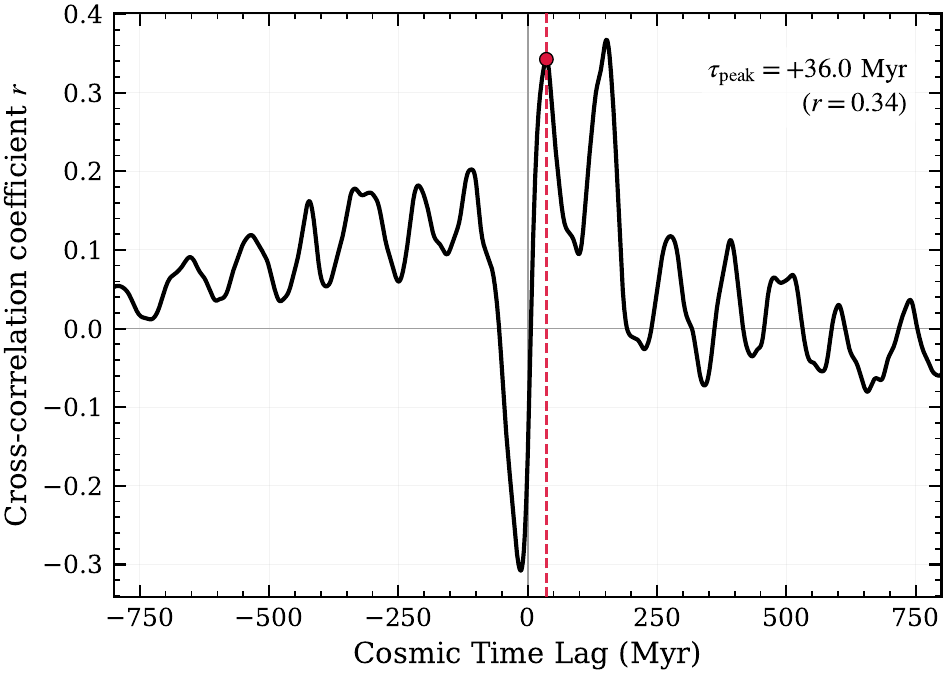}
    \caption{Lag-correlation function between the observable burstiness proxy, $\eta_{1500}$, and the LyC escape fraction for halo m12.6 at 4$\times$ resolution. The nearest positive local maximum occurs at $36.0\,\text{Myr}$ with $r_\text{peak}=0.34$. The red circle marks the selected positive peak. The H$\alpha$/UV proxy retains the delayed association with escape, with a longer lag and weaker correlation than the intrinsic burstiness parameter. A negative trough at $-14.2$\,Myr accompanies the positive peaks.}
    \label{fig:sunkit-time-lag-obs}
\end{figure}

The intrinsic burstiness parameter $\text{sfr}^{10}_{100}$ relies on SFRs averaged over fixed time windows and cannot be measured directly. We therefore use the observationally motivated proxy $\eta_{1500}$ defined in Equation~(\ref{eq:eta_1500}), which compares H$\alpha$ emission from short-lived massive stars with the longer-timescale UV continuum. If the delay identified above reflects a physical sequence from enhanced star formation to feedback-driven LyC escape, its temporal ordering should persist when burstiness is measured with these synthetic observables. For halo m12.6, the lag-correlation function has its nearest positive local maximum at $36.0\,\text{Myr}$ with $r_\text{peak}=0.34$ (Figure~\ref{fig:sunkit-time-lag-obs}). The corresponding intrinsic burstiness measurement gives a shorter lag of $15.2\,\text{Myr}$ and a stronger $r_\text{peak}=0.47$ (Figure~\ref{fig:sunkit-time-lag-B}). The H$\alpha$/UV proxy therefore retains the delayed association with LyC escape, with a longer lag and weaker correlation. The two diagnostics respond differently to the same star-formation history. The intrinsic ratio averages star formation over fixed 10 and 100\,Myr windows, while H$\alpha$ and UV luminosities weight stars continuously over different, overlapping age ranges. These different responses, together with time-dependent attenuation, change the shape and timing of the burstiness signal and can shift its cross-correlation peak with LyC escape. The measured lag therefore depends on the tracer as well as the evolution of gas transparency, and the longer lag in this example does not imply a longer gas-clearing time \citep{Endsley2025,Munoz26}.

The same curve also has a negative trough at $-14.2$\,Myr with $C^w_{AB}\simeq-0.31$ and a stronger positive peak at a later lag. Leakage can contribute to this anti-correlation: at fixed photon production, photons that escape are unavailable for hydrogen recombinations, reducing H$\alpha$. A quantitative correction must also account for absorption by dust and helium, collisional excitation, CGM absorption, stellar ages, and aperture differences, so halo-scale $f_\text{esc}^\text{LyC}$ cannot be applied as a simple $1-f_\text{esc}^\text{LyC}$ correction to H$\alpha$.

Across the halo and resolution measurements shown, the magnitude of the selected lag between $\eta_{1500}$ and the LyC escape fraction decreases with both stellar and halo mass (Figure~\ref{fig:lag_obs_power}). The residual-weighted fits described in Appendix~\ref{appendix:polar_period} give
\begin{equation} \label{eq:obs_relation}
  \frac{|\tau_{\text{peak},\,\eta_{1500}}|}{\text{Myr}} \approx
  \begin{cases}
    108.3 \left(M_\star/10^{8}\Msun\right)^{-0.204\pm0.025} \\
    146.7\left(M_\text{halo}/10^{10}\Msun\right)^{-0.303\pm0.046}.
  \end{cases}
\end{equation}
The stellar- and halo-mass projections of this trend have $R^2=0.645$ and $0.573$, respectively, with substantial object-to-object scatter. Their slopes are close to the intrinsic burstiness relations in Equation~(\ref{eq:lag_relation}), showing that the observable proxy retains the mass dependence of the delayed response. Spatial changes in the apparent distribution of recent star formation further affect how individual episodes appear in these tracers \citep{WangZoomClumps2026}. Samples spanning different burst phases should exhibit a mass-dependent association between $\eta_{1500}$ and indirect indicators of LyC leakage. Emission-line and Ly$\alpha$ diagnostics are themselves sensitive to time-dependent gas structure \citep{Barrow2020,Choustikov2024}, so jointly interpreting them with H$\alpha$/UV can distinguish a recently brightened, embedded source from a fading system whose escape channels remain open. Population-integrated tracers offer a complementary test, since delayed-supernova models predict that burstiness preferentially boosts the mean [C\,\textsc{ii}] luminosities of low-mass galaxies and adds excess large-scale clustering power to [C\,\textsc{ii}] intensity maps \citep{Liu2024BurstCII}.

\begin{figure}
    \centering
    \includegraphics[width=\columnwidth]{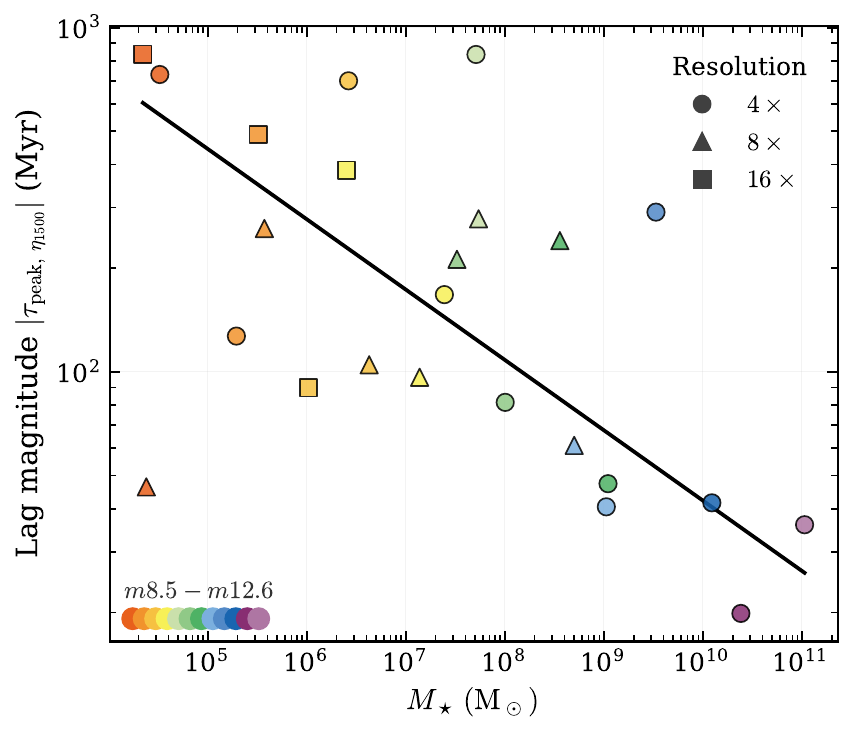}

    \includegraphics[width=\columnwidth]{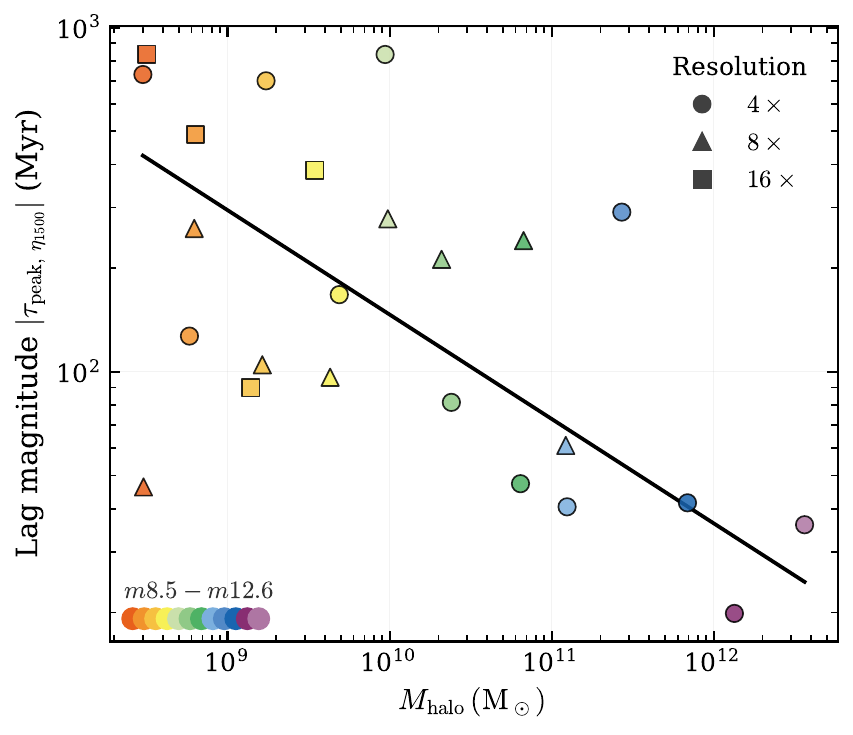}
    \caption{Magnitude of the selected lag $|\tau_{\text{peak},\,\eta_{1500}}|$ between the observable burstiness proxy $\eta_{1500}$ and the LyC escape fraction as a function of stellar mass (top) and halo mass (bottom). The 24 measurements represent target--resolution pairs. The m8.2 target is excluded from this comparison. Colours distinguish the target haloes, while circles, triangles, and squares denote the 4$\times$, 8$\times$, and 16$\times$ resolutions, respectively. The solid lines show residual-weighted power-law fits. The stellar- and halo-mass relations have $R^2=0.645$ and $0.573$, respectively.}
\label{fig:lag_obs_power}
\end{figure}

\subsection{Comparison with nearby LyC leakers}
\label{sec:nearby_comparison}

Nearby LyC emitters provide the most detailed empirical view of the conditions associated with escape. Strong leakers are frequently highly ionized, compact systems with concentrated star formation. The LyC detection rate and escape fraction both rise towards high $\text{O}_{32} \equiv [\text{O\,\textsc{iii}}]\,\lambda5007\,\text{\AA}/[\text{O\,\textsc{ii}}]\,\lambda\lambda3726,3729\,\text{\AA}$, small UV sizes, and large SFR surface densities \citep{Flury2022,Jaskot2025}. However, these properties depend on mass, metallicity, density, ionization parameter, and sightline geometry, explaining the large scatter. Spatially resolved observations reinforce the importance of morphology, with the strongest leakers being generally compact in both Mg\,\textsc{ii} and [O\,\textsc{ii}], whereas weak and non-leakers span a much broader range of gas extents \citep{Leclercq2024}.

\begin{figure*}
    \centering
    \begin{minipage}{\columnwidth}
        \centering
        \includegraphics[width=\linewidth]{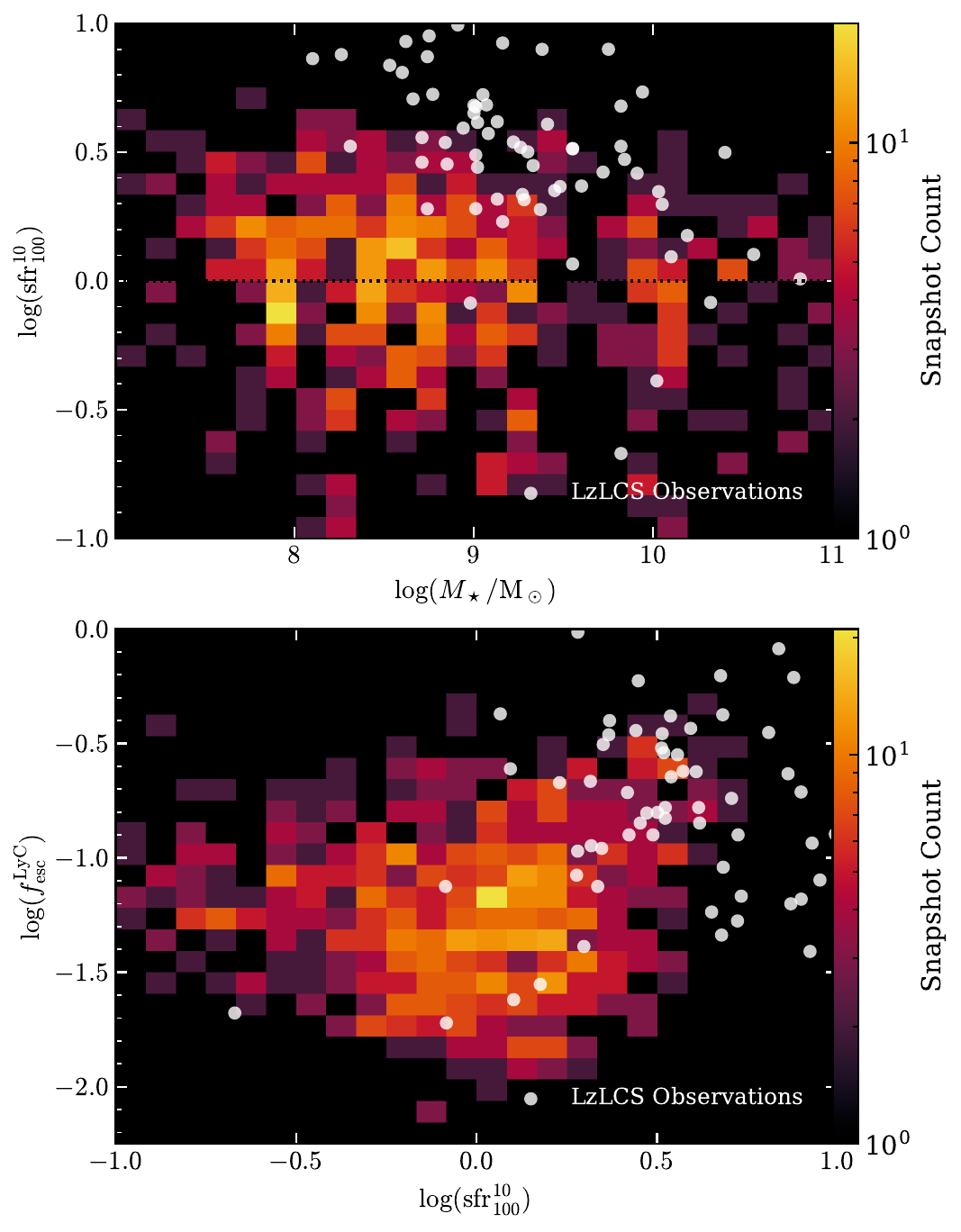}
    \end{minipage}
    \hfill
    \begin{minipage}{\columnwidth}
        \centering
        \includegraphics[width=\linewidth]{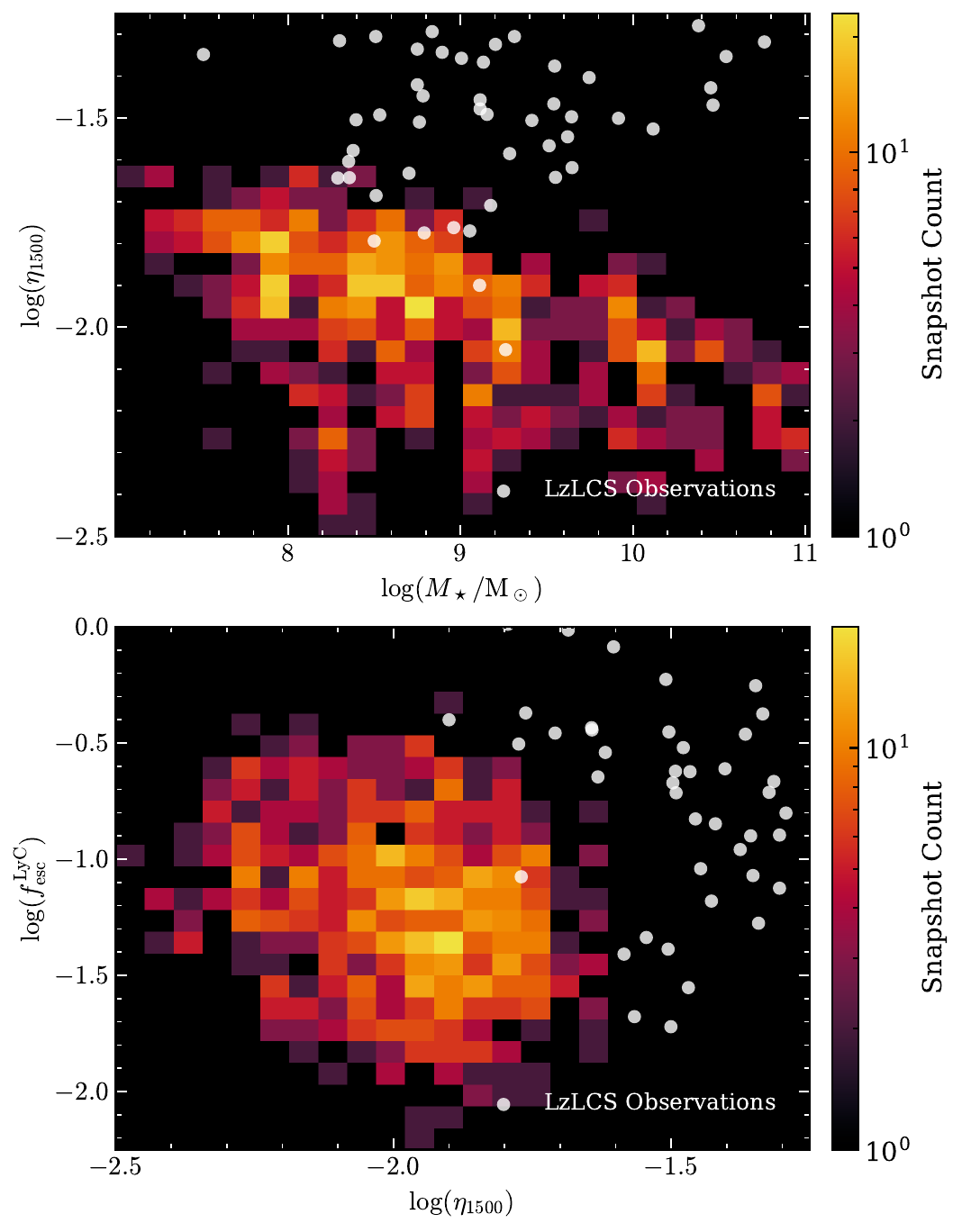}
    \end{minipage}
    \caption{Comparison of simulated galaxies with LzLCS. The left column shows $\log(\text{sfr}^{10}_{100})$ against stellar mass (top) and $\log(f_\text{esc}^\text{LyC})$ (bottom), while the right column shows the corresponding relations for $\log(\eta_{1500})$. Coloured two-dimensional histograms count target--resolution--epoch samples per bin after applying the attenuated-magnitude cut $M_{1500}<-17$. The simulated $\eta_{1500}$ uses emergent H$\alpha$ and UV luminosities within $R_\text{vir}$. Each retained snapshot contributes one count without time weighting. White circles show the LzLCS measurements \citep{Flury2022,Flury_2024,Stoffers2026}. The simulated escape fractions are angle-averaged at the virial radius, whereas the observations sample individual sightlines. }
\label{fig:LzLCS}
\end{figure*}

We make a qualitative cross-sectional comparison with the Low-redshift Lyman Continuum Survey (LzLCS) in Figure~\ref{fig:LzLCS}. The left column compares the intrinsic burstiness parameter with stellar mass and $f_\text{esc}^\text{LyC}$, while the right column compares $\eta_{1500}$ with stellar mass and $f_\text{esc}^\text{LyC}$. To reduce the weight of simulated phases in the high-$f_\text{esc}^\text{LyC}$, low-$\eta_{1500}$ tail that would probably be absent from a flux-limited sample, we include only snapshots with attenuated rest-frame UV absolute magnitude $M_{1500}<-17$.

For LzLCS, we take $M_{1500}$ and the H$\beta$-based SFR from Tables~5 and 4, respectively, of \citet{Flury2022}, including the published corrections \citep{Flury_2024}. The adopted calibration gives $\log(L_{\text{H}\alpha}) = \log(\text{SFR}_{\text{H}\beta}) + 41.27$ when luminosity is in erg\,s$^{-1}$ and SFR is in $\Msun\,\text{yr}^{-1}$. Combining this with $\log(\lambda_{1500}L_{\lambda,1500}) = 35.939 - 0.4 M_{1500}$ yields
\begin{equation}
    \log(\eta_{1500}) = \log(\text{SFR}_{\text{H}\beta}) + 5.331 + 0.4 M_{1500} \, .
\end{equation}
The optical emission-line measurements are corrected for Galactic extinction and internal reddening using the Balmer lines, whereas $M_{1500}$ is corrected for Galactic extinction only. Our synthetic luminosities include attenuation in both bands, but H$\alpha$ absorption is generally weak, with a median loss of about 5 per cent in the UV-selected sample. Correcting the simulated H$\alpha$ luminosities for attenuation while retaining the emergent UV luminosities raises the median $\log(\eta_{1500})$ of this sample by only $0.045$ dex. Individual corrections increase with H$\alpha$ attenuation and reach $0.36$ dex in the most obscured snapshot passing the UV cut. The on-the-fly dust prescription may underestimate attenuation.

The intrinsic $\text{sfr}^{10}_{100}$ values are taken directly from the Prospector fits of \citet{Stoffers2026}, rather than approximated with an H$\alpha$-to-UV ratio. The fits combine GALEX and SDSS photometry with emission-line measurements to constrain non-parametric star-formation histories in eight lookback-time bins, with the first boundaries at 5 and 10\,Myr. The model also includes a free aperture correction between the SDSS spectroscopy and photometry. Their Table~4 reports $\text{sfr}^{10}_{100}$ and its posterior uncertainty for each galaxy, providing an observational estimate of the same SFR ratio.

The LzLCS galaxies lie at systematically larger values of $\text{sfr}^{10}_{100}$ and $\eta_{1500}$ than the densest part of the simulated distribution, but both samples associate stronger recent activity with higher escape fractions. The samples cover different populations: LzLCS selects galaxies well above the typical star-forming main sequence at $z\sim0.3$, while the simulations follow a broader population at $z>3$. The simulations average the escaping flux over solid angle at $R_\text{vir}$, whereas each observation samples one direction through an anisotropic gas distribution \citep{Paardekooper2015}. A transparent sightline can yield a larger escape fraction than the angular mean of the same galaxy, while an obscured direction can yield a non-detection. Random orientations in a complete parent sample would broaden the distribution without necessarily shifting its mean. A LyC-detected subset preferentially samples transparent directions, while the bias in the full observed distribution also depends on selection and the treatment of non-detections. The unweighted snapshot counts show the range of temporal variations rather than a volume-weighted galaxy population. Figure~\ref{fig:LzLCS} thus provides a qualitative reference for interpreting burst phase and leakage across these populations, with selection, viewing geometry, and attenuation contributing to their different distributions.

Nearby studies challenge a simple picture in which LyC escape begins only after supernovae from the current burst clear the surrounding gas. \citet{Carr2025a} find that the strongest leakers in their sample are dominated by radiative feedback and generally lack cool Mg\,\textsc{ii} outflows, while systems with clearer supernova feedback tend to have weaker escape. Using six deliberately similar compact galaxies as a cross-sectional starburst sequence, \citet{Carr2025b} infer that LyC escape peaks early, before a fast galactic wind has developed, and declines as a supernova-driven outflow accelerates and accumulates neutral gas and dust. A subset of galaxies with reliable radio data ($3$--$6\,$GHz spectral indices) shows that strong leakers have flatter continua, consistent with younger bursts and less dominant supernova-powered synchrotron emission \citep{Bait2024}. However, non-thermal emission shows that supernova activity can still be present and a flat spectrum is not predictive of strong escape. In stacked spectra, populations younger than 3\,Myr distinguish the highest-escape stacks, but older populations and mechanical feedback are also common among leakers, motivating a two-stage burst scenario or multiple escape modes rather than a strictly pre-supernova sequence \citep{Flury2025,Jaskot2025}. Indirect escape estimates for the CLASSY sample suggest that single-burst systems can show early escape associated with very young populations, whereas multi-burst systems preferentially show delayed escape after older populations and supernova feedback have had time to modify the ISM \citep{Parker2026}.

J1044+0353 illustrates how successive stellar generations can separate photon production from the clearing of escape paths. Its compact $3$--$4\,$Myr starburst supplies the current ionizing photons, while a $15$--$20\,$Myr population about $1\,$kpc away drove the outflow that excavated the escape path \citep{Olivier2022,Peng2023,Martin2024}. For $M_\star \sim 10^7\,\Msun$, the population relation in Equation~(\ref{eq:lag_relation}) gives a lag of about $76$\,Myr, longer than either stellar age. The observed ages reconstruct one sequence involving spatially distinct populations, while our relation measures correlated variations across integrated halo histories. This sequence shows how a young ionizing population can illuminate channels opened by earlier star formation, so its age alone does not specify how long feedback has been shaping the escape paths.

A similar separation appears in Mrk~71/NGC~2366, an indirectly inferred LyC-escape candidate. Its very young cluster Mrk~71-A dominates the ionizing luminosity, while the older cluster Mrk~71-B is associated with a superbubble blowout \citep{Micheva2017}. The compact molecular gas around Mrk~71-A is inconsistent with a freely expanding energy-driven superwind \citep{Oey2017}, suggesting that the young ionizing source does not have to drive such a wind itself. High-resolution cloud simulations provide a complementary mechanism: radiation-driven dispersal can permit efficient escape from dense, compact star-forming regions before the ionizing population fades \citep{Menon2025}. Prompt leakage from a star-forming region can therefore coexist with escape paths shaped by earlier activity elsewhere in the galaxy.

The adopted stellar-population model affects the overlap between photon production and gas clearing. Binary interactions in BPASS can prolong ionizing emission relative to single-star models such as \textsc{starburst99}, increasing the photon supply after escape paths have opened \citep{Ma2016,Eldridge2017,Ma2020}. Changing the relative contributions of stars of different ages and their ionizing spectra can also alter the photoionization state and escape fraction, even at fixed gas density. Stellar ages inferred from the non-ionizing UV continuum of nearby leakers likewise depend on the adopted spectral models and star-formation history \citep{Flury2025}. These ages describe the populations present at one epoch, while our lags measure the evolving relation between star formation and escape across successive episodes. Comparing the two therefore requires accounting for both the stellar-population assumptions and the different timescales being measured.

The timing of LyC escape therefore depends on both the age of the ionizing population and the feedback history of the surrounding gas. Our analysis follows how photon production and gas transparency evolve together through repeated bursts in individual high-redshift galaxies, measuring the angle-averaged escaping flux at $R_\text{vir}$. The measured lags characterise the delayed association between burstiness and escape across these histories, in which different stellar generations can contribute to photon production and gas clearing. This complements the sequences inferred from nearby populations and resolved regions. Our $\approx10$\,Myr cadence captures the longer variations but does not resolve the $<5$\,Myr phase emphasised by some local studies. Following individual star clusters through to halo-scale escape at finer time resolution would connect these early and delayed modes and test how their relative importance changes with galaxy properties. Spatially resolved studies of $\text{O}_{32}$ could further clarify how ionization structure traces the direction and timing of LyC escape.

\subsection{Recurrence intervals and amplitudes of LyC escape}
\label{sec:peak-trough}

The lag and phase-space analyses indicate that recurrent feedback cycles contribute to the temporal variability of LyC escape. Peak--trough statistics provide a complementary view by asking how often fluctuations above a specified amplitude recur in each individual time series. We identify extrema in the LyC escape fraction, burstiness parameter, SFRs, and LyC photon production rates, then measure the intervals between successive peaks and troughs as described in Appendix~\ref{appendix:peak_finding}. Figures~\ref{fig:peaktime_fesc} and~\ref{fig:peaktime_burst} plot the peak-to-peak intervals against the stellar or halo mass at the intervening trough for thresholds of $0.25$, $0.5$, and $1.0$ dex. Increasing the threshold excludes smaller fluctuations and can group several modest excursions into one longer interval. These amplitude-selected recurrence times describe individual time series and can differ from the winding period of the joint phase-space trajectory.

\begin{figure*}
\centering
\begin{minipage}{0.48\textwidth}
    \centering
    \includegraphics[width=\linewidth]{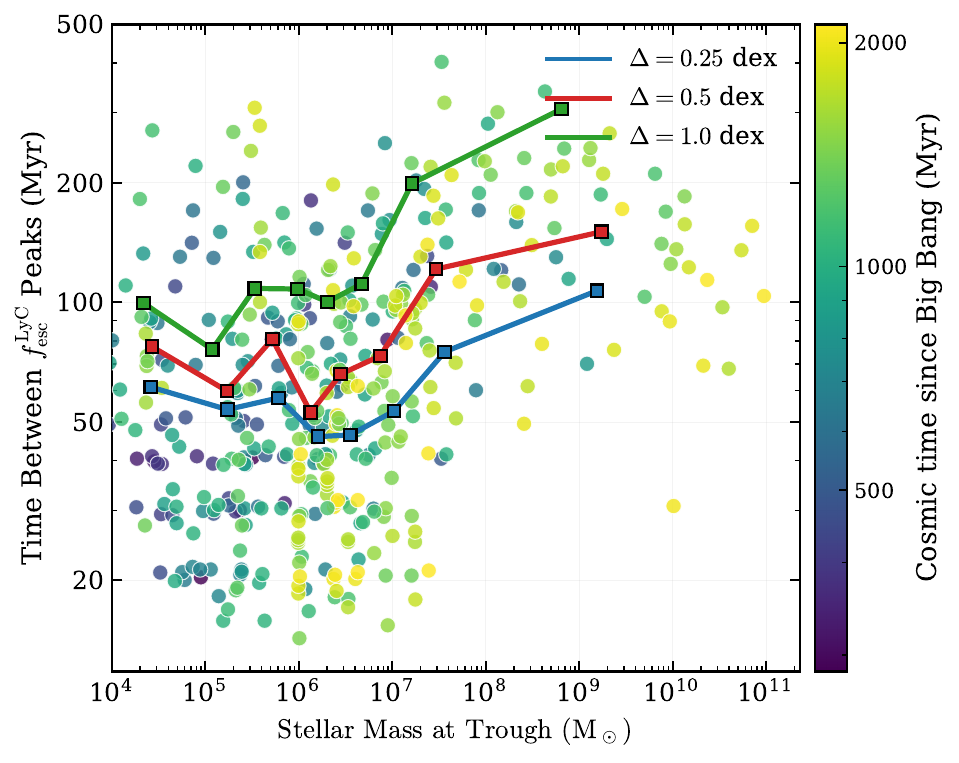}
\end{minipage}
\hfill
\begin{minipage}{0.48\textwidth}
    \centering
    \includegraphics[width=\linewidth]{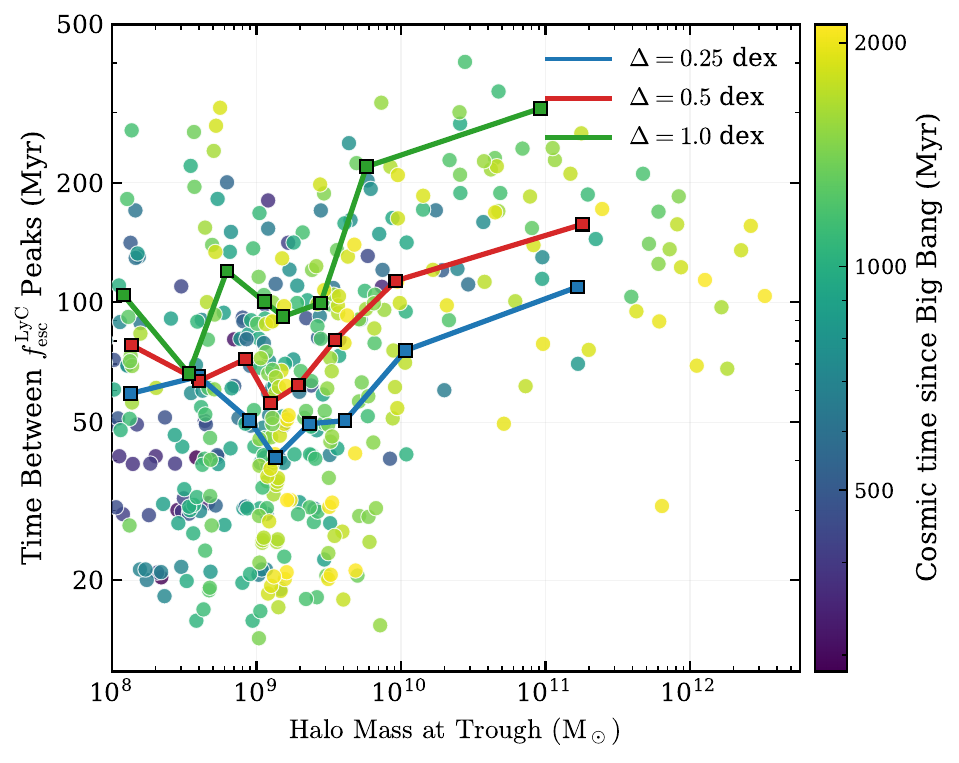}
\end{minipage}
\caption{Peak-to-peak intervals of the LyC escape fraction as functions of stellar mass (left) and halo mass (right), with each interval placed at the mass of its intervening trough. Individual points show detected intervals and are coloured by cosmic time since the Big Bang. The square-linked curves show arithmetic mean intervals in equal-count mass bins, positioned at the geometric mean mass, for peak-identification thresholds of $0.25$, $0.5$, and $1.0$ dex. Larger thresholds select rarer, higher-amplitude excursions and therefore yield systematically longer intervals. The summary trends have a broad minimum near $M_\star\sim10^6$--$10^7\,\Msun$ or $M_\text{halo}\sim10^9\,\Msun$ and rise towards higher masses, while remaining separated by threshold.}
\label{fig:peaktime_fesc}
\end{figure*}

\begin{figure*}
\centering
\begin{minipage}{0.48\textwidth}
    \centering
    \includegraphics[width=\linewidth]{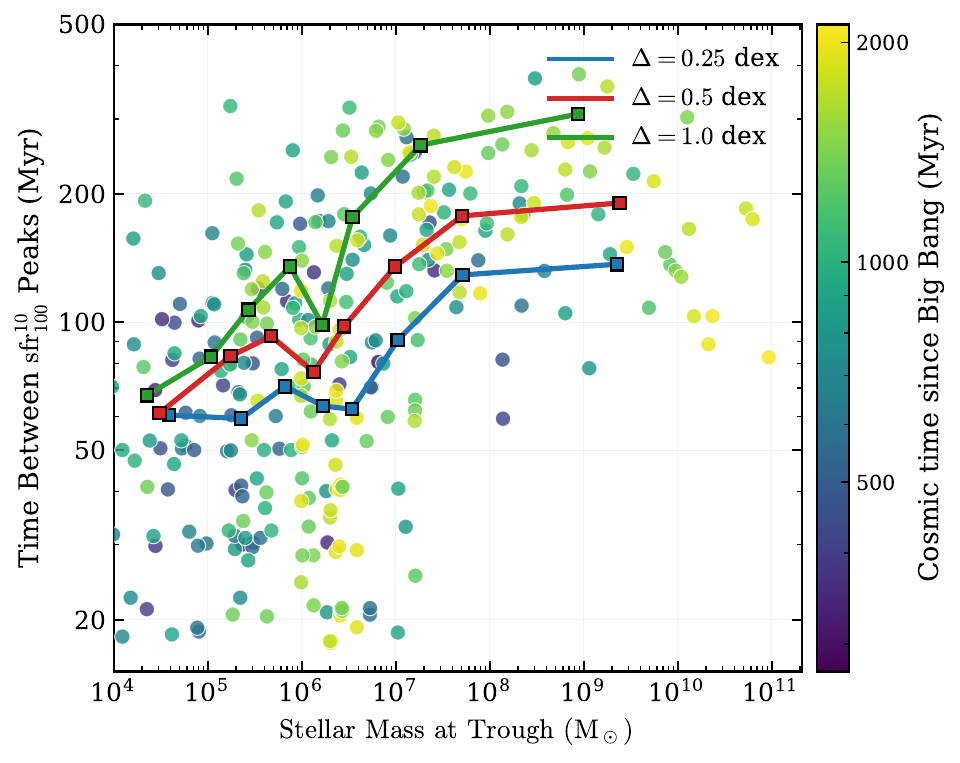}
\end{minipage}
\hfill
\begin{minipage}{0.48\textwidth}
    \centering
    \includegraphics[width=\linewidth]{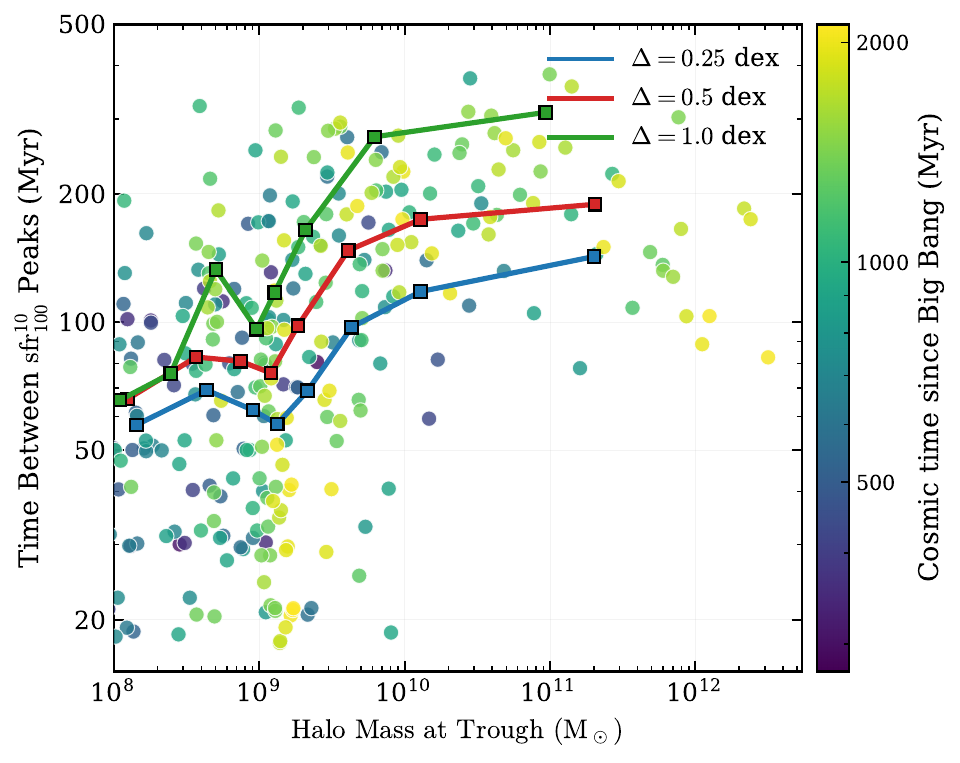}
\end{minipage}
\caption{Peak-to-peak intervals of the burstiness parameter as functions of stellar mass (left) and halo mass (right), evaluated at the mass of the intervening trough. Individual intervals are coloured by cosmic time, and the square-linked curves show the same binned means as in Figure~\ref{fig:peaktime_fesc} for thresholds of $0.25$, $0.5$, and $1.0$ dex. The threshold ordering, broad intermediate-mass minimum, and subsequent high-mass rise resemble those of the LyC escape fraction. The curves remain threshold-dependent because larger thresholds isolate less frequent, higher-amplitude changes in star-formation activity.}
\label{fig:peaktime_burst}
\end{figure*}

The escape-fraction intervals range from tens to several hundred Myr and show substantial scatter at fixed mass. Their summary trends reach a broad local minimum at intermediate masses before increasing again. This minimum may mark a transition where gravitational confinement begins to compete effectively with feedback-driven gas removal. The colour coding shows that the intervals are drawn from different cosmic epochs, and halo growth couples mass to time in this pooled sample. The apparent mass dependence may therefore contain both a genuine change in the feedback regime and secular evolution within the haloes. The persistent separation of the threshold curves is itself physically informative. LyC escape contains variability on several amplitude scales, with large excursions recurring less frequently than modest fluctuations.

The burstiness parameter exhibits the same threshold ordering, a similar intermediate-mass minimum, and an increase in the recurrence time of prominent peaks towards higher masses. This correspondence connects star-formation variability and LyC escape through a common hierarchy of temporal structure. The high-mass rise of the peak-to-peak intervals contrasts with the declining fitted phase-space period in Equation~(\ref{eq:polar_relation}). The peak analysis measures the spacing of amplitude-selected extrema in one quantity and can skip intermediate windings, whereas the polar analysis follows net winding of the joint trajectory. The two diagnostics therefore capture different aspects of the recurrent evolution.

\begin{figure*}
\centering
\begin{minipage}{0.48\textwidth}
    \centering
    \includegraphics[width=\linewidth]{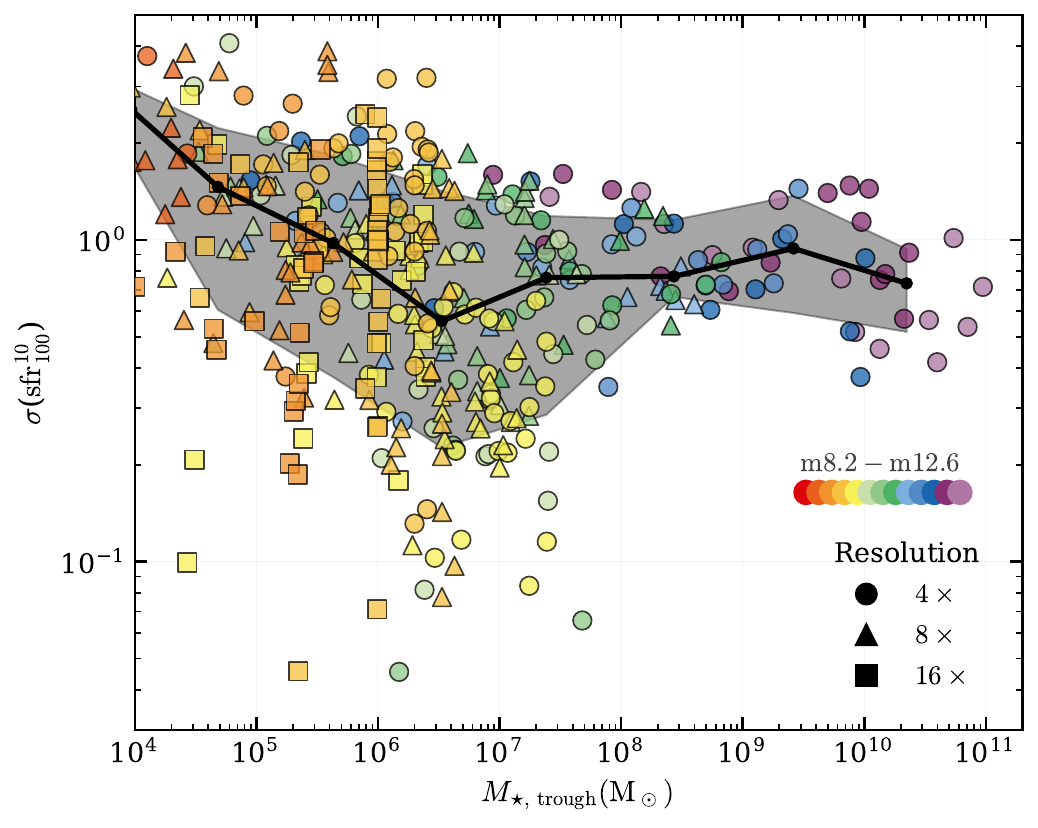}
\end{minipage}
\hfill
\begin{minipage}{0.48\textwidth}
    \centering
    \includegraphics[width=\linewidth]{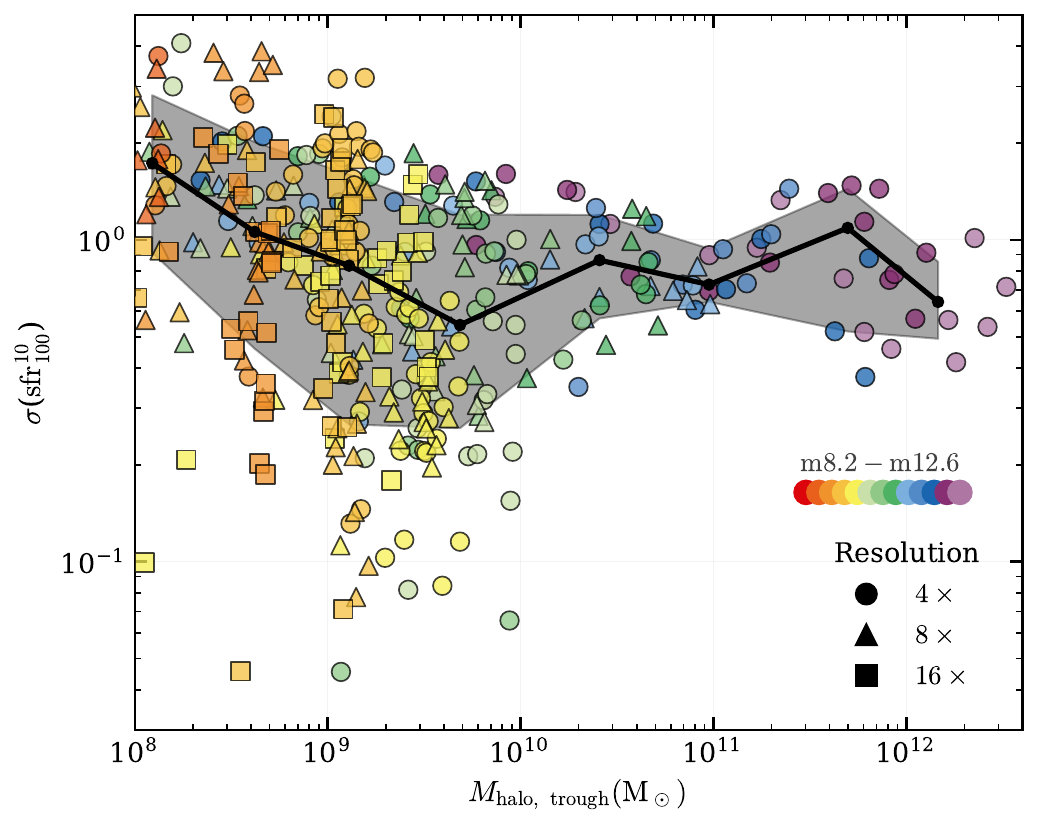}
\end{minipage}
\caption{Variability of $\text{sfr}^{10}_{100}$ within intervals bounded by consecutive $f_\text{esc}^\text{LyC}$ peaks, shown as a function of stellar mass (left) and halo mass (right). Individual measurements are separated by target halo and resolution, and the black curves show medians in mass bins, with grey shading spanning the 16th--84th percentiles of the interval measurements. Colours identify targets, while circles, triangles, and squares denote 4$\times$, 8$\times$, and 16$\times$ resolution. Masses are evaluated at the intervening trough. Binned summaries include measurements outside the displayed ranges. The wide low- and intermediate-mass distributions narrow towards high mass, although the median trend is non-monotonic.}
\label{fig:std_eta_cycle}
\end{figure*}

\begin{figure*}
\centering
\begin{minipage}{0.48\textwidth}
    \centering
    \includegraphics[width=\linewidth]{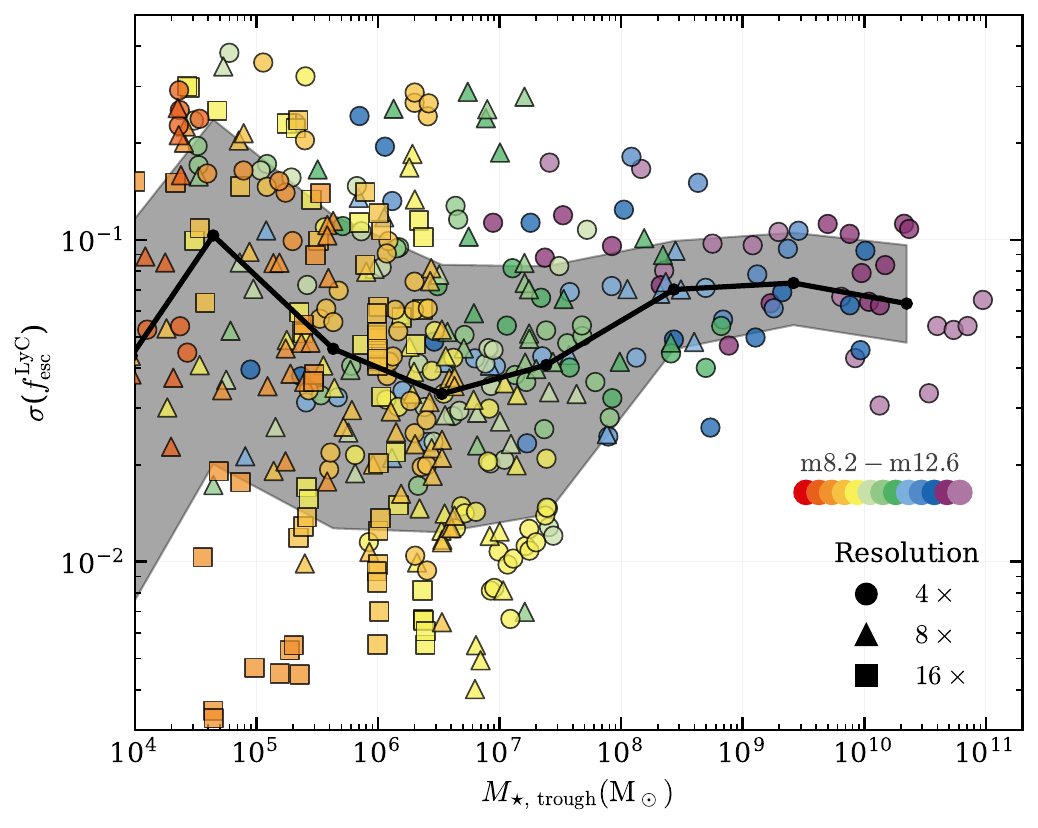}
\end{minipage}
\hfill
\begin{minipage}{0.48\textwidth}
    \centering
    \includegraphics[width=\linewidth]{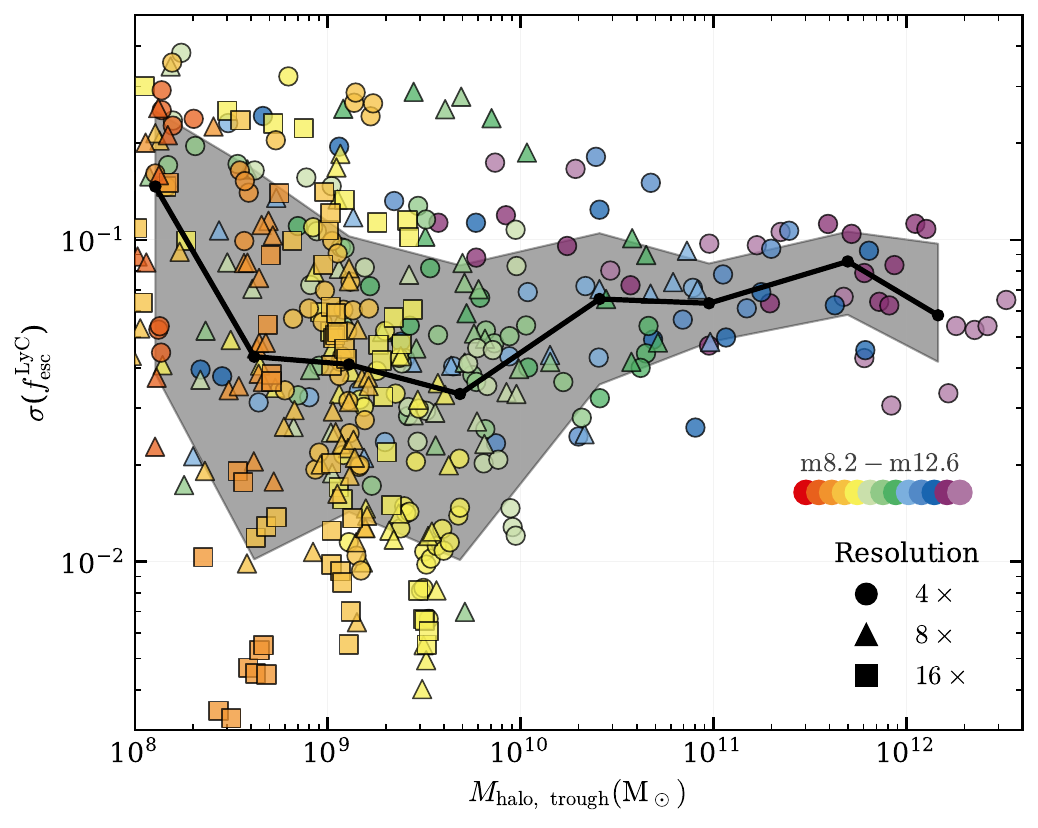}
\end{minipage}
\caption{Variability of $f_\text{esc}^\text{LyC}$ within intervals bounded by consecutive escape-fraction peaks, shown as a function of stellar mass (left) and halo mass (right). Masses are evaluated at the intervening trough, and colours and symbols follow Figure~\ref{fig:std_eta_cycle}. Larger values indicate stronger within-interval fluctuations. As for the burstiness variability in Figure~\ref{fig:std_eta_cycle}, the range of outcomes narrows towards high mass.}
\label{fig:std_fesc_cycle}
\end{figure*}

We next connect these recurrence times to the absolute escape level and the strength of each episode. Consecutive peaks delimit the intervals over which we measure variability, with each measurement placed at the mass of the intervening trough. Figures~\ref{fig:std_eta_cycle} and~\ref{fig:std_fesc_cycle} show the standard deviations of $\text{sfr}^{10}_{100}$ and $f_\text{esc}^\text{LyC}$, respectively, within intervals bounded by consecutive escape-fraction peaks. Both statistics are evaluated in linear space, with each snapshot between the bounding peaks, including the endpoints, receiving equal weight. Both quantities give broad distributions at low and intermediate masses and narrower distributions at high mass. Their median behaviour is non-monotonic, so increasing mass primarily reduces the range of interval-to-interval outcomes rather than steadily suppressing all variability.

\begin{figure*}
\centering
\begin{minipage}{0.48\textwidth}
    \centering
    \includegraphics[width=\linewidth]{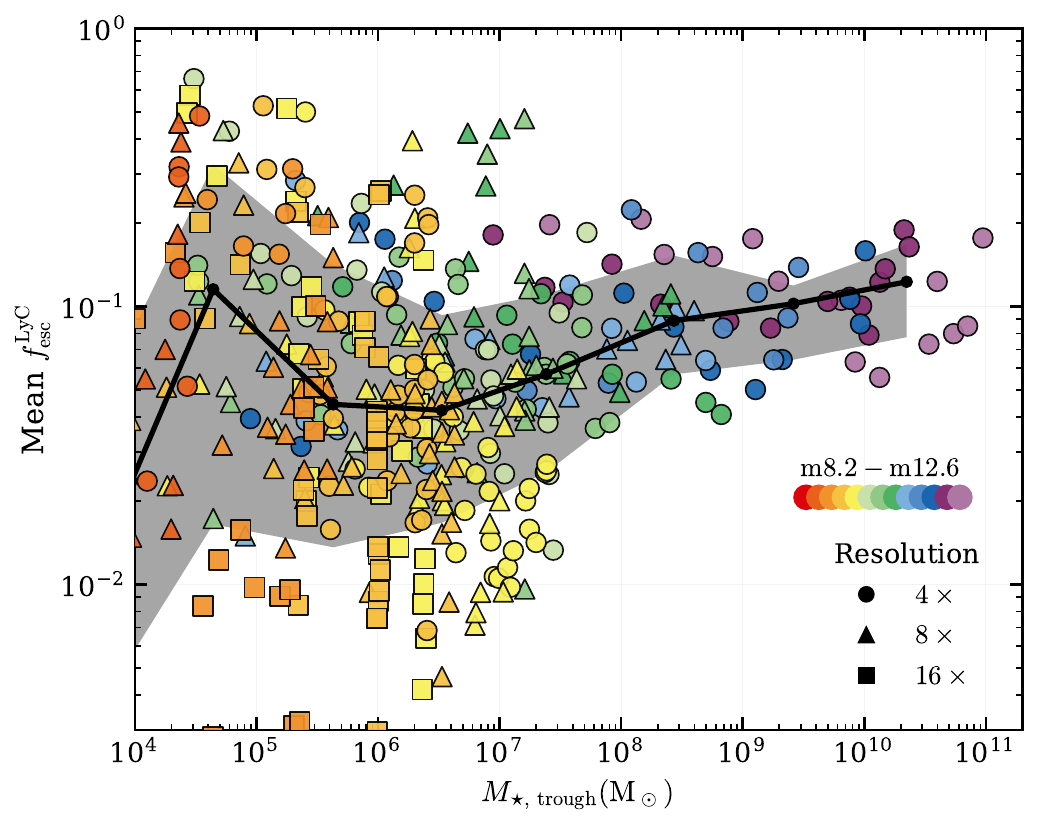}
\end{minipage}
\hfill
\begin{minipage}{0.48\textwidth}
    \centering
    \includegraphics[width=\linewidth]{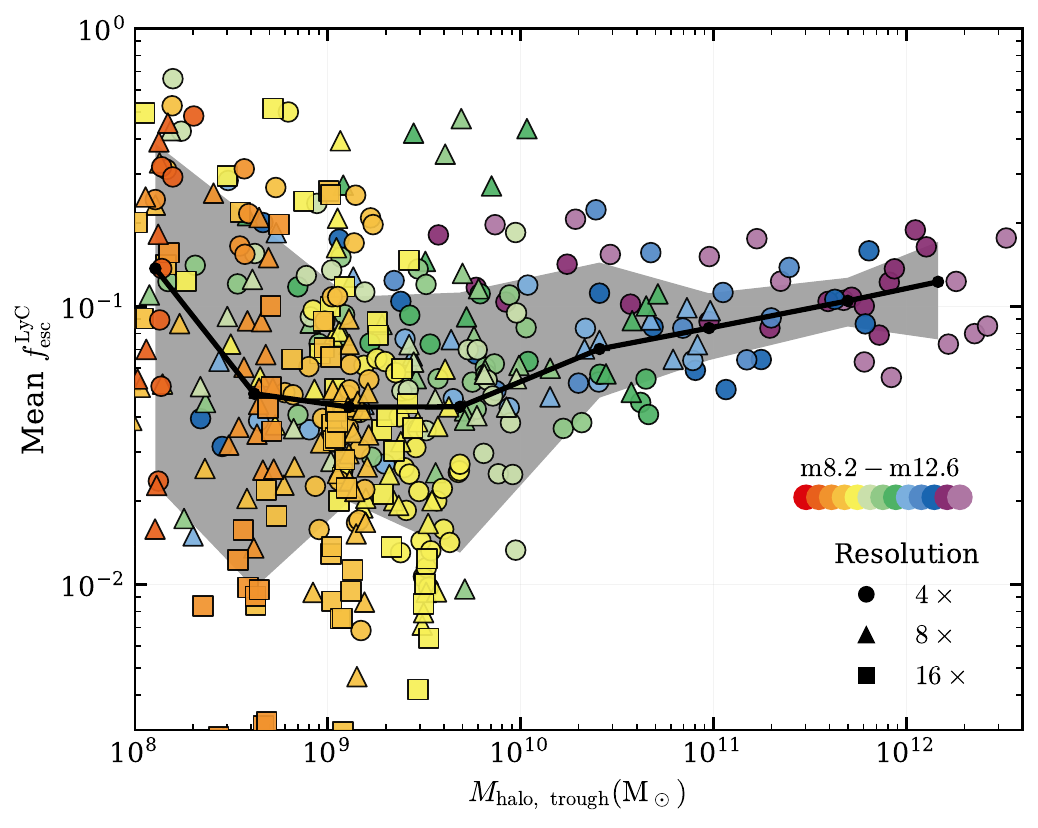}
\end{minipage}
\caption{Mean LyC escape fraction measured between consecutive escape-fraction peaks as a function of stellar mass (left) and halo mass (right), evaluated at the intervening trough. Colours and symbols follow Figure~\ref{fig:std_eta_cycle}. The broad range of mean escape levels at low and intermediate masses narrows towards the most massive systems, complementing the variability measurements in Figure~\ref{fig:std_fesc_cycle}.}
\label{fig:avg_fesc_cycle}
\end{figure*}

\begin{figure*}
\centering
\begin{minipage}{0.48\textwidth}
    \centering
    \includegraphics[width=\linewidth]{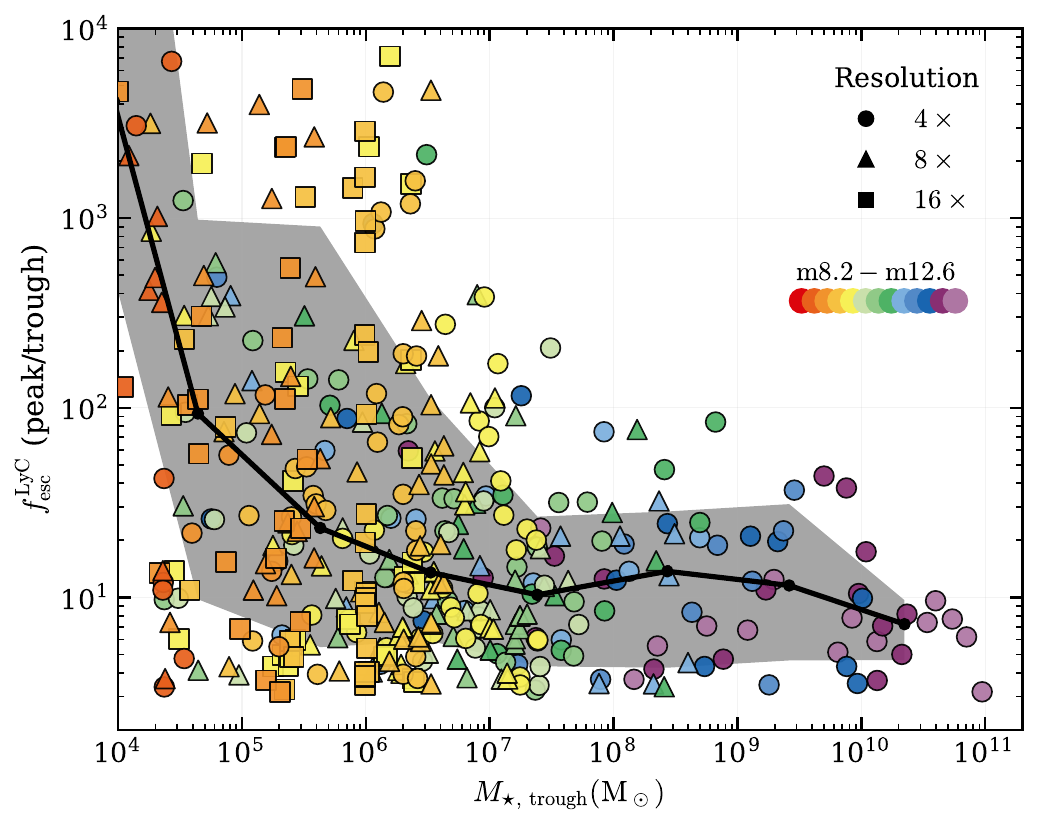}
\end{minipage}
\hfill
\begin{minipage}{0.48\textwidth}
    \centering
    \includegraphics[width=\linewidth]{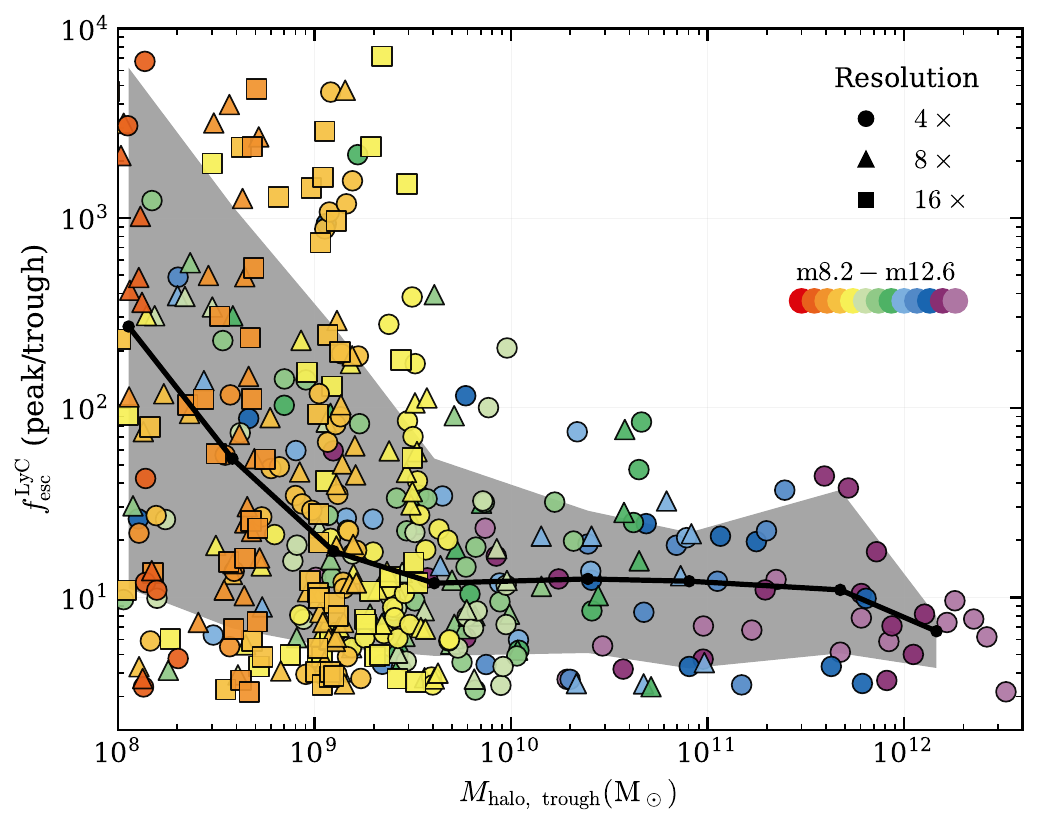}
\end{minipage}
\caption{Peak-to-trough amplitude ratios of the LyC escape fraction as a function of stellar mass (left) and halo mass (right), defined as the next peak value divided by the preceding trough value. Masses are evaluated at the trough, and colours and symbols follow Figure~\ref{fig:std_eta_cycle}. The widest range of contrasts occurs at low and intermediate masses and narrows towards high mass. A large ratio can reflect an especially low trough and does not necessarily imply a high absolute escaping emissivity. Ratios are shown for positive troughs without censoring for Monte Carlo sensitivity, so the largest contrasts remain uncertain.}
\label{fig:amp_fesc_cycle}
\end{figure*}

The mean escape fraction between consecutive $f_\text{esc}^\text{LyC}$ peaks shows a similar mass dependence (Figure~\ref{fig:avg_fesc_cycle}). Low- and intermediate-mass systems pass through intervals with a wide range of mean escape levels, whereas massive systems occupy a more restricted range. The peak-to-trough ratios span several orders of magnitude at low and intermediate masses and narrow substantially towards high mass (Figure~\ref{fig:amp_fesc_cycle}). These ratios must be interpreted together with the mean escape level because a large contrast can arise from a very low trough without implying a large absolute escaping emissivity. Near-zero troughs are sensitive to Monte Carlo sampling, so the largest ratios are not calibrated measurements of the extreme-amplitude tail. An upper limit on an unresolved trough would imply a lower limit on the peak-to-trough ratio. We use the distributions to describe the range of variability without assigning physical significance to the largest individual contrasts.

The broad range of instantaneous escape fractions at fixed mass found in SPHINX \citep{Rosdahl2022} and \thesan \citep{Yeh2023} already points to sensitivity to feedback history. Our time-domain analysis shows how this snapshot scatter is assembled from episodes with different baseline escape levels, amplitudes, and recurrence intervals. It also complements the absolute escape fractions and escaped-photon rates across the \thesanzoom population analysed by \citet{Summerfield2026}. The persistent dependence on the peak-selection threshold indicates that these histories contain variability on multiple scales, with frequent modest changes superimposed on rarer large excursions.

For reionization, the most relevant average weights transmission by the photons available to escape,
\begin{equation}
  \left\langle f_\text{esc}^\text{LyC}\right\rangle_{\dot{N}_\text{int}}
  \equiv
  \frac{\int \dot{N}_\text{esc}^\text{LyC}(t)\,\text{d}t}
       {\int \dot{N}_\text{int}^\text{LyC}(t)\,\text{d}t}
  =
  \frac{\int f_\text{esc}^\text{LyC}(t)\dot{N}_\text{int}^\text{LyC}(t)\,\text{d}t}
       {\int \dot{N}_\text{int}^\text{LyC}(t)\,\text{d}t} \, .
  \label{eq:photon_weighted_fesc}
\end{equation}
All integrals use the same time interval, and the subscript denotes weighting by the intrinsic LyC photon rate. The connection to timing follows from
\begin{equation}
\langle \dot{N}_\text{esc}^\text{LyC}\rangle_t = \langle f_\text{esc}^\text{LyC}\rangle_t\langle\dot{N}_\text{int}^\text{LyC}\rangle_t + \operatorname{Cov}_t(f_\text{esc}^\text{LyC},\dot{N}_\text{int}^\text{LyC}) \, ,
\label{eq:emissivity_covariance}
\end{equation}
where $\langle\cdot\rangle_t$ and $\operatorname{Cov}_t$ denote time averages and covariance over the same interval. Positive covariance raises the mean escaping photon rate above the product of the separate means because efficient escape tends to coincide with strong photon production, while negative covariance lowers it. The selected lag alone does not determine the sign or magnitude of this contribution. We use this relation to interpret the timing of escape without calculating an integrated photon budget. The time-weighted mean escape fractions and duty cycles in Appendix~\ref{appendix:duty_cycle} complement the recurrence statistics by distinguishing sustained transparency from intermittent high-escape phases. Together, these measurements describe the frequency and duration of escape episodes as well as their absolute escape fractions.

\subsection{Limitations}
\label{sec:limitations}

The radiative-transfer post-processing evaluates a common photoionization-equilibrium state for $f_\text{esc}^\text{LyC}$ and H$\alpha$ on the simulated density field, without solving for thermal equilibrium or updating the gas dynamics. Dense \HII regions respond rapidly to changes in the radiation field, while diffuse CGM gas can retain non-equilibrium ionization memory between bursts. Equilibrium can therefore affect the persistence of transparent channels differently from the H$\alpha$ emission weighted towards dense gas. These effects have not been quantified against the on-the-fly non-equilibrium solution. The Case~B treatment assumes local absorption of ionizing recombination radiation and omits its explicit transport. H$\alpha$ normalization also remains sensitive to the adopted thermal state and unresolved \HII structure.

Our escape fraction at $R_\text{vir}$ includes absorption through the CGM after photons leave the central star-forming region, whereas spatially resolved leakage diagnostics within the galaxy probe more local conditions. Direct LyC observations also sample individual sightlines rather than the angle-averaged escape fraction used here. Our observational comparison is therefore qualitative, with further selection and tracer differences discussed in Section~\ref{sec:nearby_comparison}. Radial escape-fraction profiles could clarify how evolving CGM opacity connects local clearing to halo-scale escape.

The $\approx10$\,Myr snapshot spacing samples the longer variations targeted here but does not resolve evolution during the first few Myr of an individual burst. Selecting positive-lag peaks isolates delayed associations but does not rule out prompt or negative-lag responses. Weak peaks and fallback estimates can also link distinct episodes. Sensitivity to coarser output cadence remains to be quantified. Higher or adaptive time sampling would better capture the sequence of radiative clearing, stellar winds, and the earliest supernovae. The internal structure of stellar clusters and their birth clouds is also incompletely resolved. Turbulent structure and radiative clearing can open early low-column-density pathways \citep{KakiichiGronke2021}, linking the timing and amplitude of escape to numerical resolution and the adopted stellar-population, feedback, and dust prescriptions.

The thirteen selected target haloes do not constitute a statistically representative galaxy population, and different resolution realizations of a target are not independent galaxies. Halo growth couples mass to epoch, while assembly history, environment, and unequal mass coverage across resolution levels can also contribute to the fitted trends. Assessing convergence requires matched-halo comparisons over common intervals, examining burstiness and escape separately. The window tests in Appendix~\ref{appendix:polar_period} preserve the negative period--mass trend, while the fixed-centroid comparison shows that net winding can depend strongly on the centre. Reversals reduce the accumulated phase, and a fitted period longer than the recorded history does not establish a completed cycle. The quoted bootstrap and window-choice errors do not quantify the full uncertainty associated with model assumptions, numerical resolution, and sample selection.

\section{Summary and Conclusions}
\label{sec:conclusions}
We have investigated the temporal regulation of LyC escape using \colt radiative-transfer post-processing of thirteen target haloes from the \thesanzoom simulation suite at all available resolution levels. Sampling their evolution at intervals of approximately $10\,\text{Myr}$ allows us to treat the escape fraction as a time-dependent galaxy property and to follow the ordering of star formation, ionizing-photon production, and photon escape. We combine lag measurements, phase-space trajectories, peak--trough statistics, prominence diagrams, and synthetic observables to distinguish changes in the ionizing sources from changes in transmission through the surrounding gas. Our principal findings are as follows.
\begin{enumerate}
\item The clearest delayed relationship is between the burstiness parameter $\text{sfr}^{10}_{100}$ and the LyC escape fraction. Massive haloes typically have lags of $\sim16\,\text{Myr}$ and $r_\text{peak}\sim0.4$--$0.6$. For the representative m12.6 halo, the nearest positive local maximum occurs at $15.2\,\text{Myr}$ with $r_\text{peak}=0.47$. The lag becomes longer and less coherent towards lower masses, consistent with a more irregular response of the gaseous environment to successive bursts.

\item Across the analysed sample, the characteristic lag scales approximately as $M_\star^{-0.19}$ and $M_\text{halo}^{-0.27}$. Repeated turns in the joint burstiness--escape trajectory reveal recurrent evolution, with shorter fitted winding periods towards higher mass. Lags and winding periods show a positive association with substantial scatter, connecting the delayed response to the broader burst--feedback sequence.

\item Recent star formation closely tracks both the intrinsic and escaping LyC photon rates. In the representative lag comparisons, $\text{SFR}_{10}$ has $r_\text{peak}=0.99$ and $0.90$ with $\dot{N}_\text{int}^\text{LyC}$ and $\dot{N}_\text{esc}^\text{LyC}$, respectively, and the two photon rates have $r_\text{peak}=0.88$ with no resolved delay. Together with the delayed burstiness--escape correlation, these results distinguish the rapid evolution of the ionizing sources from the response of the surrounding ISM and CGM. High transmission during a faint phase can still yield a small escaping photon rate.

\item Peak--trough recurrence times range from tens to several hundred Myr and remain dependent on the amplitude threshold used to identify an event. Larger thresholds isolate stronger excursions that recur less frequently, while the summary trends show a broad intermediate-mass minimum followed by a rise towards higher masses. Low- and intermediate-mass systems have the widest dispersion in recurrence times and within-interval variability. The narrower high-mass distributions retain variability across several amplitude scales. In the representative prominence diagrams, $\text{SFR}_{10}$ and $\dot{N}_\text{esc}^\text{LyC}$ span broader ranges of logarithmic peak and trough amplitudes than the bounded escape fraction and normalised burstiness ratio.

\item The synthetic H$\alpha$-to-UV burstiness proxy $\eta_{1500}$ retains the delayed association with LyC escape. For m12.6, its nearest positive local maximum occurs at $36.0\,\text{Myr}$ with $r_\text{peak}=0.34$, compared with the shorter and stronger intrinsic-burstiness result. Across the analysed sample, the selected proxy lag decreases approximately as $M_\star^{-0.20}$ and $M_\text{halo}^{-0.30}$, with substantial object-to-object scatter. The qualitative comparison with LzLCS shows a shared association between stronger recent activity and higher escape fractions, with the observed galaxies occupying a more strongly bursting region of parameter space.
\end{enumerate}

The results support a feedback-regulated sequence in which a burst rapidly increases the intrinsic LyC photon supply while young stars can remain embedded in dense, absorbing gas. The combined radiative and mechanical feedback model subsequently changes the neutral covering fraction and the geometry of escape paths, allowing the escape fraction to remain elevated as the burst declines \citep{Kimm2014,Trebitsch2017,Katz2023}. Gas re-accretion eventually restores dense star-forming conditions and closes or reorients those paths. The lag and phase winding describe the response and recurrence of escape within this sequence. Gravitational confinement and gas recycling provide a physical interpretation of the mass trends, complementing the burst--quench cycles and structural evolution measured elsewhere in \thesanzoom \citep{McClymontZoomSFMS2025,McClymontZoomSize2025}. Lower-mass haloes show longer, less regular responses, while massive haloes have shorter lags and phase-space periods together with smoother population trends. Substantial stochasticity and amplitude dependence remain at every mass. Nearby observations also show that strong LyC escape can precede a fast, mass-loaded supernova-driven wind \citep{Carr2025a,Carr2025b}, illustrating how different feedback channels and burst histories can produce early and delayed leakage.

The separation between source- and environment-regulated quantities connects galaxy feedback to the radiation reaching the IGM. Recent SFR predicts the intrinsic photon supply and much of the variability in the escaping photon rate, while burst phase and gas transparency determine when that supply can escape efficiently. A photon-weighted mean escape fraction preserves the integrated escaping output when applied to the photon production over the same interval. Following the temporal overlap of production and transmission additionally captures the delays, duty cycles, and recurrent modulation of that output. This timing is central to linking observed galaxies to their ionizing contribution and to interpreting the emerging JWST photon budget \citep{Munoz2024,Giovinazzo2025}.

The observable burstiness proxy provides a route to locating galaxies within this temporal sequence. Comparisons that incorporate sightline dependence, instrumental sensitivity, and non-detections can connect the simulated histories to populations observed at different burst phases. Relating these phases to metal-line absorption, chemical abundances, the survival of metal-free star formation, and the local ionization environment would extend this connection to the wider context explored with \thesanzoom \citep{McClymontZoomMetal2025,ZierZoomPop2025,ZierZoomUV2025,PrutoZoom2026}. Larger time-resolved samples can separate the effects of mass, epoch, assembly history, and environment, and guide escape prescriptions that preserve both the amount and timing of the radiation galaxies release.


\section*{Acknowledgements}
The authors gratefully acknowledge the Gauss Centre for Supercomputing e.V. (\url{www.gauss-centre.eu}) for funding this project by providing computing time on the GCS Supercomputer SuperMUC-NG at Leibniz Supercomputing Centre (\url{www.lrz.de}), under project pn29we.
The authors acknowledge the MIT Office of Research Computing and Data, FAS Division of Science Research Computing Group at Harvard University, and High Performance Computing at The University of Texas at Dallas (HPC@UTD) for providing resources that have contributed to the research results reported within this paper.
Support for programs JWST-AR-08709 (AS) and JWST-AR-04814 (XS, MV) was provided by NASA through a grant from the Space Telescope Science Institute, which is operated by the Association of Universities for Research in Astronomy, Inc., under NASA contract NAS 5-03127.
RK acknowledges support of the Natural Sciences and Engineering Research Council of Canada (NSERC) through a Discovery Grant and a Discovery Launch Supplement, funding reference numbers RGPIN-2024-06222 and DGECR-2024-00144.
LK acknowledges the support of a Royal Society University Research Fellowship (grant number URF$\backslash$R1$\backslash$251793).
MV acknowledges support through NASA ATP Grant 23-ATP23-149 and NSF AAG Grant AST-2307699.
This work was supported by the LOEWE programme of the State of Hesse (LOEWE-Spitzen-Professur).
This work was supported in part by a grant of access to OpenAI models through the ChatGPT for Academic Researchers program.
Generative AI tools were used to assist with language editing and consistency checks. The authors remain responsible for the scientific content and final wording of the manuscript.

\section*{Data Availability}

All simulation data, including snapshots, group catalogues, and merger trees, will be made publicly available in the near future via \url{www.thesan-project.com}. Before the public release, the data underlying this article can be shared upon reasonable request to the corresponding author(s).



\bibliographystyle{mnras}
\bibliography{biblio}




\appendix

\section{Peak identification and prominence}
\label{appendix:peak_finding}

\subsection{Peak and trough identification}
\label{sec:peak_identification}

\begin{figure*}
\centering
\includegraphics[width=\linewidth]{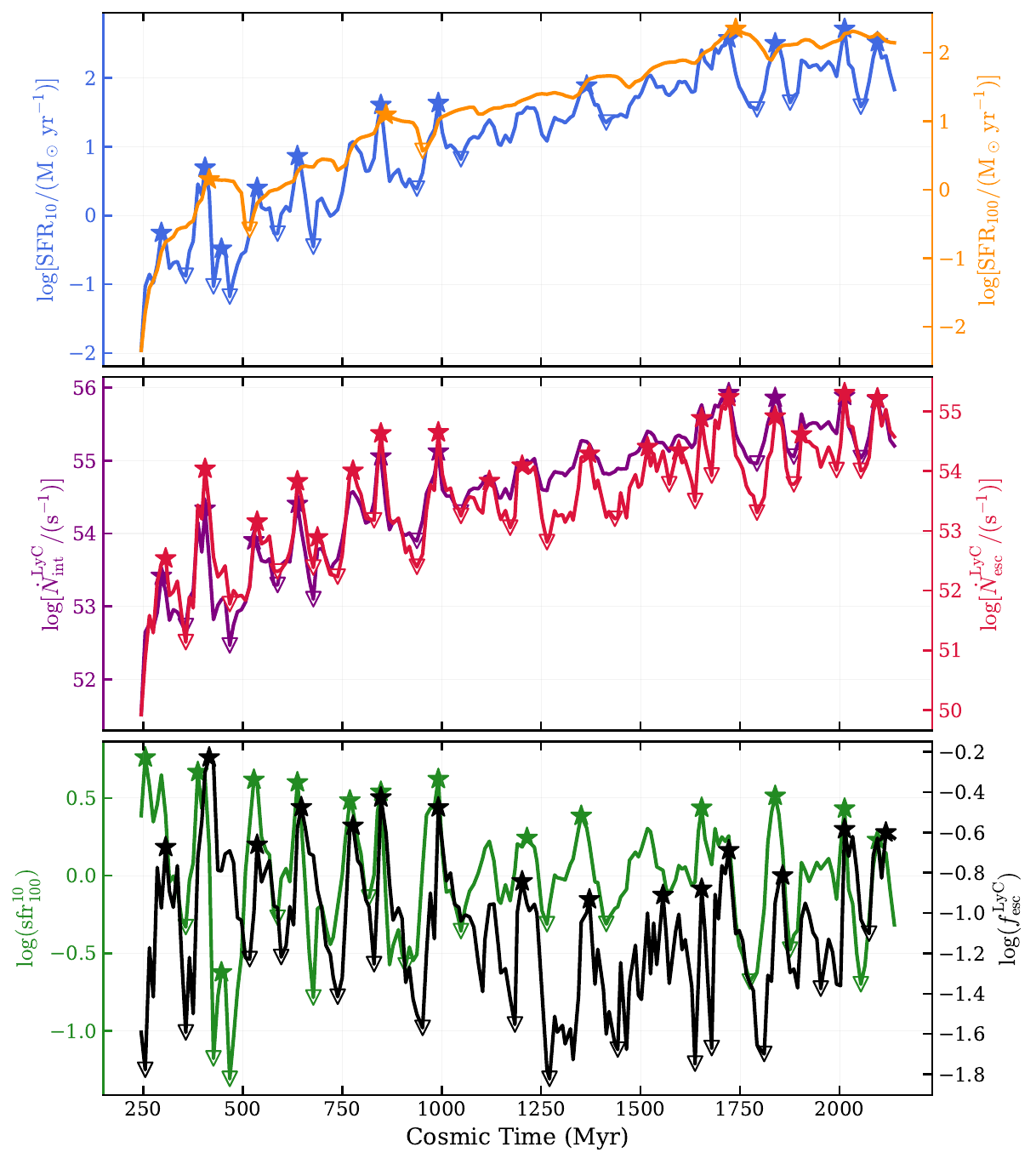}
\caption{Peak and trough identification for the representative m12.6 halo at 4$\times$ resolution using a $0.5$-dex prominence threshold. The upper panel compares $\text{SFR}_{10}$ and $\text{SFR}_{100}$ in $\Msun\,\text{yr}^{-1}$, the middle panel compares $\dot{N}_\text{int}^\text{LyC}$ and $\dot{N}_\text{esc}^\text{LyC}$ in $\text{s}^{-1}$, and the lower panel compares $\text{sfr}^{10}_{100}$ and $f_\text{esc}^\text{LyC}$. All amplitudes are logarithmic. Filled stars denote retained peaks and open inverted triangles denote retained troughs. The algorithm is applied separately to each quantity, preserving prominent variations without requiring their extrema to occur at the same time.}
\label{fig:peak_overplot}
\end{figure*}

We identify prominent peaks and troughs in the LyC escape fraction, burstiness parameter, SFRs, and LyC photon production rates. The aim is to isolate variations of comparable logarithmic amplitude across quantities with different units and dynamic ranges, then use the retained extrema to measure their recurrence times and within-interval behaviour. For a candidate peak and its paired trough, we define the logarithmic amplitude as
\begin{equation}
  \Delta_\text{dex} = \left| \log\!\left(\frac{X_\text{peak}}{X_\text{trough}}\right) \right| \, ,
\end{equation}
where $X$ denotes the quantity being analysed. The threshold on $\Delta_\text{dex}$ sets the minimum prominence required for a variation to be retained. The fiducial value of $0.5$ dex corresponds to a peak-to-trough ratio of approximately $3.2$, while the additional $0.25$- and $1.0$-dex measurements in Section~\ref{sec:peak-trough} test the sensitivity to weaker and stronger selections.

\begin{figure*}
\centering
\begin{minipage}{0.48\textwidth}
    \centering
    \includegraphics[width=\linewidth]{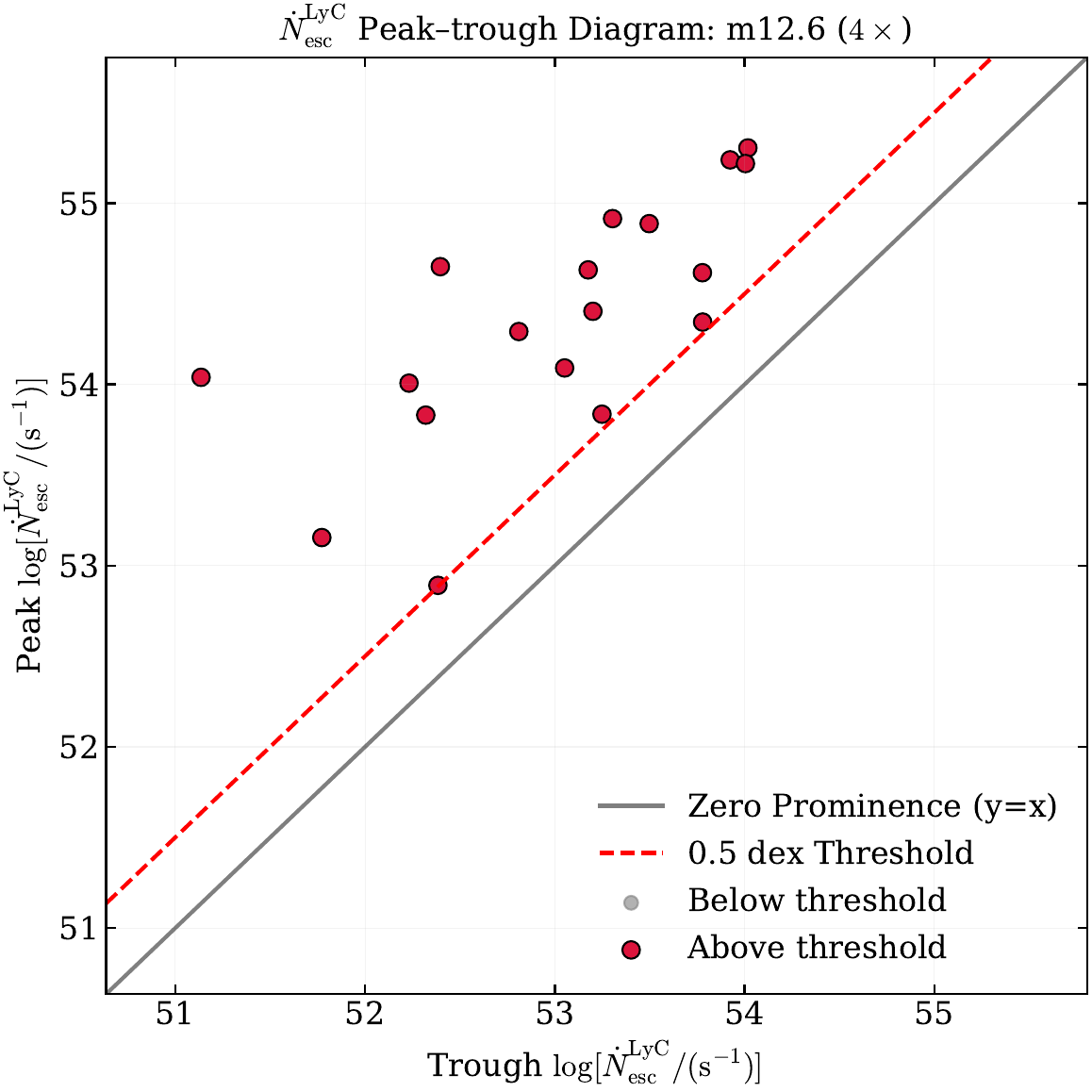}
\end{minipage}
\hfill
\begin{minipage}{0.48\textwidth}
    \centering
    \includegraphics[width=\linewidth]{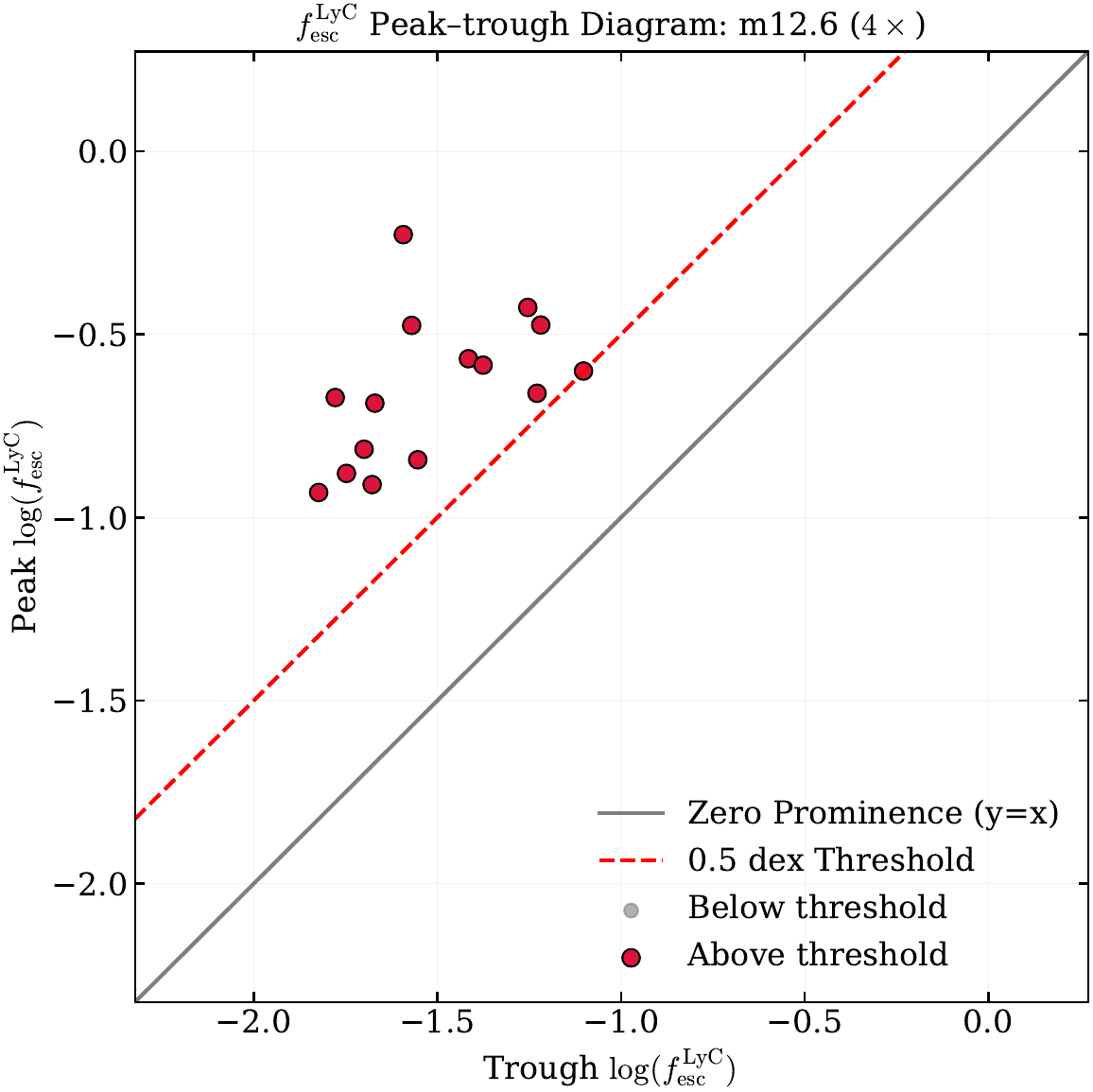}
\end{minipage}
\caption{Peak--trough prominence diagrams for the escaping LyC photon rate (left) and LyC escape fraction (right) of m12.6 at 4$\times$ resolution. Each point pairs the logarithmic amplitude of a trough on the horizontal axis with that of its associated peak on the vertical axis. The grey line $y=x$ marks zero prominence, and the dashed red line marks a prominence of $0.5$ dex. Features above the dashed line satisfy the fiducial selection used to identify prominent variations.}
\label{fig:combined_persistence1}
\end{figure*}

The algorithm starts from local extrema in the logged series and enforces alternating maxima and minima, retaining the more extreme value when consecutive candidates have the same type. It then repeatedly removes the adjacent peak--trough pair with the smallest contrast until every surviving adjacent contrast meets the threshold. Track endpoints are not treated as extrema. Increasing the threshold can combine several modest fluctuations into a broader event, explaining why the inferred recurrence interval increases with the selected amplitude scale. The retained extrema define alternating rising and declining phases. Adjacent peaks and troughs determine the amplitude and duration of an individual phase, whereas consecutive peaks or consecutive troughs delimit a full recurrence interval. In the main analysis, each peak-to-peak interval is associated with the galaxy or halo mass measured at its intervening trough.

Figure~\ref{fig:peak_overplot} illustrates the fiducial selection for the representative m12.6 halo at 4$\times$ resolution. The more rapidly varying $\text{SFR}_{10}$ is shown alongside the smoother $\text{SFR}_{100}$ in the upper panel, the intrinsic and escaping LyC photon rates appear in the middle panel, and the burstiness parameter and escape fraction appear in the lower panel. Filled stars and open inverted triangles mark the retained peaks and troughs, respectively. The selection is performed independently for each time series, so the extrema of different quantities can occur at different times. Their offsets are especially apparent for the burstiness parameter and escape fraction and are consistent with the delayed relationship quantified by the lag analysis.

\subsection{Peak--trough prominence}
\label{sec:persistence}

The prominence diagrams pair each retained peak with its nearest preceding retained trough. The horizontal and vertical coordinates give the logarithmic trough and peak amplitudes, respectively, so the vertical displacement above $y=x$ is $\log(X_\text{peak}/X_\text{trough})$ in dex. Prominence measures the amplitude of this rising phase independently of its duration.

\begin{figure*}
\centering
\begin{minipage}{0.48\textwidth}
    \centering
    \includegraphics[width=\linewidth]{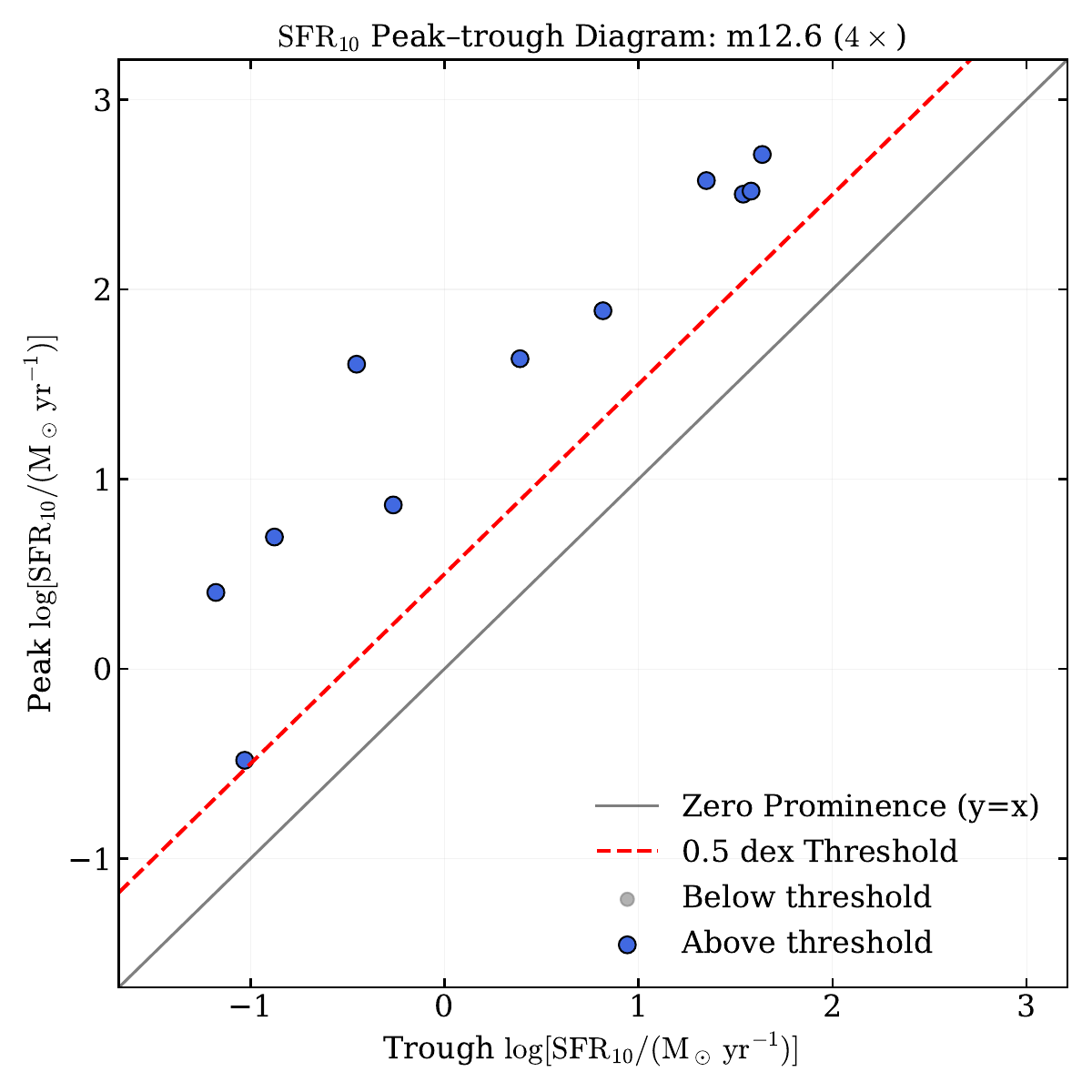}
\end{minipage}
\hfill
\begin{minipage}{0.48\textwidth}
    \centering
    \includegraphics[width=\linewidth]{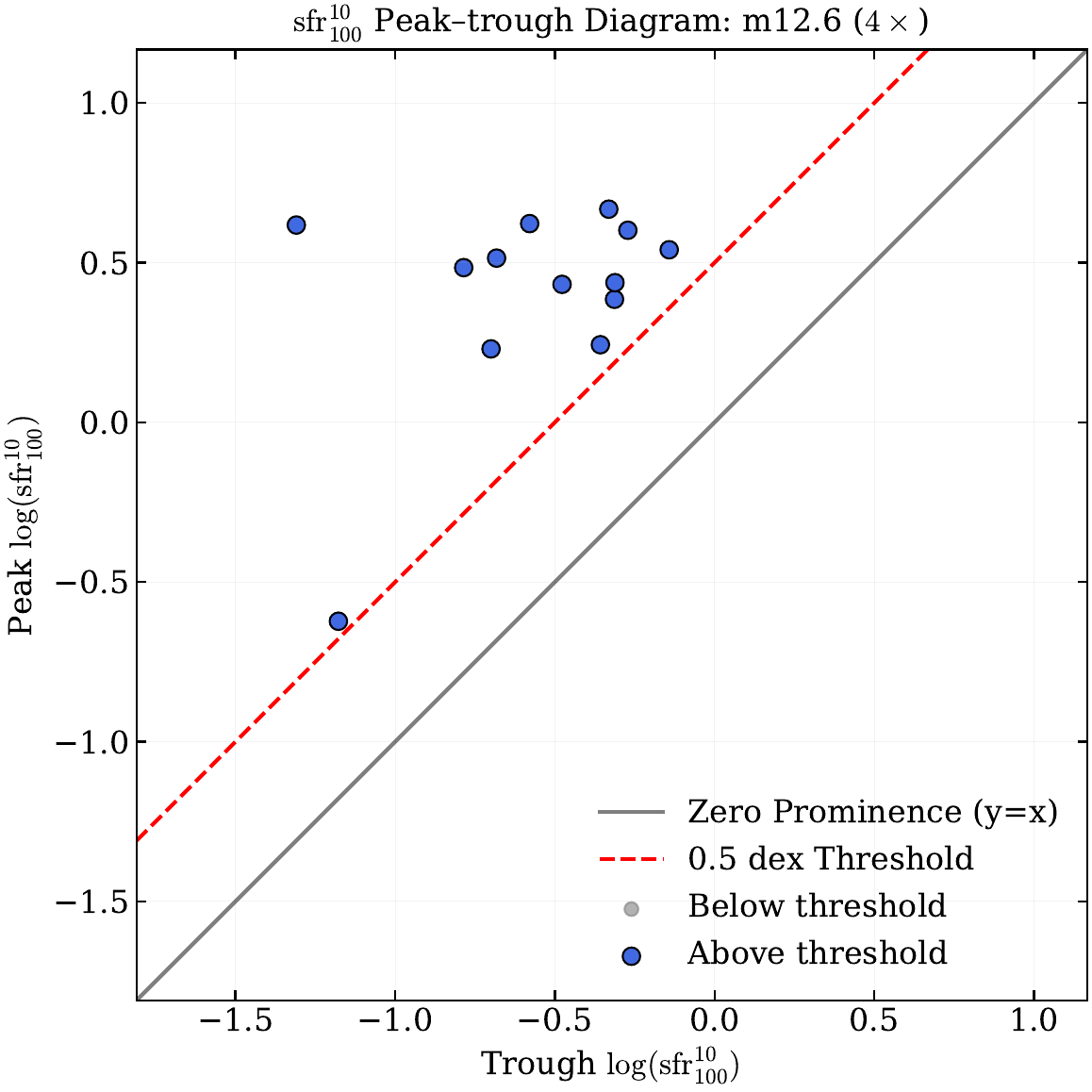}
\end{minipage}
\caption{Peak--trough prominence diagrams for $\text{SFR}_{10}$ (left) and the intrinsic burstiness parameter (right) of m12.6 at 4$\times$ resolution. The plotting convention follows Figure~\ref{fig:combined_persistence1}. Displacement above the grey diagonal measures logarithmic peak-to-trough prominence, and the dashed red line marks the fiducial $0.5$-dex selection.}
\label{fig:combined_persistence2}
\end{figure*}

Figures~\ref{fig:combined_persistence1} and~\ref{fig:combined_persistence2} show the pairs retained at the fiducial $0.5$-dex threshold. The grey diagonal marks zero prominence, and the dashed red line $y=x+0.5$ marks the selection threshold. These diagrams show the baseline levels and contrasts of the variations whose recurrence times we measure in Section~\ref{sec:peak-trough}, where we also use thresholds of $0.25$ and $1.0$ dex to compare weaker and stronger excursions.

The displayed diagrams correspond to m12.6 at 4$\times$ resolution and compare the escaping LyC photon rate, escape fraction, $\text{SFR}_{10}$, and intrinsic burstiness parameter. The $\text{SFR}_{10}$ and $\dot{N}_\text{esc}^\text{LyC}$ features span relatively broad ranges of logarithmic peak and trough amplitudes, consistent with the secular growth and superimposed fluctuations visible in Figure~\ref{fig:peak_overplot}. The $f_\text{esc}^\text{LyC}$ and $\text{sfr}^{10}_{100}$ features occupy more confined regions of their respective diagrams. Part of this compactness follows from the bounded nature of the escape fraction and the normalisation built into the burstiness ratio. Together with the lag and source--environment analyses, these distributions distinguish variations in the absolute photon supply from the relative activity and gas response traced by burstiness and escape fraction.

\section{Duty Cycle}
\label{appendix:duty_cycle}

\begin{figure*}
\centering
\begin{minipage}{0.48\textwidth}
\centering
\includegraphics[width=\linewidth]{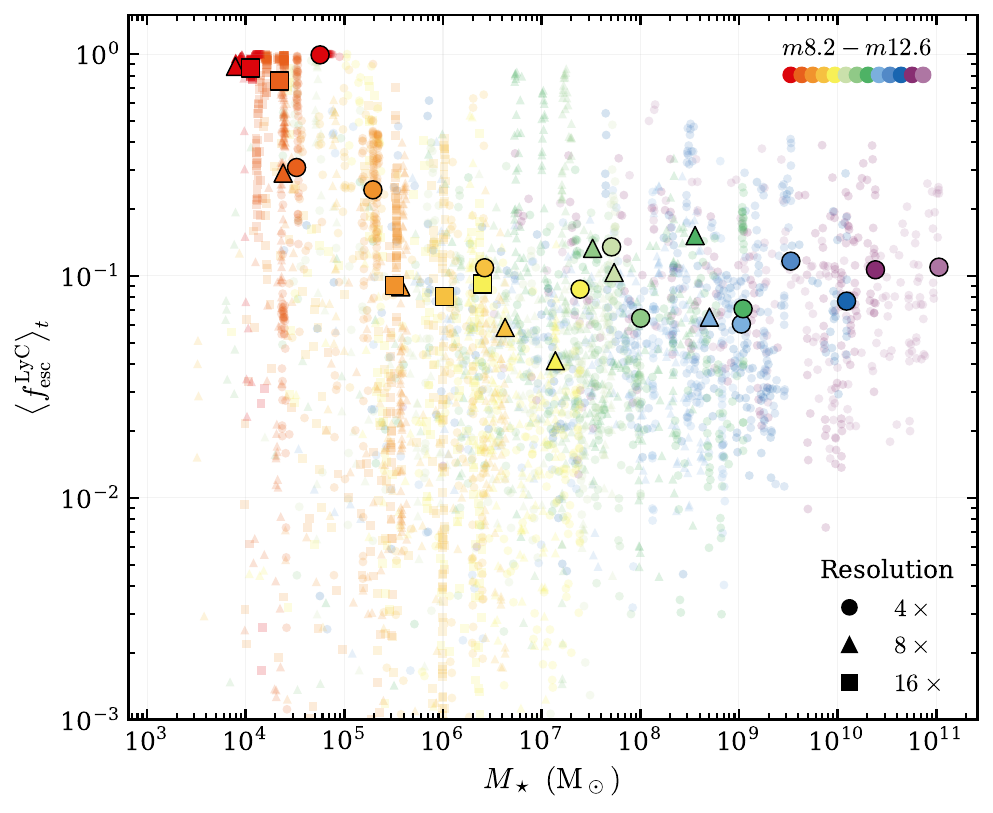}
\end{minipage}
\hfill
\begin{minipage}{0.48\textwidth}
\centering
\includegraphics[width=\linewidth]{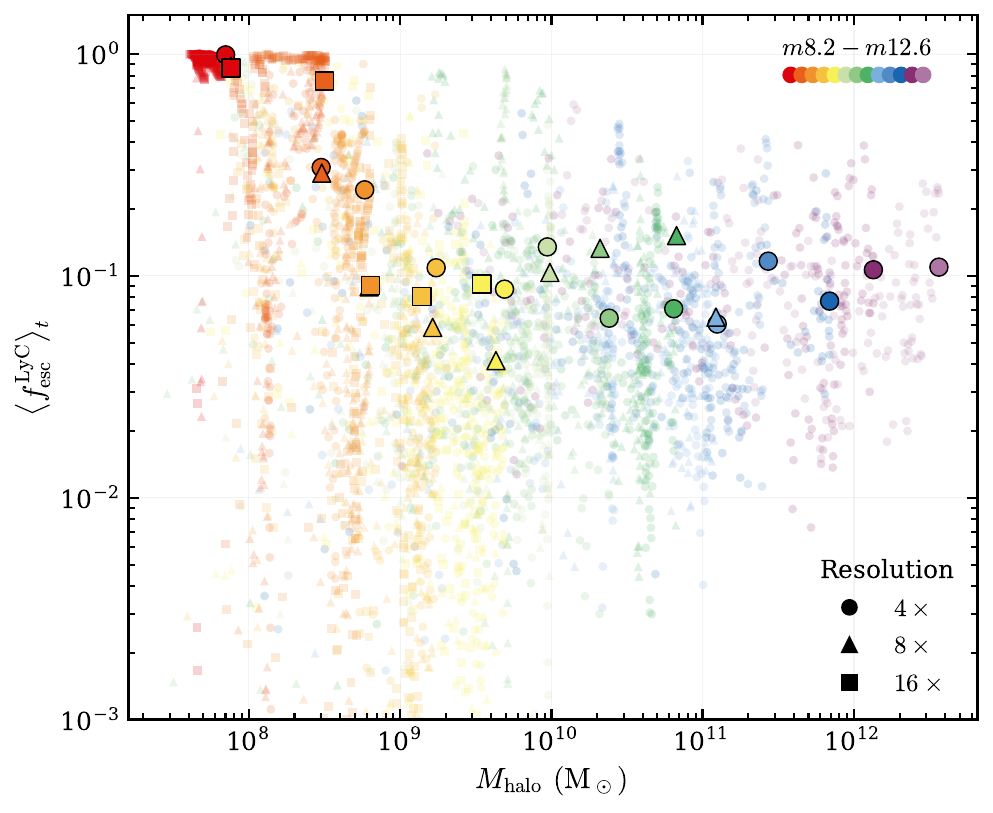}
\end{minipage}
\caption{Time-weighted mean LyC escape fraction as a function of final stellar mass (left) and halo mass (right). Large outlined symbols show the time-weighted means at the final masses, while faint points show individual snapshots at their instantaneous masses. Colours identify targets, and circles, triangles, and squares denote 4$\times$, 8$\times$, and 16$\times$ resolution. The lowest-mass targets have the highest mean escape fractions, while the higher-mass targets lie near $\langle f_\text{esc}^\text{LyC}\rangle_t\sim0.1$.}
\label{fig:fesc_variability}
\end{figure*}

\begin{figure}
\centering
\includegraphics[width=\linewidth]{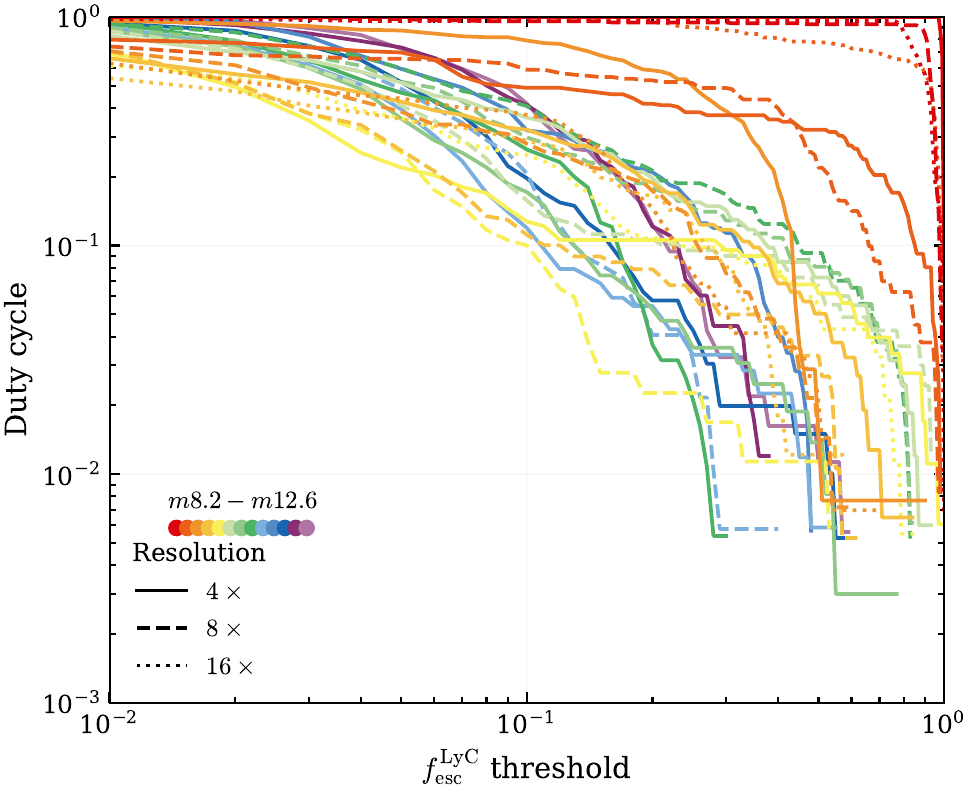}
\caption{Duty cycle of the LyC escape fraction as a function of the escape-fraction threshold. The duty cycle denotes the fraction of the total evolutionary time for which a halo has $f_\text{esc}^\text{LyC}$ at or above the specified threshold. Colours indicate halo mass, while line styles denote the simulation resolution. Low-mass haloes spend a larger fraction of their evolution at high escape fractions, whereas the duty cycle decreases towards higher masses.}
\label{fig:duty_cycle}
\end{figure}

We complement the recurrence statistics with the time-weighted mean escape fraction and the fraction of time that each halo spends above a given escape-fraction threshold. The time-weighted average is defined as
\begin{equation}
\langle f_\text{esc}^\text{LyC}\rangle_t =
\frac{\int_{t_\text{initial}}^{t_\text{final}} f_\text{esc}^\text{LyC}(t)\,\text{d}t}
{t_\text{final}-t_\text{initial}} \, ,
\end{equation}
and is evaluated over the recorded history of each halo. We show this quantity as a function of the final halo and stellar masses in Figure~\ref{fig:fesc_variability}. The corresponding mean escaping photon rate depends on the joint evolution of photon production and transmission through the covariance in Equation~(\ref{eq:emissivity_covariance}). The photon-weighted mean in Equation~(\ref{eq:photon_weighted_fesc}) accounts for that temporal overlap and connects it to the integrated escaping radiation.

The time-averaged escape fraction exhibits a clear mass dependence. The lowest-mass haloes have the highest average LyC escape fractions. In these shallow potential wells, feedback-driven outflows can efficiently remove or redistribute neutral gas along the line of sight, producing episodes of highly efficient LyC escape. As the halo mass increases, the time-averaged escape fraction decreases, before approaching a relatively shallow trend at the highest masses considered. Stronger gravitational confinement provides a physical explanation for the lower mean transmission of more massive haloes.

We additionally quantify the temporal behaviour of the escape fraction through a duty-cycle analysis. For a given threshold $f_\text{th}$, we define the duty cycle as the fraction of the total evolutionary time during which the halo satisfies
$f_\text{esc}^\text{LyC} \geq f_\text{th}$,
\begin{equation}
D(f_\text{th}) =
\frac{t(f_\text{esc}^\text{LyC}\geq f_\text{th})}
{t_\text{final}-t_\text{initial}} \, .
\end{equation}
Here the numerator is the total time spent at or above the threshold. We evaluate the time average and duty cycle with trapezoidal weights on the native cosmic-time grid, applying the threshold to each snapshot for the latter. The resulting duty-cycle curves are shown in Figure~\ref{fig:duty_cycle}. The analysis provides a complementary view of the mass dependence seen in the time-weighted averages. Low-mass haloes spend a substantial fraction of their evolution at high escape fractions, resulting in comparatively large duty cycles even for stringent $f_\text{esc}^\text{LyC}$ thresholds. The most transparent low-mass targets remain near unit escape fraction for much of their recorded evolution.

More massive haloes spend less time at high escape fractions. Their duty cycles fall rapidly as the threshold approaches unity, even though they can remain above modest thresholds for a substantial part of their evolution. The time-weighted means and duty cycles therefore distinguish sustained transparency in the lowest-mass targets from the intermittent high-escape phases of more massive systems. These statistics describe the available history of each target and complement the amplitude-selected episodes in Section~\ref{sec:peak-trough}.

\section{Phase winding and centre choice}
\label{appendix:polar_period}

\begin{figure*}
\centering
\includegraphics[width=\linewidth]{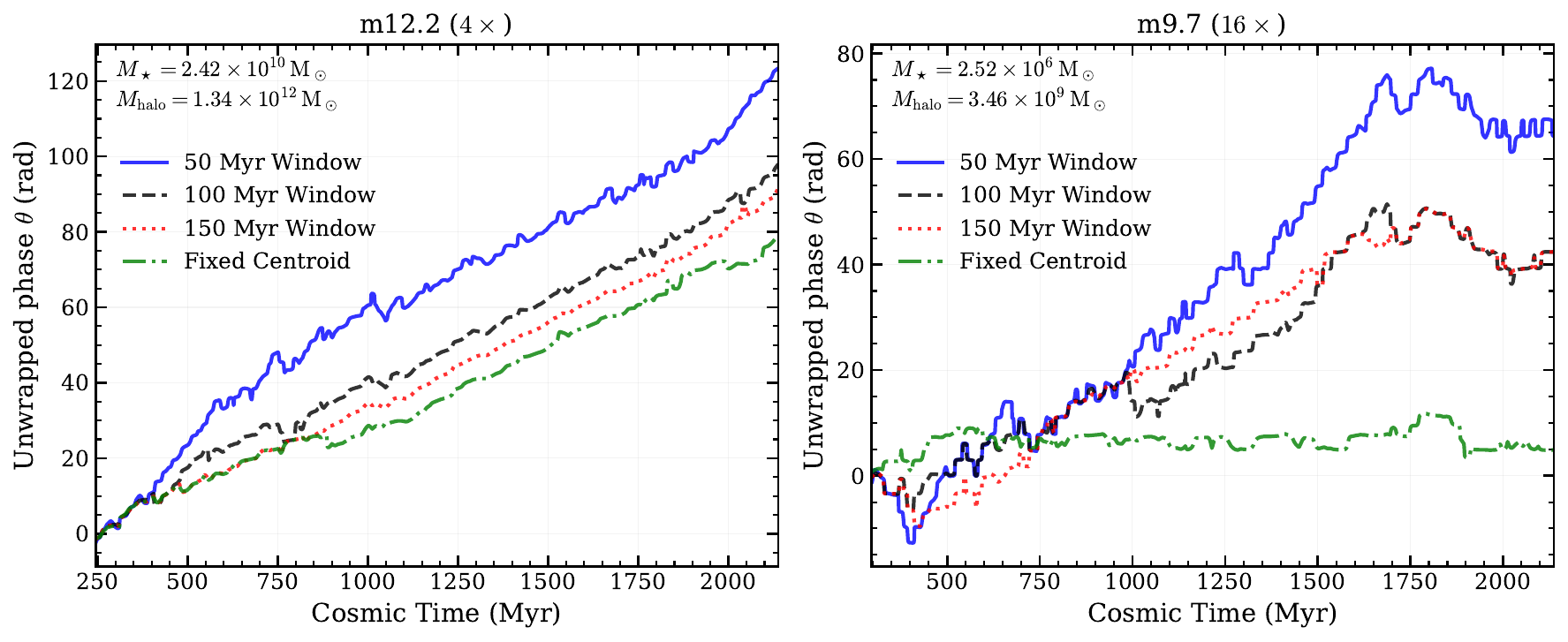}
\caption{Unwrapped phase of the $f_\text{esc}^\text{LyC}$--$\text{sfr}^{10}_{100}$ trajectory for the high-mass target m12.2 at 4$\times$ resolution (g205, left) and the lower-mass target m9.7 at 16$\times$ resolution (g137030, right). The blue solid, black dashed, and red dotted curves use centred windows of 50, 100, and 150\,Myr, respectively. The green dot-dashed curve uses the fixed mean of each logarithmic coordinate. The high-mass track accumulates phase for all four choices, while the low-mass track shows reversals and substantially less net winding about the fixed centroid.}
\label{fig:overlaid_theta}
\end{figure*}

\begin{figure}
\centering
\includegraphics[width=\linewidth]{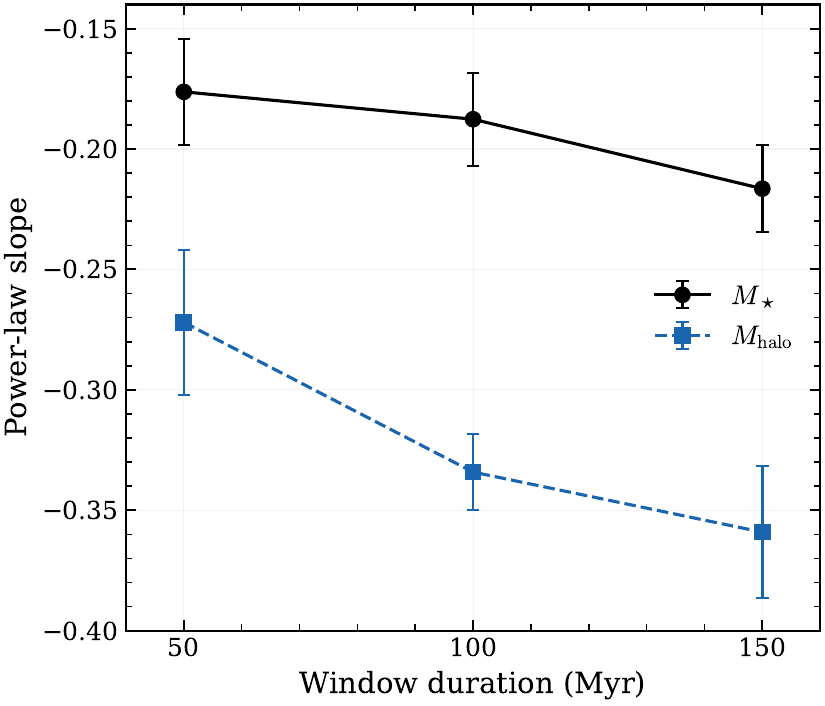}
\caption{Fitted slope of the period--mass relation for $\text{sfr}^{10}_{100}$ using 50, 100, and 150\,Myr moving centres. Black circles and blue squares show the stellar- and halo-mass relations, respectively, with masses evaluated at $z=3$. Error bars show the reported slope uncertainties. The slopes remain negative for both masses and become steeper for longer windows.}
\label{fig:window_slope}
\end{figure}

\begin{figure}
\centering
\includegraphics[width=\linewidth]{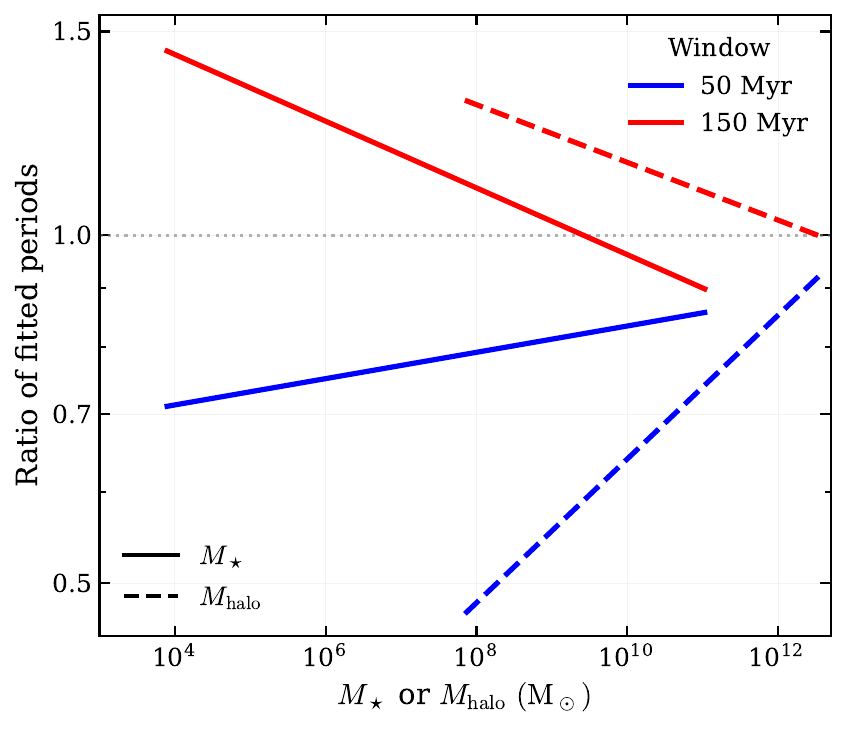}
\caption{Ratios of the fitted $\text{sfr}^{10}_{100}$ period--mass relations for 50\,Myr (blue) and 150\,Myr (red) windows to the fiducial 100\,Myr result. Solid and dashed curves show the stellar- and halo-mass relations, respectively, plotted against the corresponding mass at $z=3$. The grey horizontal line marks unity. These are ratios of population fits, rather than ratios for individual haloes. The choice of window changes both the normalization and mass dependence of the inferred periods.}
\label{fig:ratio_window}
\end{figure}

The phase analysis measures how the joint burstiness--escape trajectory winds around a reference centre. We first convert the snapshot redshifts to cosmic time, sort each history chronologically, and interpolate onto 2000 uniformly spaced times spanning the available track. The grid spacing therefore depends on the track duration and is distinct from the native snapshot cadence. We work with logarithmic coordinates,
\begin{equation}
\label{eq:log_polar}
  x(t)=\log_{10}\text{sfr}^{10}_{100}(t) \qquad \text{and} \qquad
  y(t)=\log_{10} f_\text{esc}^\text{LyC}(t) \, ,
\end{equation}
and apply the same procedure to $\text{sfr}^{10}_{30}$ for the alternative burstiness definition.

For a temporal window $T_\text{win}$, we construct a centred boxcar kernel with $N_\text{win}=\max[1,\operatorname{round}(T_\text{win}/\Delta t_\text{grid})]$ grid points and equal weights $K_j=1/N_\text{win}$. The local centre is
\begin{equation}
  \bar{x}_i=\frac{1}{N_\text{win}}\sum_{j\in\mathcal W_i}x_j \qquad \text{and} \qquad
  \bar{y}_i=\frac{1}{N_\text{win}}\sum_{j\in\mathcal W_i}y_j \, ,
\end{equation}
where $\mathcal W_i$ is centred on $t_i$. The fiducial 100\,Myr window covers approximately $t_i\pm50$\,Myr. At the ends of a track, we extend the nearest boundary value to fill the kernel. We then subtract the local centre and scale each coordinate by its standard deviation over the track,
\begin{equation}
  X_i=\frac{x_i-\bar{x}_i}{\sigma_x} \qquad \text{and} \qquad
  Y_i=\frac{y_i-\bar{y}_i}{\sigma_y} \, .
\end{equation}
The radius and wrapped angle are
\begin{equation}
  r_i=\sqrt{X_i^2+Y_i^2} \qquad \text{and} \qquad
  \phi_i=\operatorname{atan2}(Y_i,X_i) \, .
\end{equation}
The two-argument arctangent preserves the quadrant. Unwrapping the time-ordered angles removes jumps at the $-\pi/\pi$ boundary and gives the cumulative phase $\theta_i$, whose fitted slope defines $T$ through Equation~(\ref{eq:phase_period}). The radius measures displacement in standardized logarithmic coordinates and is not used as a physical variability amplitude.

The uncertainty in the mean winding rate combines bootstrap scatter with scatter from varying the local centre. We calculate the standard deviation of slopes from 1000 bootstrap resamples of the interpolated $(t_i,\theta_i)$ pairs. We then recompute the phase using running-median centres over 80, 100, and 120\,Myr windows, truncated at the track boundaries, and add the standard deviation of these fitted slopes in quadrature. This second contribution includes changes in both the window duration and the centring statistic relative to the fiducial boxcar mean. We propagate the combined slope uncertainty through $T=2\pi/|\omega|$ to obtain the period uncertainty. For a fixed centroid, only the bootstrap contribution is included. The bootstrap treats the interpolated points as independent and therefore does not account for their temporal correlations.

We fit the population period--mass relations in logarithmic coordinates, first using an unweighted line to obtain residuals $d_i$ and then refitting with residual multipliers $q_i\propto(|d_i|+\epsilon)^{-1}$, where $\epsilon=\max[0.1\,\operatorname{median}(|d_i|),10^{-5}]$. The observable lag--mass fits follow the same procedure with $q_i\propto\{\max[d_i^2,\operatorname{median}(d_i^2),10^{-6}]\}^{-1}$. Both prescriptions reduce the influence of points far from the initial relation. For the intrinsic lag--mass and lag--period fits, we use $q_i=|r_{\mathrm{peak},i}|$ to give greater weight to the more coherent responses.

In each fit, $q_i$ multiplies the residual before it is squared in the least-squares objective, and the formal slope errors follow from the scaled fit covariance. The reported $R^2$ values use $q_i$ as the weights in both the residual and total sums of squares. The covariance calculation treats the different resolution realizations of each target as independent measurements.
The intrinsic lag--mass fits contain 27 target--resolution measurements for each burstiness definition. Six 10/100 measurements and eight 10/30 measurements use the global-maximum fallback because no positive-lag peak meets the prominence criterion. Three of the latter have negative lags: m10.0 at 8$\times$, m9.7 at 4$\times$, and m9.3 at 4$\times$ resolution. Their absolute offsets enter the lag--mass fit, while the positive-lag selection leaves 27 and 24 measurements in the 10/100 and 10/30 lag--period fits, respectively. The fallback points carry small correlation weights and describe weak associations rather than established escape delays.

The phase fits span 1.02--1.89\,Gyr. Six of the 27 fiducial 10/100 winding periods exceed their track duration, as do seven of the 10/30 periods. We retain these formal winding estimates in the pooled fits, but they do not establish completed cycles. This distinction is most relevant for the weakly winding low-mass tracks.

Figure~\ref{fig:overlaid_theta} compares the phase histories obtained with 50, 100, and 150\,Myr boxcar centres and a fixed centroid given by the full-track mean of each logarithmic coordinate. For m12.2, all four choices recover sustained winding, with different rates of phase accumulation. For m9.7, the moving-centre tracks share an extended rise followed by a partial reversal, while the fixed-centroid phase remains much closer to its starting value. The fixed centre measures circulation about the global mean, whereas the moving centre follows circulation relative to the evolving local state. Their difference is especially pronounced when the trajectory drifts substantially through phase space.

The population fits retain a negative period--mass slope for all three moving windows (Figure~\ref{fig:window_slope}). For windows of 50, 100, and 150\,Myr, the halo-mass slopes are $-0.272\pm0.030$, $-0.334\pm0.016$, and $-0.359\pm0.027$, respectively. The corresponding stellar-mass slopes are $-0.176\pm0.022$, $-0.188\pm0.019$, and $-0.216\pm0.018$. The shorter-period trend at higher mass therefore persists across the tested windows, while its slope and normalization vary. Figure~\ref{fig:ratio_window} compares the fitted periods directly: the 50\,Myr window gives shorter periods than the fiducial window across the plotted mass range, while the 150\,Myr window raises the fitted periods at low mass and slightly lowers them at high mass.

We retain 100\,Myr as the common reference window for the main results and use these comparisons to show how the phase measurement responds to the choice of centre. The fitted winding period is distinct from a gas recycling time or the recurrence interval of an individual SFR tracer. Local phase reversals and periods longer than the recorded track can produce poorly constrained mean winding times, even when a linear slope can be fitted. Section~\ref{sec:limitations} discusses these cases and the associated uncertainties and limitations of the pooled sample.


\label{lastpage}

\bsp	
\end{document}